\RequirePackage{rotating}
\documentclass[fleqn,usenatbib]{mnras}

\usepackage{newtxtext}
\usepackage[varg,varvw,smallerops]{newtxmath}
\usepackage[utf8]{inputenc}

\usepackage[T1]{fontenc}
\usepackage{ae,aecompl}
\usepackage{xcolor}
\usepackage{multirow}

\usepackage{graphicx}	

\usepackage{amsmath}	
\usepackage{amssymb}	
\usepackage{hyperref}
\usepackage{adjustbox}
\usepackage{subcaption}
\usepackage{rotating}

\usepackage{savesym}
\savesymbol{tablenum}
\usepackage{siunitx}
\usepackage{ulem}
\restoresymbol{SIX}{tablenum}
\newcounter{ionstage}
\renewcommand{\ion}[2]{\setcounter{ionstage}{#2}%
  \ensuremath{\mathrm{#1\,\scriptstyle\Roman{ionstage}}}}

\DeclareSIUnit\msun{\text{M\ensuremath{_\odot}}}
\DeclareSIUnit\mearth{\text{M\ensuremath{_\oplus}}}
\DeclareSIUnit\lsun{\text{L\ensuremath{_\odot}}}

\title[DEep Spectra of Ionized Regions Database (DESIRED)]{Density biases and temperature relations for DESIRED HII regions}

\author[J. E. M\'endez-Delgado et al.]
{J. E. M\'endez-Delgado$^{1}$ \thanks{E-mail: jemd@uni-heidelberg.de},
C. Esteban$^{2,3}$, J. Garc{\'{\i}}a-Rojas$^{2,3}$,  K. Z. Arellano-C\'ordova$^{4}$,
\newauthor 
K. Kreckel$^{1}$, V. Gómez-Llanos$^{2,3}$, O. V. Egorov$^{1}$, M. Peimbert$^{5}$ and M. Orte-Garc\'ia$^{3}$.
\\
$^{1}$Astronomisches Rechen-Institut, Zentrum f\"ur Astronomie der Universit\"at Heidelberg, Mönchhofstraße 12-14, D-69120 Heidelberg, Germany\\
$^{2}$Instituto de Astrof\'isica de Canarias (IAC), E-38205 La Laguna, Spain\\
$^{3}$Departamento de Astrof\'isica, Universidad de La Laguna, E-38206 La Laguna, Spain\\
$^{4}$Department of Astronomy, The University of Texas at Austin, 2515 Speedway, Stop C1400, Austin, TX 78712, USA\\
$^{5}$Instituto de Astronom\'ia, Universidad Nacional Aut\'onoma de M\'exico, Apdo. Postal 70-264 Ciudad Universitaria, M\'exico\\}

\date{Accepted XXX. Received YYY; in original form ZZZ}

\pubyear{2020}
\begin{document}
\label{firstpage}
\pagerange{\pageref{firstpage}--\pageref{lastpage}}
\maketitle

\begin{abstract}

We present a first study based on the analysis of the DEep Spectra of Ionized REgions Database (DESIRED). This is a compilation of 190 high signal-to-noise ratio optical spectra of \ion{H}{2} regions and other photoionized nebulae, mostly observed with 8-10m telescopes and containing $\sim$29380 emission lines. We find that the electron density --$n_{\rm e}$-- of the objects is underestimated when [\ion{S}{2}] $\lambda6731/\lambda6716$ and/or [\ion{O}{2}] $\lambda3726/\lambda3729$ are the only density indicators available. This is produced by the non-linear density dependence of the indicators in the presence of density inhomogeneities. The average underestimate is $\sim 300$ cm$^{-3}$ in extragalactic \ion{H}{2} regions, introducing systematic overestimates of $T_{\rm e}$([\ion{O}{2}]) and $T_{\rm e}$([\ion{S}{2}]) compared to $T_{\rm e}$([\ion{N}{2}]). The high-sensitivity of [\ion{O}{2}] $\lambda\lambda7319+20+30+31/\lambda\lambda3726+29$ and [\ion{S}{2}] $\lambda\lambda4069+76/\lambda\lambda6716+31$ to density makes them more suitable for the diagnosis of the presence of high-density clumps. If $T_{\rm e}$([\ion{N}{2}]) is adopted, the density underestimate has a small impact in the ionic abundances derived from optical spectra, being limited to up to $\sim$0.1 dex when auroral [\ion{S}{2}] and/or [\ion{O}{2}] lines are used. However, these density effects are critical for the analysis of infrared fine structure lines, such as those observed by the JWST in local star forming regions, implying strong underestimates of the ionic abundances. We present temperature relations between $T_{\rm e}$([\ion{O}{3}]), $T_{\rm e}$([\ion{Ar}{3}]), $T_{\rm e}$([\ion{S}{3}]) and $T_{\rm e}$([\ion{N}{2}]) for the extragalactic \ion{H}{2} regions. We confirm a non-linear dependence between $T_{\rm e}$([\ion{O}{3}])-$T_{\rm e}$([\ion{N}{2}]) due to a more rapid increase of $T_{\rm e}$([\ion{O}{3}]) at lower metallicities. 
\end{abstract}




\section{Introduction}
\label{sec:introduction}

The determination of chemical abundances from emission line spectra of ionized nebulae is an essential tool for studying the chemical composition and evolution of the Universe, from the Milky Way to high-redshift galaxies. In ionized nebulae, the total abundance of heavy elements, the metallicity, is traced by the O/H abundance, as it comprises $\sim 55$ per cent of the total metal content \citep{Peimbert:2007}. This information can be used to explore the nucleosynthesis of chemical elements and the galaxy formation and evolution. In fact, the mean metallicity of the galaxies and the shape of radial abundance gradients depend on their masses, the star formation history and the relative importance of the gas inflows/outflows across their discs \citep[e.g.][]{Tinsley:1980,Prantzos:2008,Matteucci:2014}. 

The chemical abundances of elements heavier than He can be derived from bright collisionally excited lines (CELs) in the emission line spectra of ionized nebulae. In the optical range, the emissivity of CELs is exponentially dependent on the electron temperature, $T_{\rm e}$, being a critical physical parameter for obtaining accurate abundance values. This is the basis of the so-called direct method for determining  chemical abundances \citep[e.g.][]{Dinerstein:1990,Peimbert:2017,Perez-Montero:2017}. Moreover, recently \citet{mendezdelgado:2023} demonstrated the presence of temperature inhomogeneities within the highly ionized gas as theorized by \citet{Peimbert:1967}. The existence of such spatial temperature variations introduces a systematic bias towards lower abundances that can reach errors as high as $\sim 0.5$ dex in the O/H abundance \citep{mendezdelgado:2023}. On the other hand, the fine structure CELs in the infrared (IR) range that arise from atomic transitions of low energy levels ($\Delta$ E$<<$ 1 eV) have a smaller temperature-dependence \citep{Osterbrock:2006}. However, in these cases the electron density, $n_{\rm e}$, is a fundamental parameter to accurately determine chemical abundances, as the critical densities of these low-energy levels are smaller than those involved in the emission of optical CELs \citep{Osterbrock:2006}.

With the advent of optical spectroscopic surveys using large Integral Field Units (IFU), data for myriads of \ion{H}{2} regions in large samples of external spiral galaxies have become available \citep[e.g.][]{sanchez2012, bryant2015, bundy2015, emsellem2022}. However, it is common that most of the spectra of extragalactic \ion{H}{2} regions in these surveys are not deep\footnote{With the concept of "deep spectrum" we mean a long-exposure time spectrum with a high signal-to-noise ratio where the main purpose is the detection of weak emission lines, such as auroral CELs or RLs.} enough to detect the faint auroral CELs necessary to determine $T_{\rm e}$ or the even fainter recombination lines (RLs) of heavy-element ions. When the gas temperature is not available one has to rely on the so-called strong-line methods to estimate the gas-phase metallicity, which are based on calibrations of the O/H ratio ---the  proxy for metallicity when analyzing nebular spectra--- built with observed intensity ratios of bright nebular CELs \citep[e.g.][]{Pagel:1979,pilyugin2010, pilyugin2012, marino2013, pilyugin2016} or on photoionization models \citep[e.g.][]{mcgaugh1991, Kewley2002, Kobulnicky2004, tremonti2004}. Comparing the different calibrations available in the literature, one can find very large differences between the O/H ratios for the same set of observations, differences that can amount to 0.2-0.7 dex \citep[e.g.][]{kewley2008, lopez-sanchez2012,Groves:2023}. From the available strong-line methods, only those of \citet{pena-guerrero:2012} take into account the presence of temperature inhomogeneities. 

The large amount of data generated by big surveys that one can gather from the literature permit us to explore, constrain and minimize the effects of statistical errors in the estimate of metallicities of \ion{H}{2} regions in a given galaxy or a group of similar galaxies \citep[e.g][]{Sanchez:2015,Ho:2019,Kreckel:2019,Metha:2021}. However, only detailed studies of deep spectra of \ion{H}{2} regions allow us to adequately explore and constrain the effects of systematic errors in the determination of physical conditions and ionic and total abundances. On this matter, there are previous works dedicated to collect auroral CELs from the most commonly studied ions ([\ion{O}{2}], [\ion{O}{3}], [\ion{S}{2}], [\ion{S}{3}], [\ion{N}{2}]) \citep{pilyugin2012,Croxall:2016,Berg:2020,Rogers:2021,Rogers:2022, Zurita:2021}. However, with some notable exceptions where recombination lines were considered \citep{Peimbert:2005,Guseva:2011,pena-guerrero:2012-2,Valerdi:2019,Skillman:2020}, most previous studies are limited to the CELs of few ions, which do not provide the complete picture of the physics of the ionized gas.

Since the beginning of this century, our group has gathered a large number of intermediate spectral resolution longslit or high spectral resolution echelle spectra for a large number of Galactic and extragalactic \ion{H}{2} regions as well as Galactic planetary nebulae (PNe) and ring nebulae (RNe) around massive Wolf-Rayet and Of stars. This collection of data is what we call DESIRED (DEep Spectra of Ionized Regions Database, see Section~\ref{sec:DESIRE} for references and a description of the data). The vast majority of the data have been obtained with large-aperture (8-10m) telescopes and the observations were designed to detect very faint emission lines. As a result of the remarkable signal-to-noise ratio of our collection of nebular spectra, each individual object counts with tens or even hundreds of emission lines, showing good measurements of all or some of these: (a) one or several faint $T_{\rm e}$-sensitive auroral CELs, (b) several density indicators based on the intensity ratios of CELs, (c) RLs of one or some heavy-element ions and (d) sets of rare faint lines as those of [\ion{Fe}{2}] and/or [\ion{Fe}{3}] or fluorescence ones, useful for detailed studies on the internal physics of the ionized gas.

The DESIRED papers seek to analyze global properties of the ionized gas in unprecedented detail, detecting and describing phenomena that have --or might have-- an impact on interpretations of large-scale studies based on solid observational evidence. The present work is dedicated to the study of physical conditions ($T_{\rm{e}},n_{\rm{e}}$) of the ionized gas, including information about their internal structures and the temperature relations. The prescriptions, warnings and relations of this study are intended to consider different types of ionized regions and can be used both in studies of individual objects and in large-scale studies.

\section{Description of DESIRED}
\label{sec:DESIRE}

\begin{figure*}
\centering
\includegraphics[width=\textwidth]{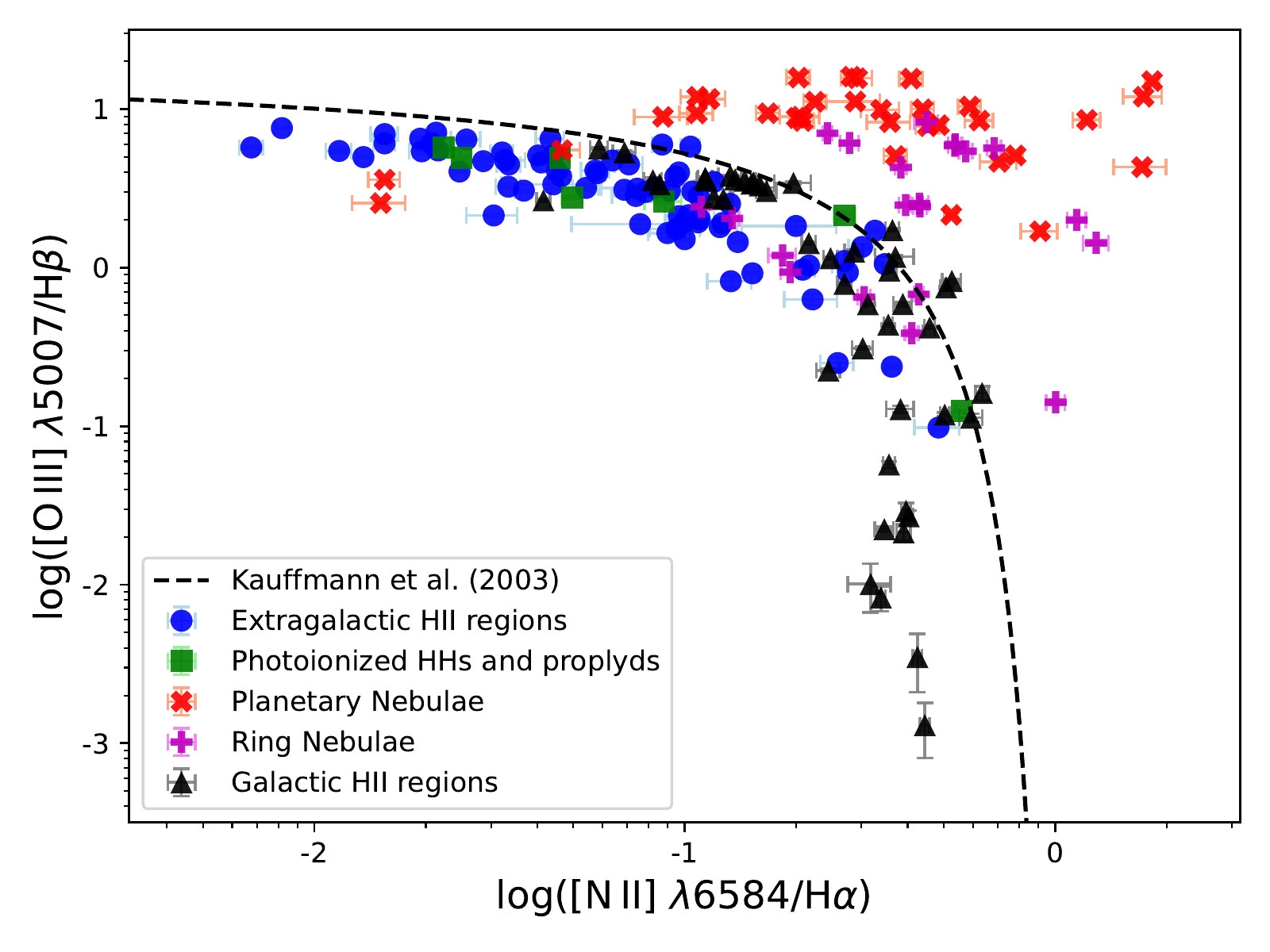}
\caption{ BPT diagram of the DESIRED spectra. The dashed line represents the boundaries between star-forming regions (to the left and below the line) and regions with harder ionizing sources (generally associated to Active Galactic Nuclei) \citep{Kauffmann:2003}.} 
\label{fig:BPT}
\end{figure*}

DESIRED comprises a set of 190 spectra, 72 of them correspond to 68 extragalactic \ion{H}{2} regions, 56 spectra of 41 Galactic \ion{H}{2} regions, 34 Galactic PNe, 21 spectra of 7 Galactic RNe as well as 6 spectra of 5 photoionized Herbig-Haro objects (HHs) and 1 protoplanetary disk (proplyd) of the Orion Nebula. 
References to the spectra are shown in Tables \ref{tab:photoionizedHHs}, \ref{tab:RNs_references}, \ref{tab:extragalactic_references}, \ref{tab:PNes_references} and \ref{tab:Galactic_references}. 
All the spectra have been observed by our group except those of the Galactic PNe IC~418, IC~2501, IC~4191 and NGC~7027 \citep{Sharpee:2003,Sharpee:2007}. We decided to include these data in DESIRED as they show an analogous level of depth and quality as the rest of the objects included in Table~\ref{tab:PNes_references} \citep[see the comparative analysis performed  by][]{Rodriguez:2020}.
The database contains 29380 emission line detections, associated with 2486 transitions of 148 ionic species\footnote{In this context, permitted and forbidden transitions are considered independently. For instance, [\ion{O}{3}] and \ion{O}{2} are counted as different ionic species.}. Of that total number of detections, 8715 are forbidden lines, while 18986 are permitted ones and 1679 remain unidentified or with doubtful identifications. From the detected permitted lines, 7836 are associated to metals. A number of 851 forbidden lines correspond to $T_{\rm e}$-sensitive auroral transitions [\ion{O}{2}] $\lambda \lambda 7319+20+30+31$, [\ion{S}{2}] $\lambda \lambda 4069+76$, [\ion{N}{2}] $\lambda 5755$, [\ion{S}{3}] $\lambda 6312$, [\ion{Ar}{3}] $\lambda 5192$ and [\ion{O}{3}] $\lambda 4363$, that can be used for $T_{\rm e}$ determinations. 

The remarkably high signal-to-noise ratio of the DESIRED spectra can be verified in any of the  published reference articles. We can highlight fig.~1 of \citet{Esteban:2014b} or fig.~7 of \citet{mendezdelgado:2021a} in the case of the Orion Nebula; fig.~3 of \citet{dominguez:2022} for extragalactic \ion{H}{2} regions in the Magellanic Clouds; fig.~3 of \citet{Esteban:2016} for the RN NGC~6888 and fig.~4 of \citet{garciarojas:2018} for a group of PNe. 

The observations have been taken from 2002 to date with the spectrographs and telescopes shown in Table~\ref{tab:telescopes_refs}\footnote{The spectra of \citet{Sharpee:2003,Sharpee:2007} were taken between 2001 and 2003.}. The spectra were reduced and calibrated manually following a consistent procedure, using IRAF routines \citep{Tody:1993}, Python codes and some tasks from the ESO UVES pipeline \citep{Ballester:2000}. The flux, wavelength and FWHM of the lines were measured manually using the IRAF task SPLOT, individually estimating the continuum. 

Echelle spectra were not corrected from telluric emissions, since the slit does not usually cover sky areas. However, the high spectral resolution permits us to separate the doppler shifted nebular emissions from the sky contaminations. Sky-blended lines are identified and their use has been ruled out in this work. In most of the spectra the telluric absorption bands were not corrected. This potentially affects several wavelength ranges as $\lambda \lambda 7600-7700$\AA, $\lambda \lambda 9000-10000$\AA, where atmospheric O$_2$ and H$_2$O bands are strong and dense \citep{Stevenson:1994}. UVES spectra may have optical reflections within the second dichroic of the blue arm ($\lambda \lambda 3750-4995$\AA). The wavelength position of these spurious ``ghosts'' can be determined directly from the echellograms as they cross the different observed orders. The use of these lines is also discarded along with those with detected individual spurious effects. 

Intermediate spectral resolution spectra ($R \sim$ 3000-4000) were mostly taken with long-slit two-arms spectrographs. We verified the accuracy of the relative flux calibration between the bluest and reddest wavelength ranges. The sky emission was removed in the case of the smaller angular size nebulae (most of the Galactic \ion{H}{2} regions observed with OSIRIS at the 10.4m GTC telescope, the extragalactic ones and PNe), this was not possible in the case of IC~5146 and M43, extended Galactic \ion{H}{2} regions observed with ISIS at the 4.2m WHT telescope. 

The spectra were corrected for interstellar extinctions and underlying stellar absorptions following the iterative process described by \citet{lopez-sanchez:2006}, which is based on the results of \citet{Mazzarella:1993} and the observed \ion{H}{1} Balmer and Paschen decrements, when available. No corrections for underlying stellar absorptions were made to the \ion{He}{1} lines. However, the Galactic objects did not require such corrections \citep{mendezdelgado:2020}. The detailed procedure for each object is described in the reference articles.

In Fig.~\ref{fig:BPT}, we show a BPT diagram \citep{Baldwin:1981} of all DESIRED spectra distinguishing their corresponding types of nebulae. The dashed line indicates the separation between the \ion{H}{2} regions and active galactic nuclei (AGNs) as defined by the empirical equation~(1) of \citet{Kauffmann:2003}. All Galactic and extragalactic \ion{H}{2} regions as well as photoionized HH objects and the proplyd are located in the zone of star forming regions. This is consistent with gas photoionized by O or early B type stars. PNe and RNe are present both in the star forming region zone and in the area usually associated with AGNs. RNe associated with Wolf-Rayet stars are located within the AGN zone whereas those associated to Of stars are together with the \ion{H}{2} regions. This is due both to a harder ionizing spectrum from Wolf-Rayet stars and to a larger contribution from shocks, associated with stellar feedback \citep{Esteban:2016}. Most PNe are located well above the \ion{H}{2} regions line \citep[e.g.][]{Kniazev:2008}, as expected from their harder ionizing sources. However, Abell~46, Abell~63 and Ou5 \citep{Corradi:2015} fall within the area of \ion{H}{2} regions. This interesting result seems linked to the fact that these 3 regions have the largest abundance discrepancy factor (ADF) between the O$^{2+}$/H$^{+}$ abundances derived with both RLs and CELs of the whole sample. This is in agreement with the scenario where these PNe contain metal-rich cold inclusions within the ionized gas, enhancing the emission of the \ion{H}{1} RLs, as proposed by several authors \citep{Corradi:2015,garciarojas:2022}.

The metallicity range, expressed by 12+log(O/H), determined from CELs and assuming no temperature fluctuations, covered by the sample objects goes from 7.72 and 8.70 in the case of \ion{H}{2} regions (including both Galactic and extragalactic) and from 7.76 and 8.80 in the case of PNe. It should be noted that, due to the requirements of DESIRED observations (relatively bright spectra and high probability of detecting RLs of heavy element ions), the number of \ion{H}{2} regions with 12+log(OH) below 8.0 is rather limited. A drawback that could be corrected with observations with the future very large aperture telescopes.

\section{Physical conditions}
\label{sec:pc}

The determination of the chemical composition of photoionized nebulae requires, as a first step, accurate calculations of $n_{\rm e}$ and $T_{\rm e}$. DESIRED objects potentially comprise a wide range of densities from $n_{\rm e} \sim 10^{2} \text{ cm}^{-3}$ for some extragalactic \ion{H}{2} regions to $n_{\rm e}>10^{5} \text{ cm}^{-3}$ for HHs and the photoevaporating proplyd 170-337 of the Orion Nebula. Therefore, it is possible to explore relations between several density diagnostics. To derive $n_{\rm e}$, we test the [\ion{S}{2}] $\lambda 6731/\lambda 6716$, [\ion{O}{2}] $\lambda 3726/\lambda 3729$, [\ion{Cl}{3}] $\lambda 5538/ \lambda 5518$, [\ion{Fe}{3}] $\lambda 4658/ \lambda 4702$ and [\ion{Ar}{4}] $\lambda 4740/ \lambda 4711$ line intensity ratios. To solve the statistical equilibrium equations, we use PyNeb 1.1.13 \citep{Luridiana:2015} and the transition probabilities and collision strengths given in Table~\ref{tab:atomic_data}. We use the {\it getCrossTemDen} task of PyNeb to simultaneously derive $T_{\rm e}$ and $n_{\rm e}$, cross matching the aforementioned density diagnostics with the $T_{\rm e}$-sensitive [\ion{N}{2}] $\lambda 5755/ \lambda 6584$, [\ion{O}{3}] $\lambda 4363/ \lambda 5007$, [\ion{Ar}{3}] $\lambda 5192/ \lambda 7135$ and [\ion{S}{3}] $\lambda 6312/ \lambda 9069$ line intensity ratios. Finally, we average the density values obtained with each cross-match to obtain a representative value of $n_{\rm e}$ for each tested density diagnostic. For the objects with reliable detections of density diagnostics but not of the aforementioned temperature diagnostics, we derive the density by assuming $T_{\rm e}=10000 \pm 1000 \text{ K}$. There are only 5 objects in this last case: three slit positions of M\,43 \citep[observed by][]{Simon-Diaz:2011}, two \ion{H}{2} regions of M\,33 and another two of NGC~300 \citep[observed by][]{Toribio:2016}. The temperature dependence of the density diagnostics is negligible in these cases. All these objects show $n_{\rm e} <$ 1000 cm$^{-3}$. We analyze the $n_{\rm e}$ determinations in Section~\ref{sec:density_structure}, defining a clear criteria to adopt its final representative value for each object. Finally, once $n_{\rm e}$ is fixed, $T_{\rm e}$ is calculated by using the {\it getTemDen} task of PyNeb.

The near infrared lines [\ion{S}{3}] $\lambda 9069, 9531$ can be  affected by the telluric absorption bands \citep{Stevenson:1994,Noll:2012}, potentially introducing spurious results in $T_{\rm e}$([\ion{S}{3}]) if there is no strict control over this issue. Usually, the most affected line is [\ion{S}{3}] $\lambda 9531$, which lies in a wavelength zone more contaminated by telluric absorption bands, although this effect may vary depending on internal gas velocities, as in the Orion Nebula, where [\ion{S}{3}] $\lambda 9069$ is usually the most contaminated one \citep{Baldwin:1991,mendezdelgado:2021a}. We have tried to have a strict control on the telluric absorptions, discarding the use of the affected lines, in order to avoid spurious $T_{\rm e}$([\ion{S}{3}]) determinations. As a second check, in those objects where both lines were detected, we test the [\ion{S}{3}] $\lambda 9531/ \lambda 9069$ line intensity ratio. Both lines arise from the same atomic $^1D_2$ upper level, therefore their relative intensity must be equal to 2.47 \citep{FroeseFischer:2006}, regardless of the physical conditions of the gas. We discard those objects where [\ion{S}{3}] $I(\lambda 9531)/I(\lambda 9069)$ > 2.47 beyond the observational uncertainties, as it indicates a possible effect on [\ion{S}{3}] $\lambda 9069$. However, since no telluric corrections of any kind were made, except in the \cite{mendezdelgado:2021a,mendezdelgado:2021b,mendezdelgado:2022b} spectra, we cannot guarantee that {\it all} DESIRED spectra are free of telluric absorption effects on their [\ion{S}{3}] $\lambda 9069, 9531$ lines.

Although the [\ion{O}{2}] $\lambda \lambda$7319+20+30+31/$\lambda \lambda$3726+29 and/or  [\ion{S}{2}] $\lambda \lambda$4069+76/$\lambda \lambda$6716+31 line ratios were measured in many objects, we prefer not using them in the determination of the final adopted $T_{\rm e}$ of each object. As we discuss in Section~\ref{subsec:TO2_TS2}, those line ratios are very sensitive to $n_{\rm e}$ and 
the inferred $T_{\rm e}$([\ion{O}{2}]) and $T_{\rm e}$([\ion{S}{2}]) are affected by the presence of high-density clumps within the ionized nebulae, as will be discussed in Section~\ref{subsec:TO2_TS2}.

\section{Photoionization models}
\label{sec:phot_models}

To explore the theoretical temperature relations in the absence of temperature fluctuations ($t^2=0$), we select the photoionization models of giant \ion{H}{2} regions from the Mexican Million Models database\footnote{https://sites.google.com/site/mexicanmillionmodels/} \citep{Morisset:2015}, built for the BOND project \citep{ValeAsari:2016} using Cloudy v17.01 \citep{Ferland:2017}. We adopt the same selection criteria as \citet{Amayo:2021}, which considers startburst ages lower than 6 Myr, ionization-bounded and density-bounded selected by a cut of 70 per cent of the H$\beta$ flux and a selection of realistic N/O, $U$ and O/H values \citep{ValeAsari:2016}. We also adopt the same BPT-cut defined by \citet{Amayo:2021} in their equation~(3). Since we do not intend to study the temperature relations in PNe or RNe beyond analyzing their differences with the results of \ion{H}{2} regions, we do not adopt any additional set of models.

\section{The density structure of ionized nebulae}
\label{sec:density_structure}

\begin{figure*}
\begin{minipage}{\textwidth}
\centering
\includegraphics[width=.48\textwidth]{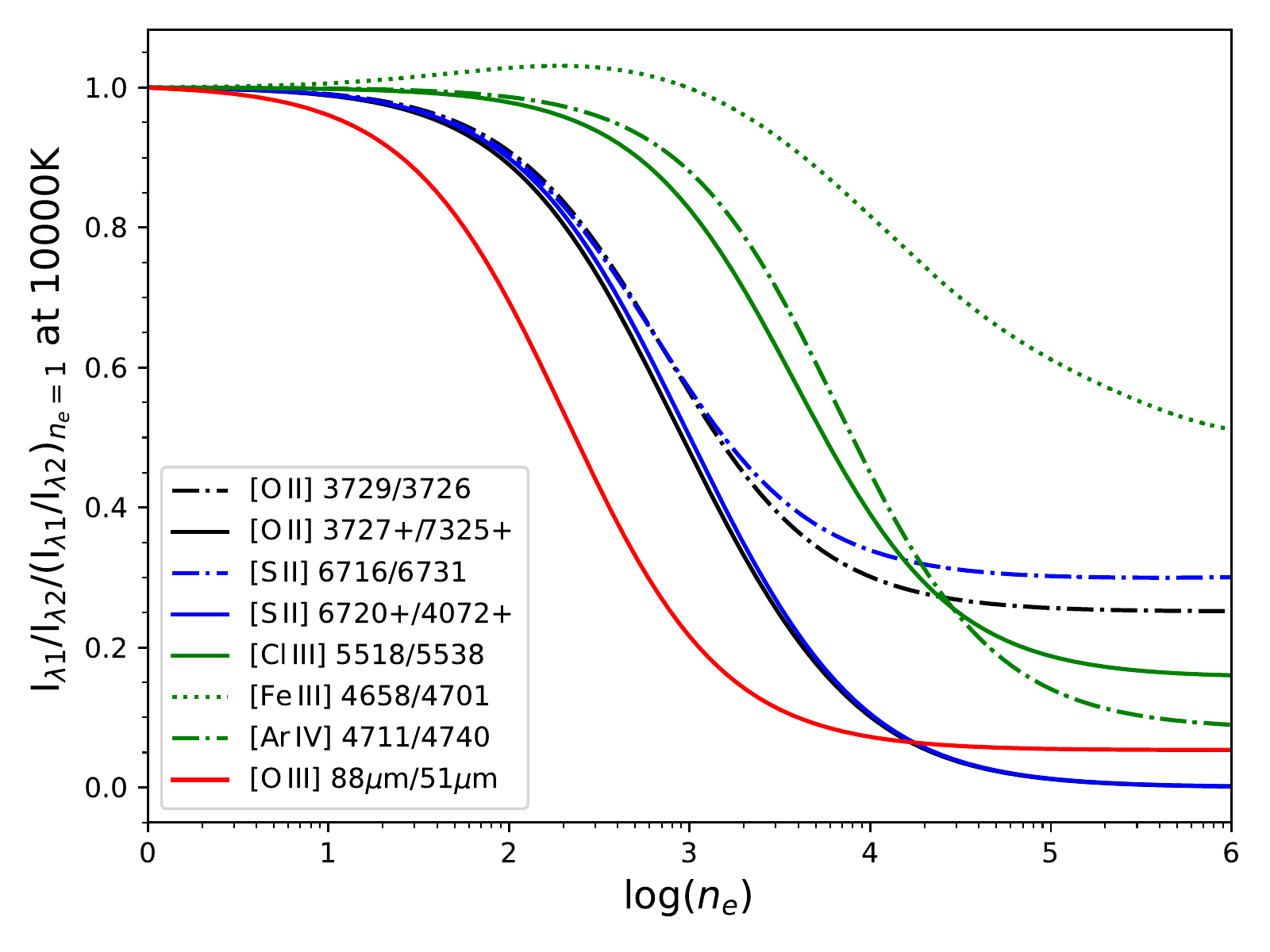}
\includegraphics[width=.48\textwidth]{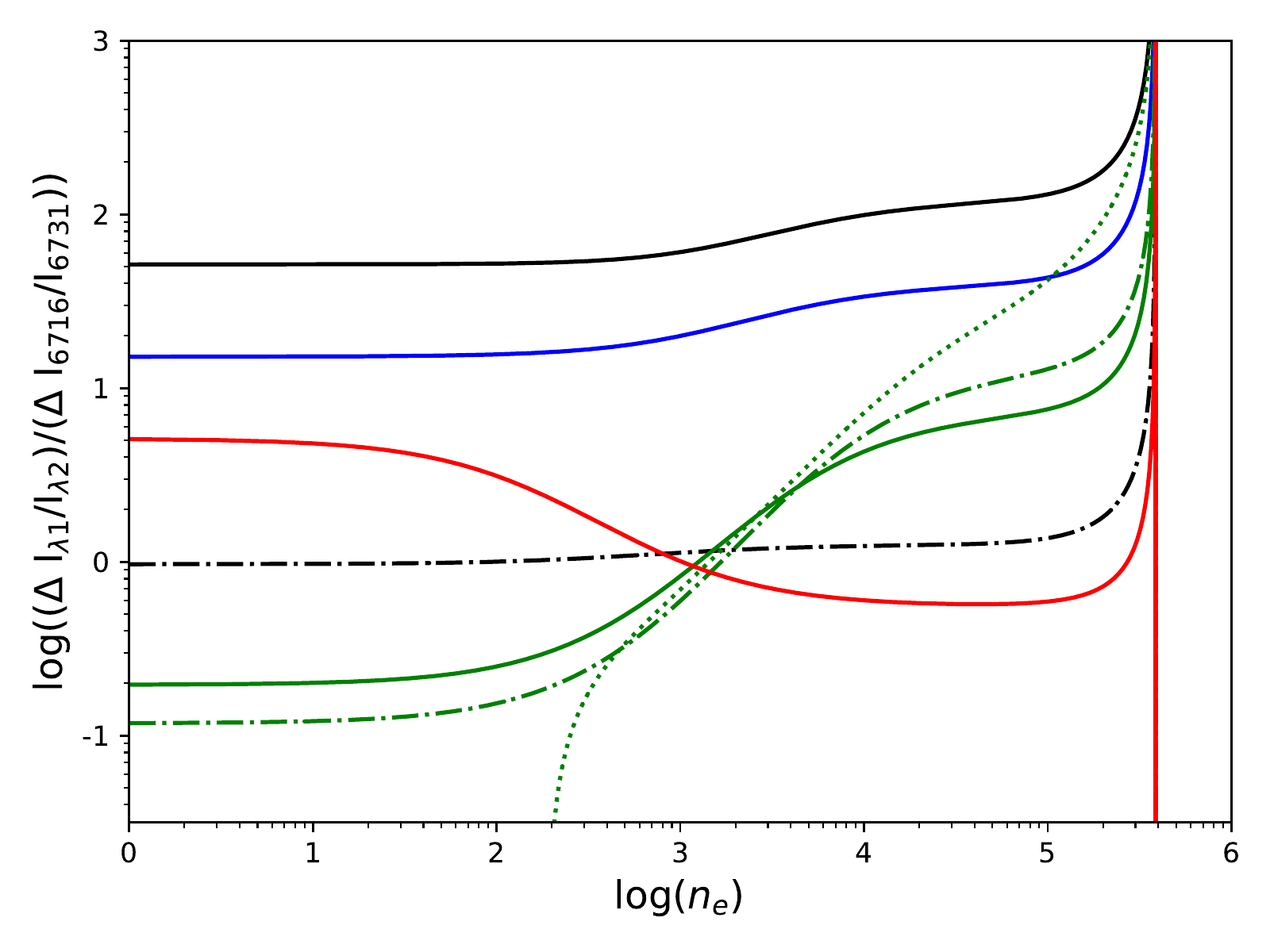}
\end{minipage}
\caption{Left panel: Dependence of different line intensity ratios with the electron density --$n_{\rm e}$--, considering a $T_{\rm e}=10000\text{ K}$ and the atomic data from Table~\ref{tab:atomic_data}. The line intensity ratios have been normalized with the expected values at $n_{\rm e}=1 \text{ cm}^{-3}$. Right panel: comparison between the density-sensitivity of the different line intensity ratios and that of [\ion{S}{2}] $\lambda 6716/\lambda 6731$, considering a $T_{\rm e}=10000\text{ K}$. The density sensitivity is defined as $\frac{\Delta\left( I_{\lambda_1}/I_{\lambda_2}\right)}{\Delta n_{\rm e}}$. When $n_{\rm e}\approx 10^{5.3} \text{ cm}^{-3}$, $\frac{\Delta\left( I_{\lambda_{6716}}/I_{\lambda_{6731}}\right)}{\Delta n_{\rm e}} \approx 0$, inducing an asymptote.
} 
\label{fig:density_diagnostics}
\end{figure*}

Several line intensity ratios emitted from atomic levels close in energy are sensitive to $n_{\rm e}$ due to their different collisional excitation and  deexcitation rates. As shown in left panel of Fig.~\ref{fig:density_diagnostics}, the density dependence of several optical and infrared line ratios is not linear and they have different ranges of validity. The [\ion{S}{2}] $\lambda 6731/\lambda 6716$ line intensity ratio is one of the most used density diagnostics in the literature due to its observational accessibility. Therefore, it will be used in this work as the main reference in the comparisons with other density diagnostics. In order to estimate the utility of a density diagnostic, it is convenient to study its sensitivity. We define this quantity as the variation of the line intensity ratio with $n_{\rm e}$, being mathematically represented with the derivative of the diagnostic with respect to the density. 

The sensitivity of the $n_{\rm e}$-diagnostics and, in general, the relationship between the inferred physical conditions and the observed line intensity ratios depend on the atomic transition probabilities and collision strengths. Several studies have a\-na\-ly\-zed the behavior of these parameters with optical spectra \citep{stasinska:2013,JuandeDios:2017,Morisset:2020,JuandeDios:2021,Mendoza:2023}. After detecting and discarding discrepant data sets, \citet{Morisset:2020} and \citet{Mendoza:2023} estimate uncertainties of $\sim 10$ per cent in the radiative atomic rates for ions like [\ion{O}{2}], [\ion{S}{2}], [\ion{Fe}{3}], [\ion{Cl}{3}] and [\ion{Ar}{4}]. We minimize the presence of errors in the atomic data by considering the results of the aforementioned studies, avoiding the use of discrepant atomic data sets. However, the impact of potential errors cannot be completely neglected, since the available calculations are few in number in the case of some ions.

A comparison of the relative sensitivity of different density diagnostics with respect to the widely used [\ion{S}{2}] $\lambda 6731/\lambda 6716$ is shown in the right panel of Fig.~\ref{fig:density_diagnostics}. The first notable result is that [\ion{O}{2}] $\lambda 3726/\lambda 3729$ and [\ion{S}{2}] $\lambda 6731/\lambda 6716$ are equivalent diagnostics in terms of sensitivity, without significant differences. This figure also shows that [\ion{Cl}{3}] $\lambda 5538/ \lambda 5518$, [\ion{Fe}{3}] $\lambda 4658/ \lambda 4702$ and [\ion{Ar}{4}] $\lambda 4740/ \lambda 4711$ are not sensitive diagnostics when $n_{\rm e}<10^3 \text{ cm}^{-3}$. However, beyond this threshold, the aforementioned diagnostics are comparatively more and more sensitive, since [\ion{S}{2}] $\lambda 6731/\lambda 6716$ decrease its sensitivity. In contrast [\ion{O}{3}] $\lambda 88\mu\text{m}/\lambda 51\mu\text{m}$ shows higher sensitivity when $n_{\rm e}<10^3 \text{ cm}^{-3}$, but beyond this value, its sensitivity decreases to a greater extent than that of [\ion{S}{2}] $\lambda 6731/\lambda 6716$. When $n_{\rm e}\approx 10^{5.3} \text{ cm}^{-3}$, then $\Delta\left( I_{\lambda_{6716}}/I_{\lambda_{6731}}\right)/\Delta n_{\rm e} \approx 0$. On the other hand, at fixed temperature, [\ion{O}{2}] $\lambda \lambda$7319+20+30+31/$\lambda \lambda$ 3726+29 and [\ion{S}{2}] $\lambda \lambda$4069+76/$\lambda \lambda$6716+31 have a very high density-sensitivity over the entire range from $10^2 \text{ cm}^{-3}<n_{\rm e}<10^6 \text{ cm}^{-3}$. These line intensity ratios will be discussed in detail in Section~\ref{subsec:TO2_TS2}. 

If the nebulae have homogeneous density, the different diagnostics should converge to the same value if they are in their density-sensitive range. However, the emissions of the different ions can come from different volumes of ionized gas and the nebulae may contain density inhomogeneities. In fact, the presence of high-density clumps has been revealed by high-resolution images in several nearby photoionized nebulae \citep[see e.~g.][]{Borkowski:1993,Odell:1996, Odell:2002}. Besides filamentary structures, jets of matter and gas flows due to photoionization are capable of compressing the gas, increasing the local density. Within the \ion{H}{2} regions, ongoing star formation can give rise to HHs \citep{Herbig:1950,Haro:1952} and proplyds \citep{Odell:1993}, which are associated with clumps of ionized gas that can reach density values of up to $\sim 10^6 \text{ cm}^{-3}$ \citep{Henney:1999}. Although the high-density inclusions may represent a small fraction of the gas volume, the different collisional deexcitation rates of the diagnostics can bias them towards higher or lower values depending on their particular density-sensitivity regime. Moreover, since the refractory elements such as Fe are mostly depleted into dust grains within the ionized environments, the [\ion{Fe}{3}] $\lambda 4702/\lambda 4658$ ratio may be more easily detected in shock-compressed higher-density areas where the dust destruction is taking place, such as HH objects \citep{mendezdelgado:2021a}.

\begin{figure*}
\begin{minipage}{\textwidth}
\centering
\includegraphics[width=.48\textwidth]{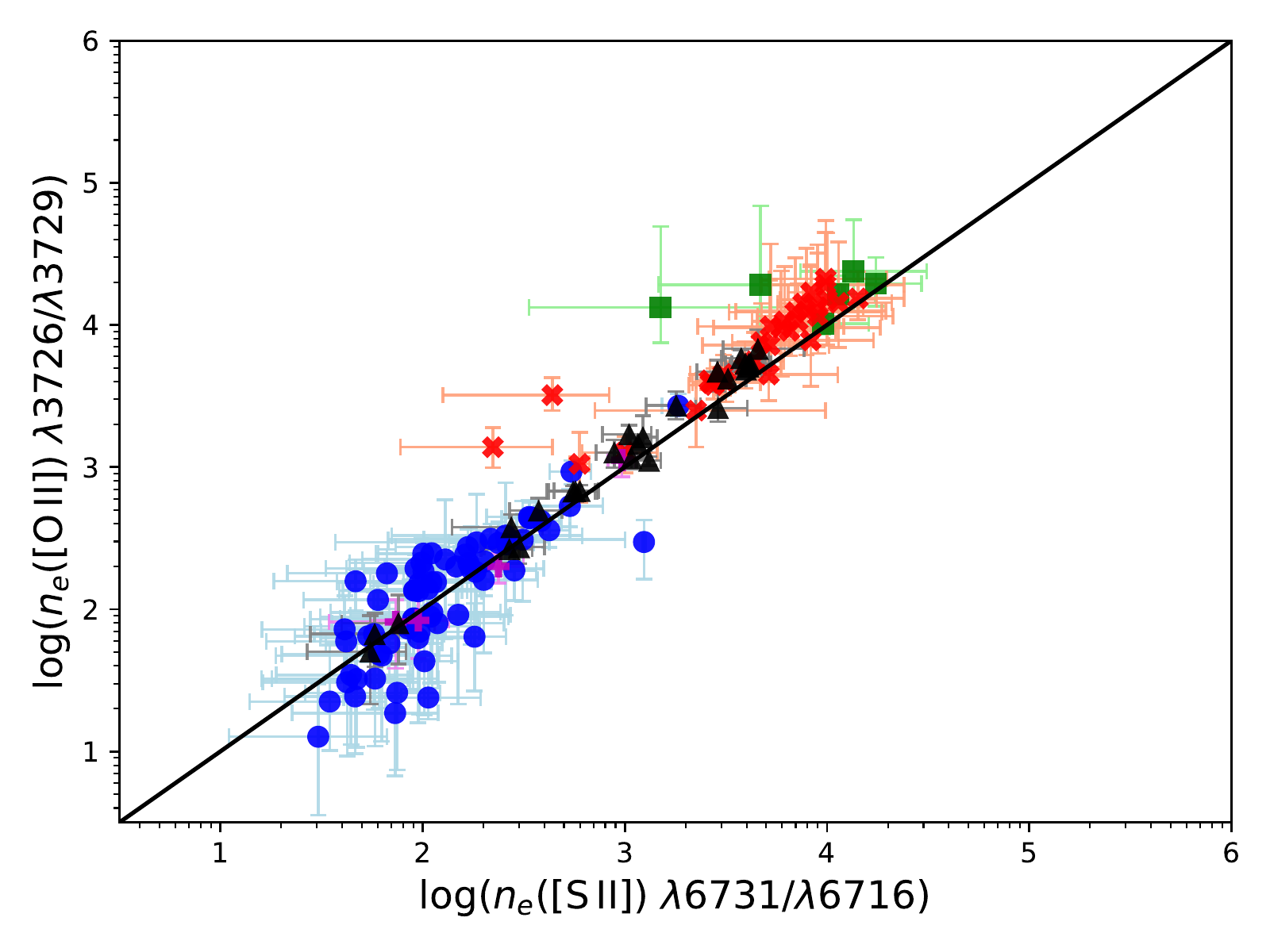}
\includegraphics[width=.48\textwidth]{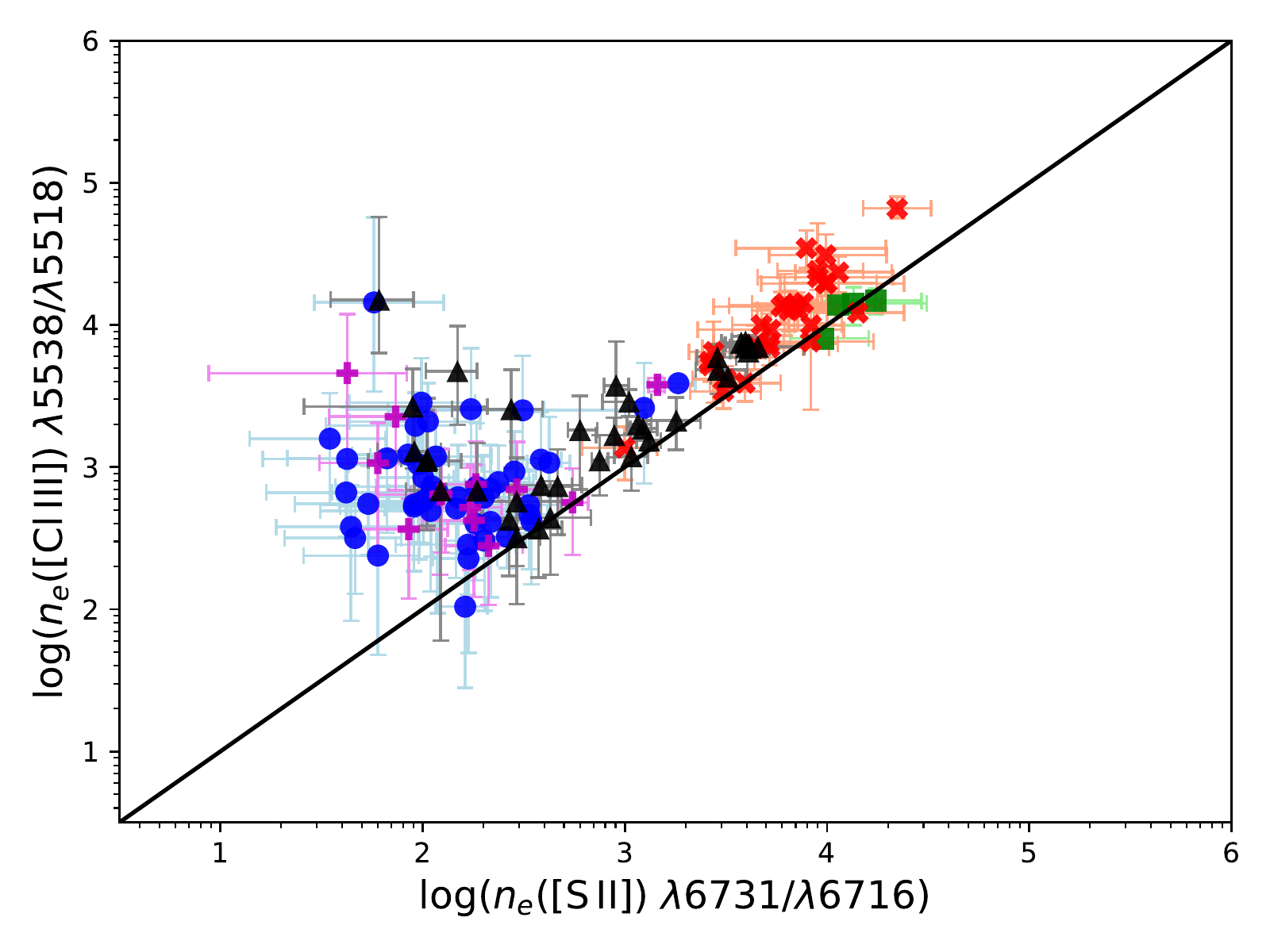}
\includegraphics[width=.48\textwidth]{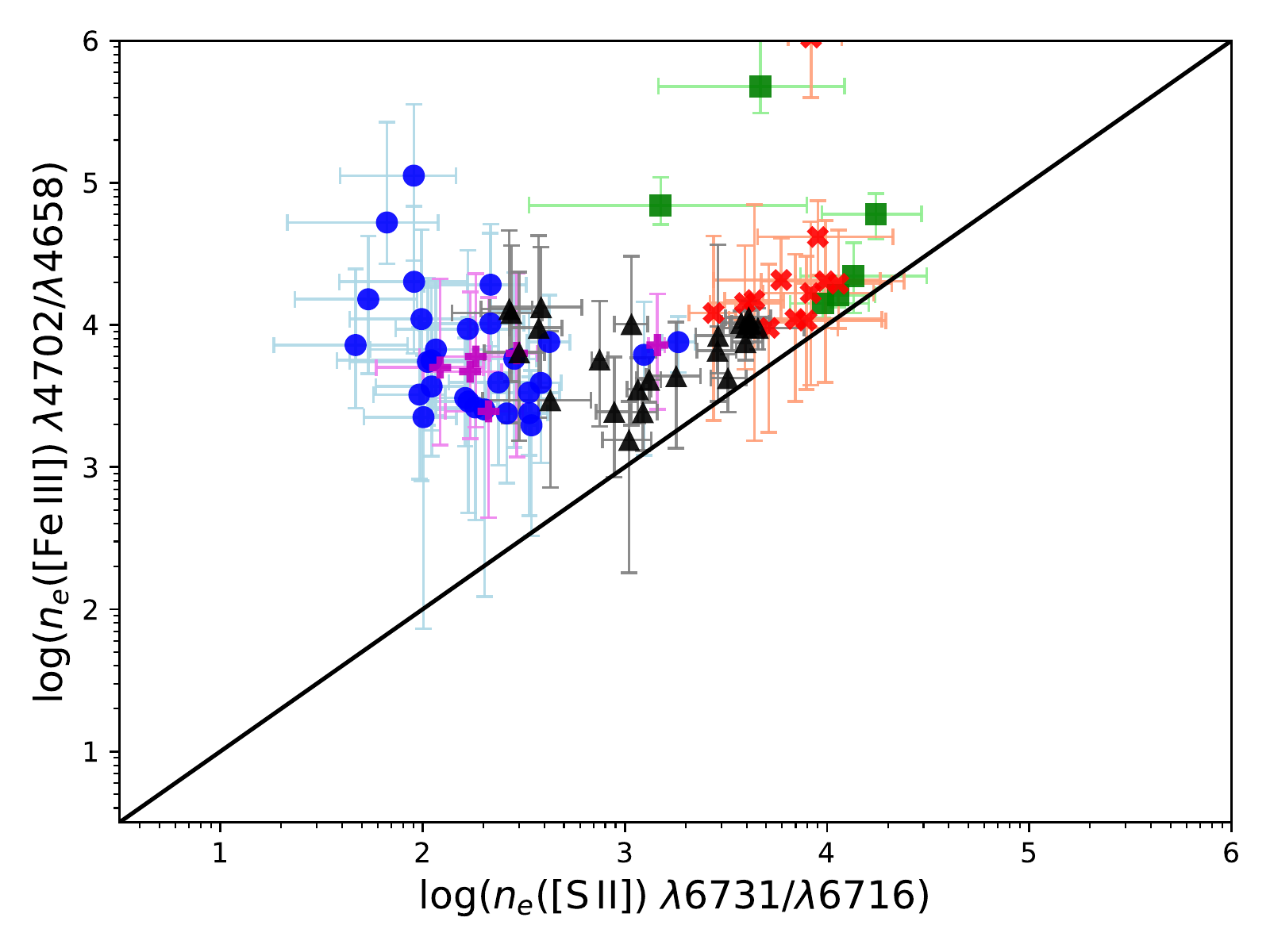}
\includegraphics[width=.48\textwidth]{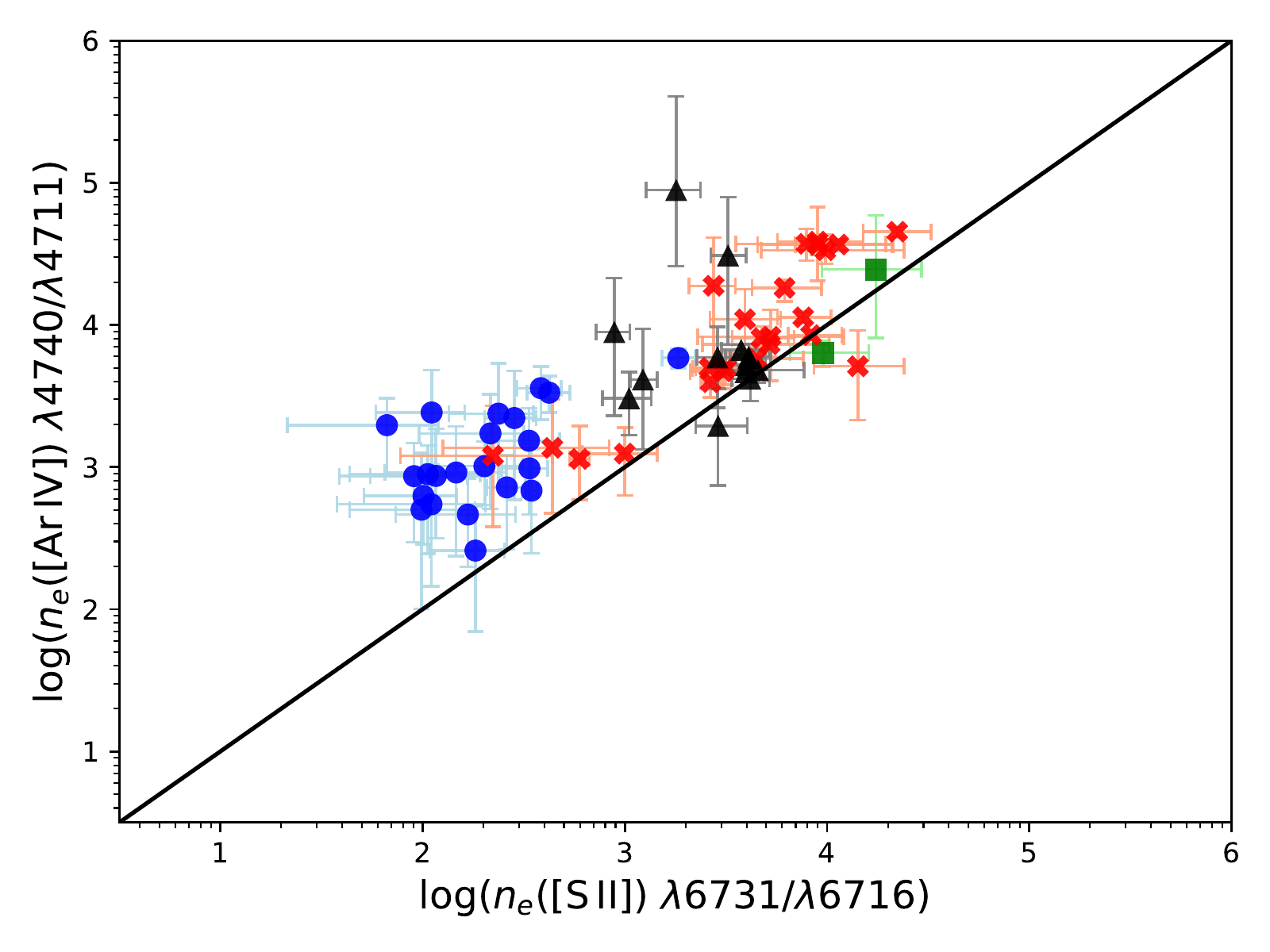}
\includegraphics[width=.48\textwidth]{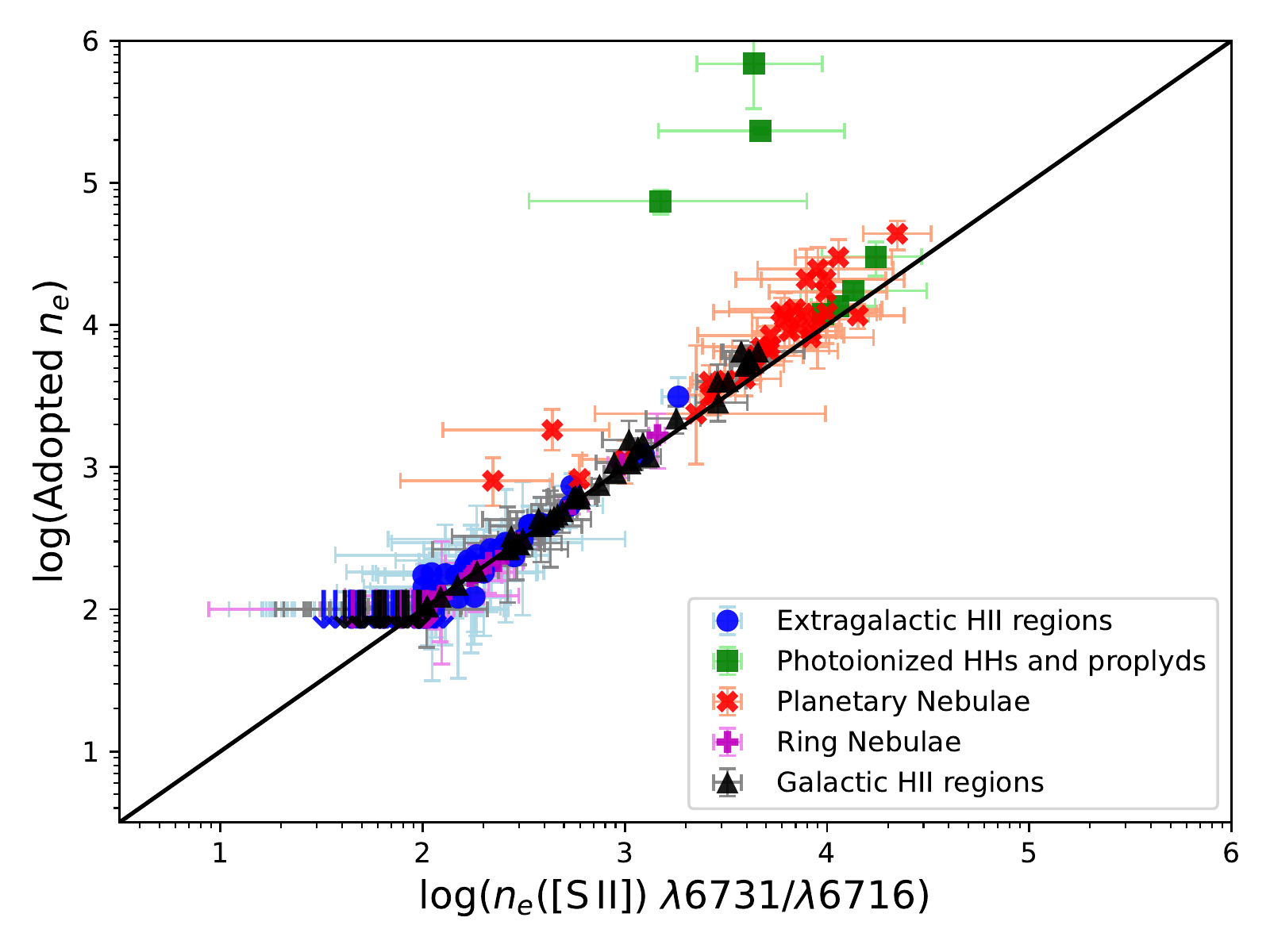}
\end{minipage}
\caption{Comparison between the density derived from the [\ion{S}{2}] $\lambda 6731/\lambda 6716$ line intensity ratio and from the rest of the diagnostics, including the average density estimated from the adopted criteria (bottom panel). The solid line represents a 1:1 linear relation. Down arrows indicate the upper limit when the value is at the low density limit ($n_{\rm e}<100$ cm$^{-3}$).} 
\label{fig:densities_plot}
\end{figure*}

Fig.~\ref{fig:densities_plot} compares 
$n_{\rm e}$([\ion{S}{2}]) and the $n_{\rm e}$ values obtained using the rest of diagnostics for all the DESIRED nebulae. The [\ion{S}{2}] $\lambda 6731/\lambda 6716$ and [\ion{O}{2}] $\lambda 3726/\lambda 3729$ diagnostics show an excellent agreement for the whole sample. This is not surprising since both O$^{+}$ and S$^{+}$ ions coexist in the volume of low degree of ionization and both show essentially the same sensitivity (see the right panel of Fig.~\ref{fig:density_diagnostics}). In some PNe, due to the possible existence of cold clumps of high metallicity \citep{Liu:2000,Liu:2006,garciarojas:2016,garciarojas:2022,Richer:2022}, we may expect an important contribution of recombination in the observed [\ion{O}{2}] lines \citep{Barlow:2003,Wesson:2018}. This is especially important in the cases of Ou5 and Abell\,46 \citep[][]{Corradi:2015}, the PNe with the largest ADF from the whole sample and the only ones with ADF$>$5, and where the density obtained from [\ion{O}{2}] $\lambda 3726/\lambda 3729$ is clearly higher than that obtained from [\ion{S}{2}] $\lambda 6731/\lambda 6716$. In the rest of the photoionized nebulae in the database, this phenomenon, if present, actually has a negligible impact on these density diagnostics.

\begin{figure*}
\begin{minipage}{\textwidth}
\centering
\includegraphics[width=.48\textwidth]{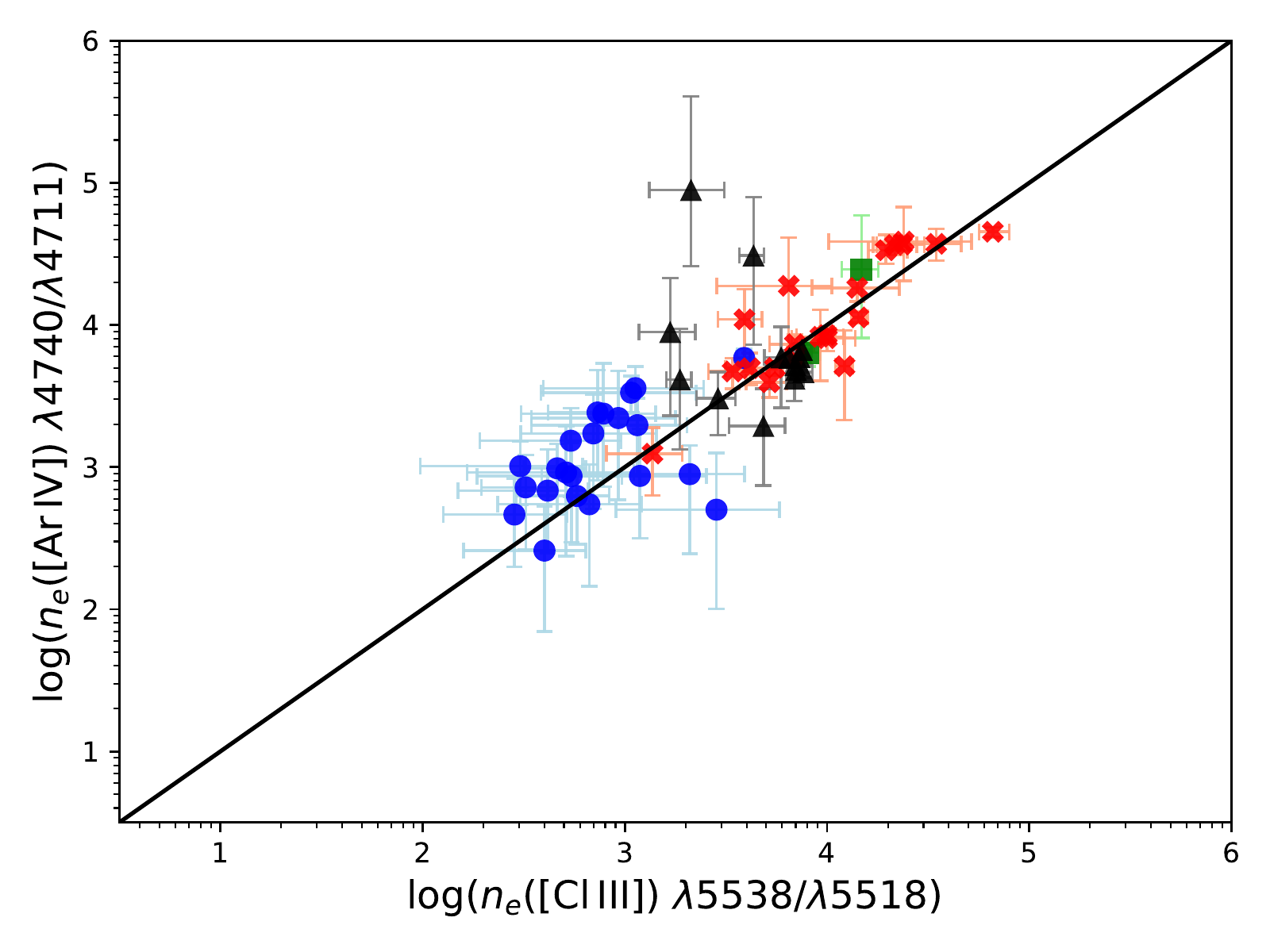}
\includegraphics[width=.48\textwidth]{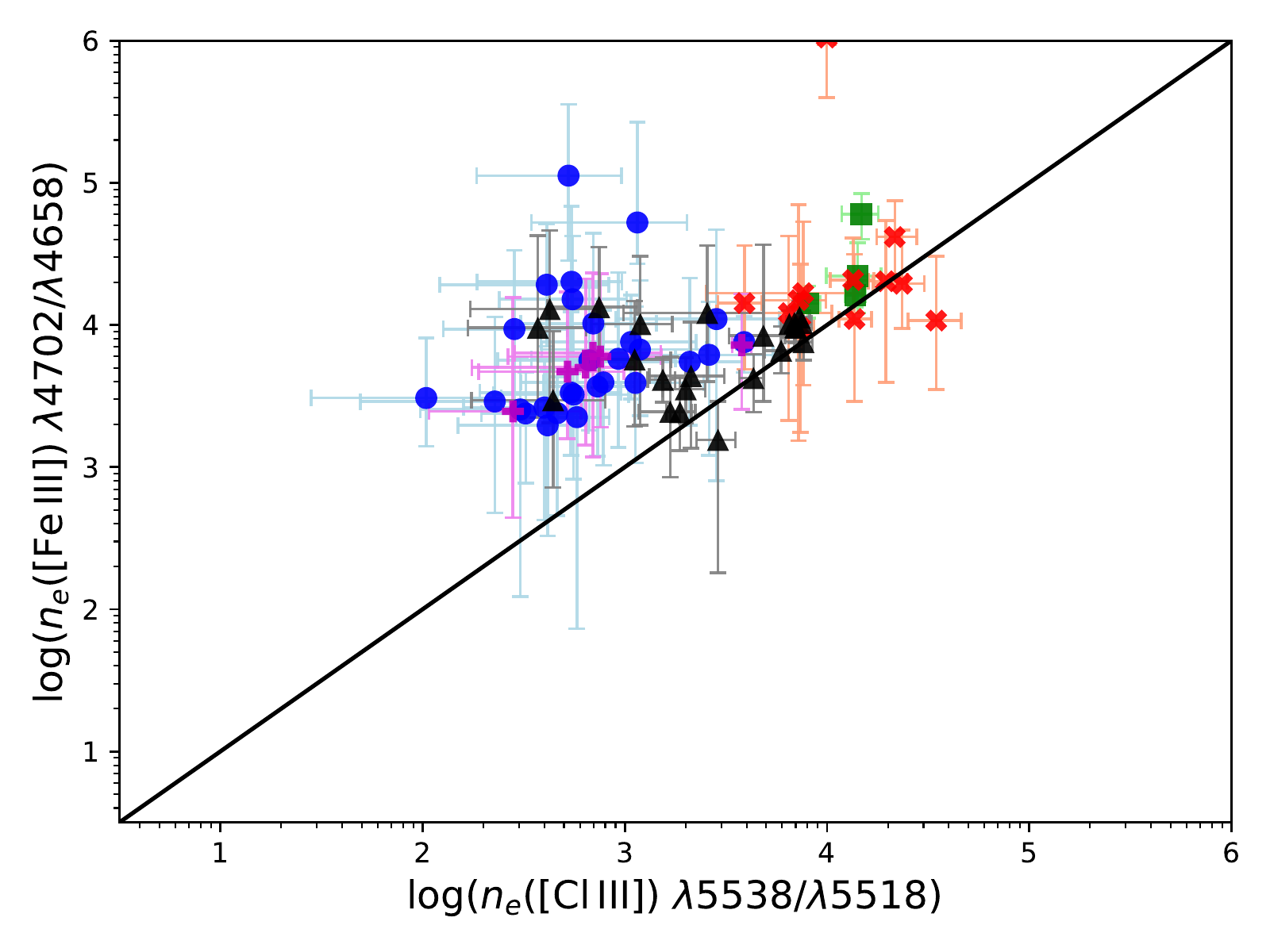}
\end{minipage}
\caption{Comparison between the density derived from the [\ion{Cl}{3}] $\lambda 5538/\lambda 5518$ line intensity ratio and those from [\ion{Ar}{4}] $\lambda 4740/\lambda 4711$ and [\ion{Fe}{3}] $\lambda 4702/\lambda 4658$. The symbols code is the same as in Fig.~\ref{fig:BPT}. It should be noticed the good agreement that exist when considering regions with $n_{\rm e}$>1000 cm$^{-3}$  (which leaves out most extragalactic \ion{H}{2} regions and RNs, blue dots and magenta crosses, respectively), as they are in their optimal sensitivity range, regardless of the degree of ionization of the ion.} 
\label{fig:densities_plot_cl3}
\end{figure*}

The comparison of the [\ion{S}{2}] $\lambda 6731/\lambda 6716$ values and those of [\ion{Cl}{3}] $\lambda 5538/\lambda 5518$, [\ion{Fe}{3}] $\lambda 4702/\lambda 4658$ and [\ion{Ar}{4}] $\lambda 4740/ \lambda 4711$ reveals significant deviations from a 1:1 relation in those objects where the first diagnostic gives $n_{\rm e}<1000 \text{ cm}^{-3}$. As expected from Fig~\ref{fig:density_diagnostics}, this is because the aforementioned diagnostics are at their low density limit, where the sensitivity is practically negligible. In the low density limit, the line intensity ratios should converge to constant values mainly fixed by the atomic collisional strengths. From the DESIRED data we obtain [\ion{Cl}{3}] $\lambda 5538/\lambda 5518=0.74 \pm 0.05$, [\ion{Fe}{3}] $\lambda 4702/\lambda 4658=0.26 \pm 0.04$ and [\ion{Ar}{4}] $\lambda 4740/ \lambda 4711=0.79 \pm 0.07$, in consistency with the predictions of the selected atomic data (see Table~\ref{tab:atomic_data}),  discarding significant errors in them. 

[\ion{Cl}{3}] $\lambda 5538/\lambda 5518$, [\ion{Fe}{3}] $\lambda 4702/\lambda 4658$ and [\ion{Ar}{4}] $\lambda 4740/ \lambda 4711$  line ratios become good density indicators for $n_{\rm e} > 10^3 \text{ cm}^{-3}$,  showing higher sensitivity than [\ion{S}{2}] $\lambda 6731/\lambda 6716$ (See Fig.~\ref{fig:density_diagnostics}). For this range of $n_{\rm e}$, Fig.~\ref{fig:densities_plot} shows a general offset between [\ion{S}{2}] $\lambda 6731/\lambda 6716$ and the rest of the aforementioned diagnostics. This is due to the combination of two phenomena, [\ion{S}{2}] $\lambda 6731/\lambda 6716$ is more sensitive in areas of lower density within the nebulae while the rest of indicators behave inversely. Furthermore, as density increases, for $n_{\rm e}>10^4 \text{ cm}^{-3}$, the accuracy of [\ion{S}{2}] $\lambda 6731/\lambda 6716$ decreases, amplifying the size of error bars. It is noticeable that [\ion{Fe}{3}], [\ion{Cl}{3}] and [\ion{Ar}{4}] density diagnostics show rather consistent trends, despite arising from the low, intermediate and very high ionization volumes. This shows that the different $n_{\rm e}$-sensitivity range of the diagnostics dominates over the possible density stratification in the nebulae, except for few dispersed objects of the sample, as it is also shown in Fig.~\ref{fig:densities_plot_cl3}.

Considering the previous discussion and in agreement with \citet{mendezdelgado:2023}, we propose the following criteria to adopt a representative density for chemical abundance determinations using optical spectra:

\begin{enumerate}
    \item If $n_{\rm e}$([\ion{S}{2}]) < 100 cm$^{-3}$, we adopt the low density limit ($n_{\rm e}<100$ cm$^{-3}$).
    \item If 100 cm$^{-3}$ < $n_{\rm e}$([\ion{S}{2}]) < 1000 cm$^{-3}$, we adopt the average value of $n_{\rm e}$([\ion{S}{2}]) and $n_{\rm e}$([\ion{O}{2}]).
    \item If $n_{\rm e}$([\ion{S}{2}]) $>$ 1000 cm$^{-3}$, we take the average values of $n_{\rm e}$([\ion{S}{2}]), $n_{\rm e}$([\ion{O}{2}]), $n_{\rm e}$([\ion{Cl}{3}]), $n_{\rm e}$([\ion{Fe}{3}]) and $n_{\rm e}$([\ion{Ar}{4}]) when available.
    \item For the HH objects we adopt $n_{\rm e}$([\ion{Fe}{3}]), while in the case of the proplyd 170-337, we adopt the reference value derived from the [\ion{S}{2}] $\lambda 4069/\lambda 4076$ line intensity ratio. 
\end{enumerate}

The resulting representative density values are shown in the bottom panel of Fig.~\ref{fig:densities_plot}. As discussed in Section~\ref{subsec:TO2_TS2}, this criteria is far from perfect, but it is accurate enough to determine chemical abundances based on optical spectra. However, we discourage its use for the determination of feedback-related pressure terms or abundances based on infrared fine structure lines without further analysis.

In criterion (i) we consider the fact that all the density diagnostics analyzed in this work are insensitive at such low densities. If the average electron density is actually in this range of values, its impact is negligible in the determination of temperature and chemical abundances \citep{Osterbrock:2006}. However, the consideration of another method is recommended for those who require precise determinations of the gas pressure (dependent on density) in low-density \ion{H}{2} regions, relevant to some phenomena such as stellar feedback \citep[e.g.][]{McLeod:2020,Barnes:2021}. As a suggestion, considering the radiative and collisional atomic transitions from \citet{Bautista:2015}, the [\ion{Fe}{2}] $\lambda 8617/\lambda 9267$ line intensity ratio should vary from a value $\sim 110$ at $n_{\rm e}=1 \text{ cm}^{-3}$ to a value $\sim 54$ at $n_{\rm e}=100 \text{ cm}^{-3}$ when assuming $T_{\rm e}=10000 \text{ K}$. These lines arise from the Fe$^{+}$ lower quartet levels, and should not have significant fluorescence contributions \citep{Baldwin:1996,Verner:2000}. \citet{mendezdelgado:2021b,mendezdelgado:2022b} have checked the adequacy of this density diagnostic in higher density regions. However, this is highly dependent on the atomic data used \citep{Mendoza:2023}.

Criterion (ii) is based on the fact that [\ion{Cl}{3}] $\lambda 5538/\lambda 5518$, [\ion{Fe}{3}] $\lambda 4702/\lambda 4658$ and [\ion{Ar}{4}] $\lambda 4740/ \lambda 4711$ are quite insensitive to densities smaller than 1000 cm$^{-3}$. In the presence of high-density inclusions within the nebulae, densities adopted under this criterion are underestimated as well as those of  criterion (i). This will be demonstrated in Section~\ref{subsec:TO2_TS2}. The impact of such underestimate is rather limited in optical studies, being constrained up to $\sim 0.1$  dex when using [\ion{O}{2}] $I(\lambda \lambda$7319+20+30+31) to estimate the O$^{+}$/H$^{+}$ abundance. However, this can introduce large systematic errors when using IR fine structure CELs, where a density underestimate of $\sim 300 $ cm$^{-3}$ can affect $T_{\rm e}$ determinations by several thousand Kelvin \citep[see fig.~3 from ][]{Lamarche:2022}. 

Criterion (iii) allows us to obtain more precise values of electron density. Although the use of $n_{\rm e}$([\ion{S}{2}] $\lambda 6731/\lambda 6716$) or $n_{\rm e}$([\ion{O}{2}] $\lambda 3726/\lambda 3729$) as single diagnostic is consistent with the adopted value in most of the denser nebulae within the error bars --given the high quality of the DESIRED spectra-- the uncertainty of these diagnostics becomes larger as the density increases. As shown in the bottom panel of Fig.~\ref{fig:densities_plot}, a systematic underestimate of the median values of density is noticeable when $n_{\rm e}$([\ion{S}{2}] $\lambda 6731/\lambda 6716$) approaches values $\sim 10^4$ cm$^{-3}$, concerning especially to PNe. It is difficult to establish whether this behaviour is linked to a density stratification in these objects, as some works suggest \citep[see e.~g.][]{Rauber:2014}, or it is just a consequence of the different sensitivity of the compared diagnostics. Since this affects some HHs as well, this seems to indicate that the different sensitivity of the diagnostics dominates the observed trend. In this range of densities, $T_{\rm e}$([\ion{N}{2}] $\lambda 5755/\lambda 6584$) depends appreciably on $n_{\rm e}$. Therefore, having large error bars in $n_{\rm e}$ gives rise to obtain inaccurate values of $T_{\rm e}$([\ion{N}{2}] $\lambda 5755/\lambda 6584$) and, finally, of the ionic abundances.

Criterion (iv) is applied to photoionized HHs because indicators based on [\ion{Fe}{3}] lines are sensitive to very high densities, but also because the destruction of Fe-bearing dust particles by shocks enhances the emission of [\ion{Fe}{3}] lines. In these cases, we adopt the values obtained with a maximum-likelihood procedure using several [\ion{Fe}{3}] lines. This method provides values fully consistent with $n_{\rm e}$([\ion{Fe}{3}] $\lambda 4702/\lambda 4658$). In Fig.~\ref{fig:densities_plot}, we can see that density determinations based on [\ion{Fe}{3}] lines --although showing larger error bars-- are marginally consistent with $n_{\rm e}$([\ion{S}{2}] $\lambda 6731/\lambda 6716$) in most of the cases except for HH~514, the proplyd 170-337 \citep{mendezdelgado:2022b} and NGC~7027 \citep{Sharpee:2007}. In these objects, the electron density is so high that a large fraction of the emission in CELs is produced through the much weaker auroral lines instead of the nebular ones. Unfortunately, because of the large dust depletion and low ionization degree of proplyd 170-337, $n_{\rm e}$([\ion{Cl}{3}] $\lambda 5538/\lambda 5518$), $n_{\rm e}$([\ion{Fe}{3}] $\lambda 4702/\lambda 4658$) and $n_{\rm e}$([\ion{Ar}{4}] $\lambda 4740/ \lambda 4711$) can not be derived for this object. However, the density can be determined from the [\ion{S}{2}] $\lambda 4069/\lambda 4076$ ratio in this case.

\section{temperature structure}
\label{sec:temps_relations}

In this section, we analyze the temperature relations for the different ionization zones in extragalactic \ion{H}{2} regions. Firstly, we will start by investigating  the dependence of the low ionization temperature diagnostics $T_{\rm e}$([\ion{O}{2}]), $T_{\rm e}$([\ion{S}{2}]) and $T_{\rm e}$([\ion{N}{2}]) on the electron density. Secondly, we will study the temperature relations obtained directly from the observations. In all figures of this section, we use the parameter $P$ defined by \citet{Pilyugin:2001}: $\frac{[\ion{O}{3}] I(5007+4959)}{[\ion{O}{3}] I(5007+4959)+[\ion{O}{2}] I(3726+3729)}$ as a proxy of the ionization degree of the gas.

\subsection{$T_{\rm e}$([\ion{O}{2}]), $T_{\rm e}$([\ion{S}{2}]), and $T_{\rm e}$([\ion{N}{2}])}
\label{subsec:TO2_TS2}

Based on the results of photoionization models \citep{Campbell:1986,Pilyugin:2006}, it is generally assumed that $T_{\rm e}([\ion{O}{2}])\approx T_{\rm e}([\ion{S}{2}]) \approx T_{\rm e}([\ion{N}{2}])$. This is also predicted by the BOND models (Sec~\ref{sec:phot_models}). However, this is rarely satisfied observationally in extragalactic \ion{H}{2} regions \citep{perezmontero:2003,Kennicutt:2003,Hagele:2006,Hagele:2008,Esteban:2009,Bresolin:2009,Berg:2015}. $T_{\rm e}([\ion{O}{2}])$ and $T_{\rm e}([\ion{S}{2}])$ are estimated from [\ion{O}{2}] $\lambda \lambda$7319+20+30+31/$\lambda \lambda$ 3726+29 and [\ion{S}{2}] $\lambda \lambda$4069+76/$\lambda \lambda$6716+31 line intensity ratios, which can be affected by several observational effects. The first line intensity ratio is highly dependent on the reddening correction as well as the quality of the flux calibration of the spectrum given the wide wavelength separation between the nebular and auroral lines. Moreover, $\lambda \lambda$7319+20+30+31 can be contaminated by telluric emissions. In the case of the latter line intensity ratio, the [\ion{S}{2}] auroral lines can be blended with \ion{O}{2} $\lambda \lambda 4069.62,4069.88,4072.15,4075.86$ lines, which can represent more than 10 per cent of the total flux in some nebulae.

In addition to the possible observational effects commented on above, some other physical phenomena have been invoked to explain the discrepancies between $T_{\rm e}([\ion{O}{2}])$, $T_{\rm e}([\ion{S}{2}])$ and $T_{\rm e}([\ion{N}{2}])$, such as: 
\begin{itemize}
    \item Mismatch between the temperature of the volumes of O$^+$, N$^+$ and S$^+$.
    \item Recombination contribution to the CELs.
    \item Temperature fluctuations.
    \item Density variations.
\end{itemize}

The high quality of the DESIRED spectra permits us to minimize the effect of observational errors on $T_{\rm e}$ determinations and to explore other physical phenomena that may cause the discrepancies. We derive $T_{\rm e}([\ion{O}{2}])$, $T_{\rm e}([\ion{S}{2}])$ and $T_{\rm e}([\ion{N}{2}])$ adopting the density criteria mentioned in Section~\ref{sec:density_structure}, actually criteria (i) or (ii) in most cases. The adoption of $n_{\rm e}$([\ion{S}{2}] $\lambda 6731/\lambda 6716$) or $n_{\rm e}$([\ion{O}{2}] $\lambda 3726/\lambda 3729$) is the standard procedure for the analysis of extragalactic \ion{H}{2} regions and therefore our results can be directly compared with other works.

\begin{figure*}
\begin{minipage}{\textwidth}
\centering
\includegraphics[width=.48\textwidth]{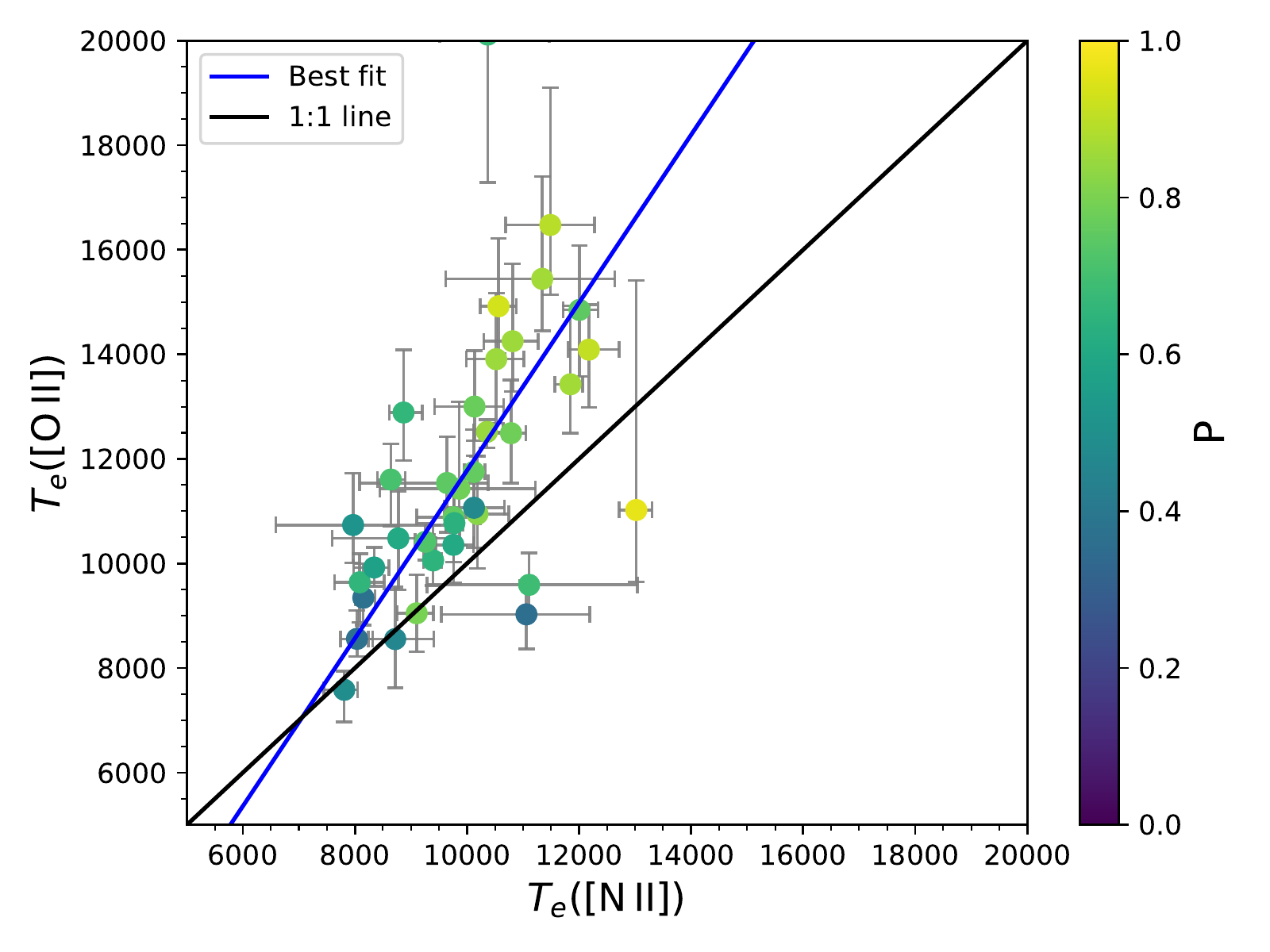}
\includegraphics[width=.48\textwidth]{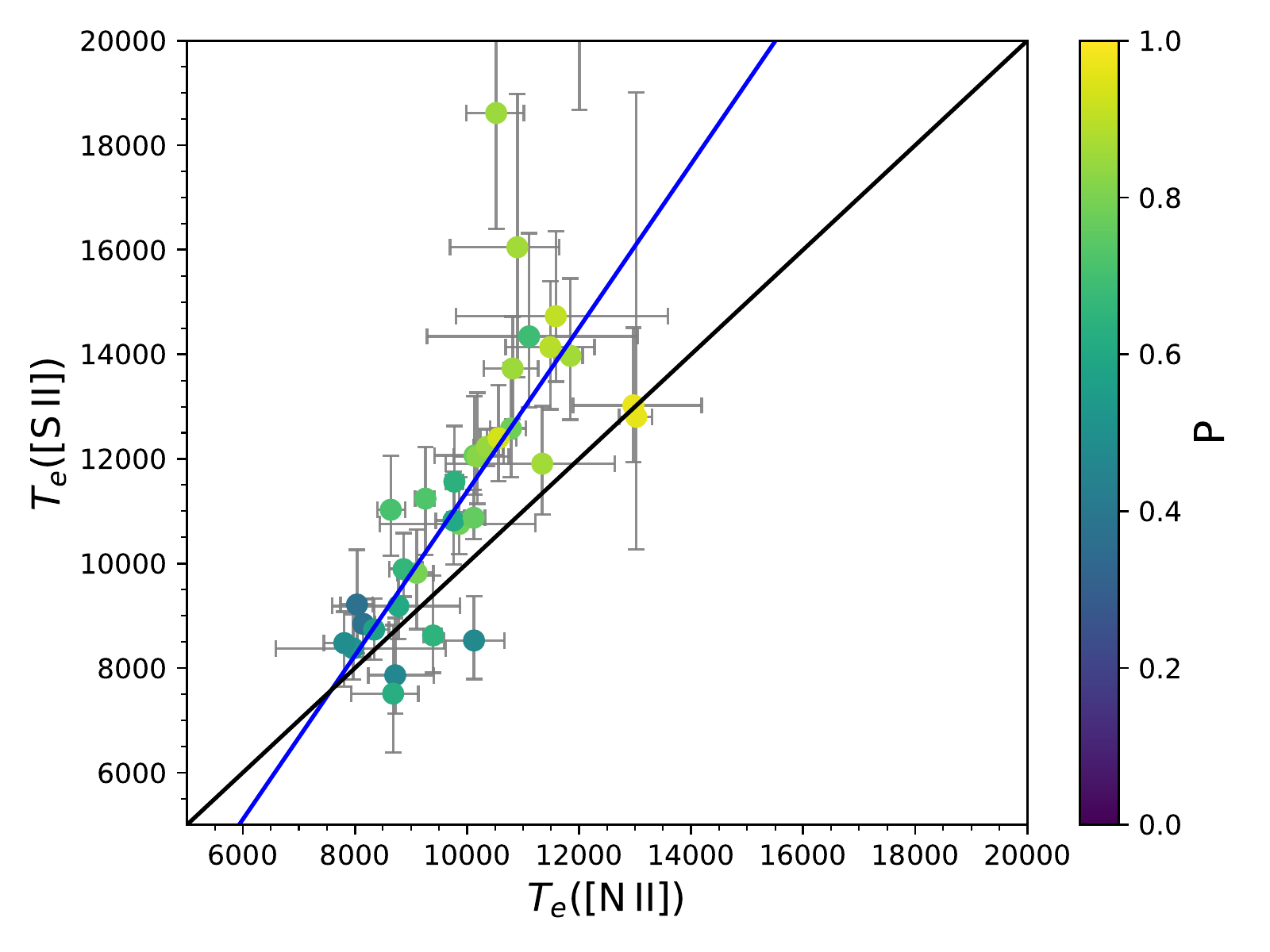}
\includegraphics[width=.48\textwidth]{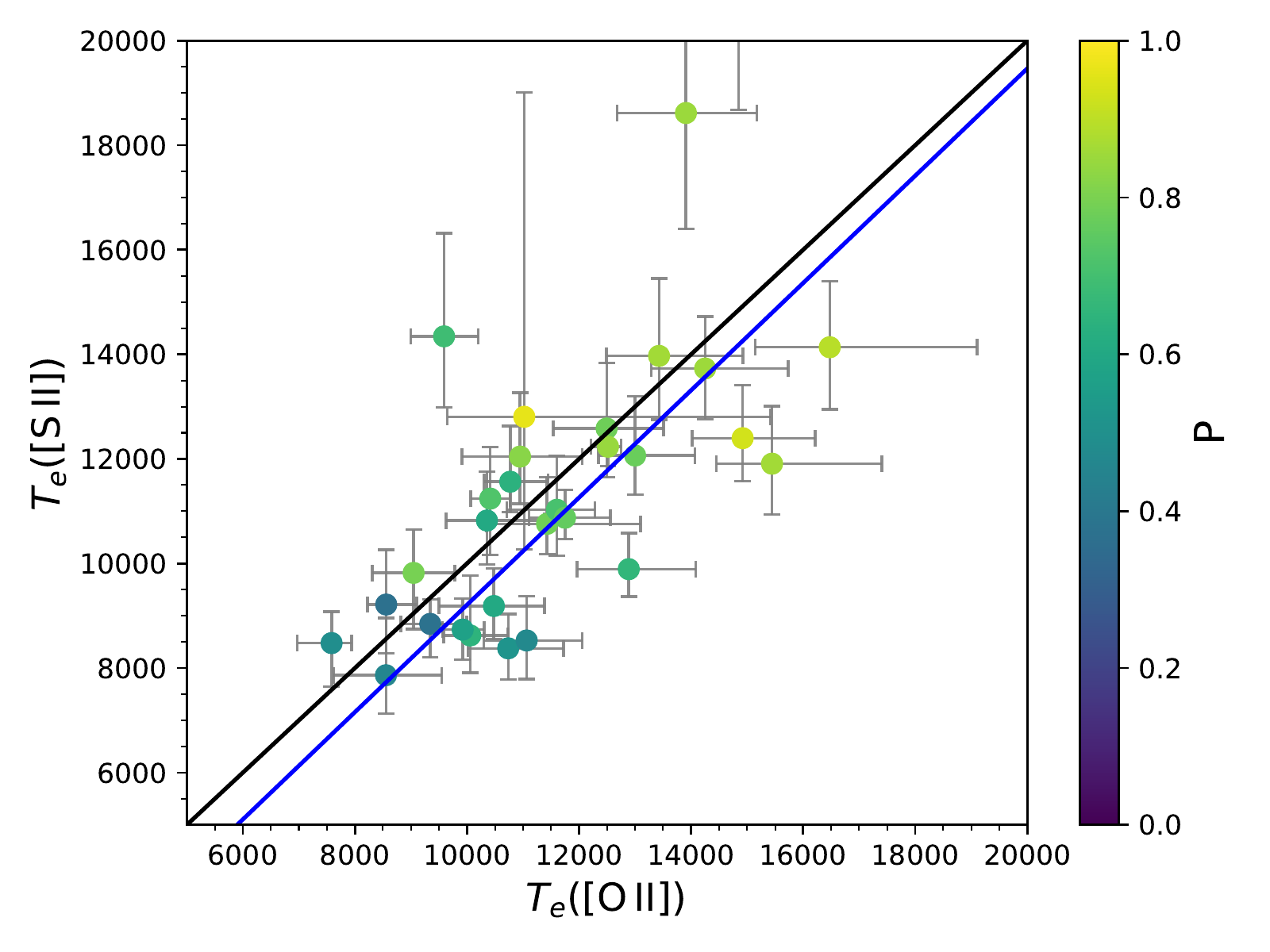}
\end{minipage}
\caption{Relations between $T_{\rm e}([\ion{O}{2}])$, $T_{\rm e}([\ion{S}{2}])$ and $T_{\rm e}([\ion{N}{2}])$ derived by adopting  $n_{\rm e}$([\ion{S}{2}] $\lambda 6731/\lambda 6716$) and $n_{\rm e}$([\ion{O}{2}] $\lambda 3726/\lambda 3729$) in the extragalactic \ion{H}{2} regions of the sample. The color of the points represents the value of their $P$ parameter \citep[][ see text]{Pilyugin:2001} that can be used as proxy of the ionization degree of the nebulae.  The blue solid line represents the linear fit to the data. The black solid line represents a 1:1 linear relation. }
\label{fig:TN2S2O2}
\end{figure*}

\begin{table}
\caption{Linear fits between $T_{\rm e}$([\ion{O}{2}]), $T_{\rm e}$([\ion{S}{2}]) and $T_{\rm e}$([\ion{N}{2}]). $\sigma$ represent the standard deviation between the linear fit and the calculated temperature values. N is the number of regions considered.}
\label{tab:app_fits}
\begin{tabular}{cccccc}
\hline
Linear fit (K) & $\sigma$ (K)& N\\
\hline

$T_{\rm e}([\ion{O}{2}])=1.60 (\pm 0.20) T_{\rm e}([\ion{N}{2}]) -4270 (\pm 1870) $ & 1210 &32 \\

$T_{\rm e}([\ion{N}{2}])=0.62 (\pm 0.08) T_{\rm e}([\ion{O}{2}]) +2660 (\pm 840) $ & 950 &32 \\

$T_{\rm e}([\ion{S}{2}])=1.57 (\pm 0.17) T_{\rm e}([\ion{N}{2}]) -4290 (\pm 1620) $ & 900 &30 \\

$T_{\rm e}([\ion{N}{2}])=0.64 (\pm 0.07) T_{\rm e}([\ion{S}{2}]) +2740 (\pm 740) $ & 730 & 30 \\

$T_{\rm e}([\ion{S}{2}])=1.03 (\pm 0.11) T_{\rm e}([\ion{O}{2}]) -1050 (\pm 1180) $ & 1280 & 39 \\

\hline
\end{tabular}
\end{table}

Fig.~\ref{fig:TN2S2O2} shows the comparison between $T_{\rm e}([\ion{O}{2}])$, $T_{\rm e}([\ion{S}{2}])$ and $T_{\rm e}([\ion{N}{2}])$ derived with the standard procedure of adopting $n_{\rm e}$([\ion{S}{2}] $\lambda 6731/\lambda 6716$) and $n_{\rm e}$([\ion{O}{2}] $\lambda 3726/\lambda 3729$) as the representative electron density. The blue lines correspond to the best linear fits, which are presented in Table~\ref{tab:app_fits}. Despite of the quality of the data, there are few outlier regions: NGC~5471, H~37 \citep{Esteban:2020}, N~66A \citep{dominguez:2022} and H~II-2 \citep{lopezsanchez:2007} with very high values of $T_{\rm e}([\ion{S}{2}])$ and NGC~2363 \citep{Esteban:2009} with an extremely high value of $T_{\rm e}([\ion{O}{2}])$. These regions may have particular physical phenomena or some non-identified contamination in the auroral lines, such as those described previously. Although they are included in Fig.~\ref{fig:TN2S2O2}, they are not considered in the linear fits shown in Table~\ref{tab:app_fits}. We will focus on the global trends.

As it can be seen in Fig.~\ref{fig:TN2S2O2}, $T_{\rm e}([\ion{O}{2}])$ and $T_{\rm e}([\ion{S}{2}])$ are higher than $T_{\rm e}([\ion{N}{2}])$ for most values of this last parameter, in agreement with previous findings \citep{Esteban:2009,Bresolin:2009,Rogers:2021}. It should be noted that $T_{\rm e}([\ion{O}{2}])$-$T_{\rm e}([\ion{N}{2}])$ and $T_{\rm e}([\ion{S}{2}])$-$T_{\rm e}([\ion{N}{2}])$ increase as a function of temperature, as shown in the fit parameters given in Table~\ref{tab:app_fits}. On the other hand, $T_{\rm e}([\ion{S}{2}])$ versus $T_{\rm e}([\ion{O}{2}])$ fits an almost 1:1 relation, with a slight offset to higher $T_{\rm e}([\ion{O}{2}])$ values. 

\subsubsection{Are these temperatures different?}

The differences between the temperatures determined from CELs of different ions are usually explained because they are representative of zones with different ionization conditions. This may result from differences in the ionization potentials, in the spectral distribution of the ionizing radiation and sometimes on the absorption edges on the ionizing radiation and on the presence of charge exchange and dielectronic recombination contributions \citep{stasinska:1980,Garnett:1992}. Although there are small differences in the ionization energy ranges of S$^{+}$, O$^{+}$ and N$^{+}$ and some other properties, photoionization models predict that this should not have relevant effects on the difference between $T_{\rm e}([\ion{S}{2}])$, $T_{\rm e}([\ion{O}{2}])$ and $T_{\rm e}([\ion{N}{2}])$. The exceptions may be very high metallicity regions, where the internal temperature gradients can be very marked \citep{stasinska:2005}. However, as can be inferred from  Fig.~\ref{fig:TN2S2O2}, the differences in the top panels are higher for the regions of higher degree of ionization, which is most typical case at lower metallicities. Furthermore, although the coexisting volumes of S$^{+}$ and O$^{+}$ usually differ much more than those of N$^{+}$ and O$^{+}$ \citep[e.g. see fig.~2 from][]{Levesque:2010}, the first pair of ions shows better consistency between their respective $T_{\rm e}$ values, as it is shown in the bottom panel of Fig.~\ref{fig:TN2S2O2}. This result suggests that the difference between the ionization structure alone does not explain the differences between $T_{\rm e}([\ion{S}{2}])$, $T_{\rm e}([\ion{O}{2}])$ and $T_{\rm e}([\ion{N}{2}])$.

It is sometimes argued in the literature that part of the optical [\ion{S}{2}] emission can be originated in the photodissociation region (PDR) where H and He are mostly neutral, discarding $T_{\rm e}([\ion{S}{2}])$. This argument has also been found together with the adoption of $n_{\rm e}$([\ion{S}{2}] $\lambda 6731/\lambda 6716$) as valid density estimator, even for the entire nebula \citep[e.g.][]{Esteban:2020}. It is clear that this can not be an explanation of the differences between $T_{\rm e}([\ion{S}{2}])$ and $T_{\rm e}([\ion{N}{2}])$ since, although there may be some S$^{+}$ in the volume where H and He are neutral, the emission of [\ion{S}{2}] lines requires numerous collisions with free electrons that can only be supplied in sufficient quantities by the ionization of H and He \citep{Odell:2023}. Therefore, [\ion{S}{2}] emission should arise from the ionized volume and the surrounding areas of the ionization front. Exceptions may appear when there are shocks in the ionization front.  

\subsubsection{Recombination contributions?}

\citet{Rubin:1986} pointed out the possibility of significant recombination contributions to the atomic levels that produce optical CELs of N and O ions.  At first order, one would expect recombinations to be more important in regions of higher metallicity \citep{stasinska:2005}, where the temperature is lower. This is the opposite behavior of our observations shown in Fig.~\ref{fig:TN2S2O2}. A complex question is to know in what proportion the recombinations affect the inferred $T_{\rm e}([\ion{S}{2}])$, $T_{\rm e}([\ion{O}{2}])$ and $T_{\rm e}([\ion{N}{2}])$ in the analyzed extragalactic \ion{H}{2} regions, as recombination contributions can affect both the auroral and nebular [\ion{O}{2}] lines whereas it is expected to only affect the auroral [\ion{N}{2}] line. To clarify this, we use the photoionization models described in Section~\ref{sec:phot_models}. These models consider the recombination contributions in the [\ion{O}{2}] and [\ion{N}{2}] lines using the recombination coefficients calculated by \citet{Pequignot:1991}, \citet{Fang:2011,Fang:2013} and \citet{Storey:2017}. 

\begin{figure}
\includegraphics[width=.48\textwidth]{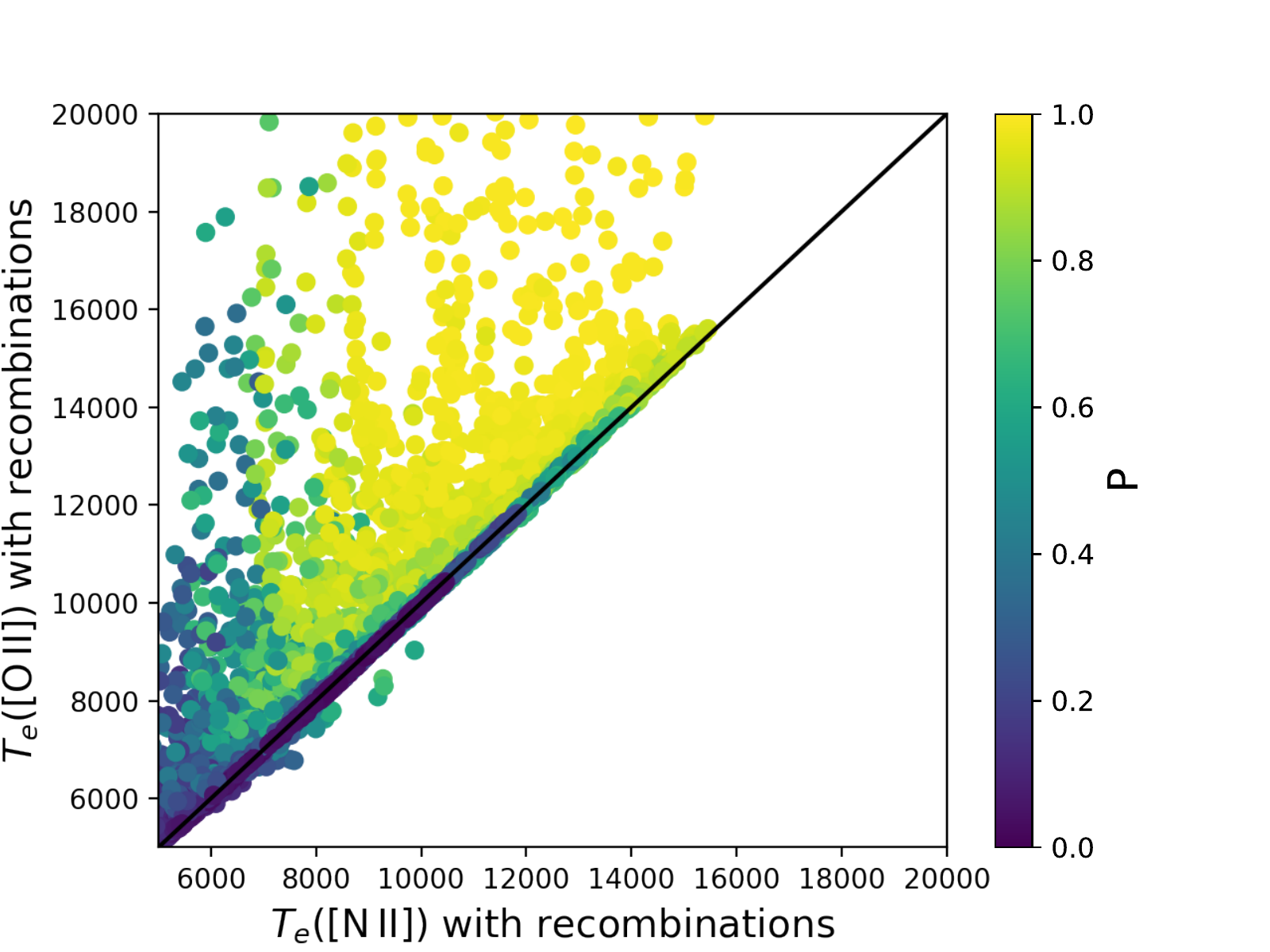}
\caption{Comparison of the impact of recombination contributions on $T_{\rm e}([\ion{O}{2}])$ and $T_{\rm e}([\ion{N}{2}])$. These predictions are based on photoionization models described in Section~\ref{sec:phot_models}. The color of the symbols represents the value of their $P$ parameter (as in Fig~\ref{fig:TN2S2O2}). The black solid line represents a 1:1 linear relation. In case of recombination contributions, $T_{\rm e}([\ion{O}{2}])>T_{\rm e}([\ion{N}{2}])$ for most cases.}
\label{fig:recs_contr}
\end{figure}

In Fig.~\ref{fig:recs_contr} we show that when the recombination contribution is relevant, the measured [\ion{O}{2}] $\lambda \lambda$7319+20+30+31/$\lambda \lambda$3726+29 line intensity ratio tends to be comparatively more enhanced than [\ion{N}{2}] $\lambda$5755/$\lambda\lambda$6548+84. This implies that, if there are recombination contributions (dielectronic plus radiative) to the [\ion{O}{2}] and [\ion{N}{2}] CELs, we would expect $T_{\rm e}([\ion{O}{2}])>T_{\rm e}([\ion{N}{2}])$ in most cases.

\begin{figure}
\includegraphics[width=.48\textwidth]{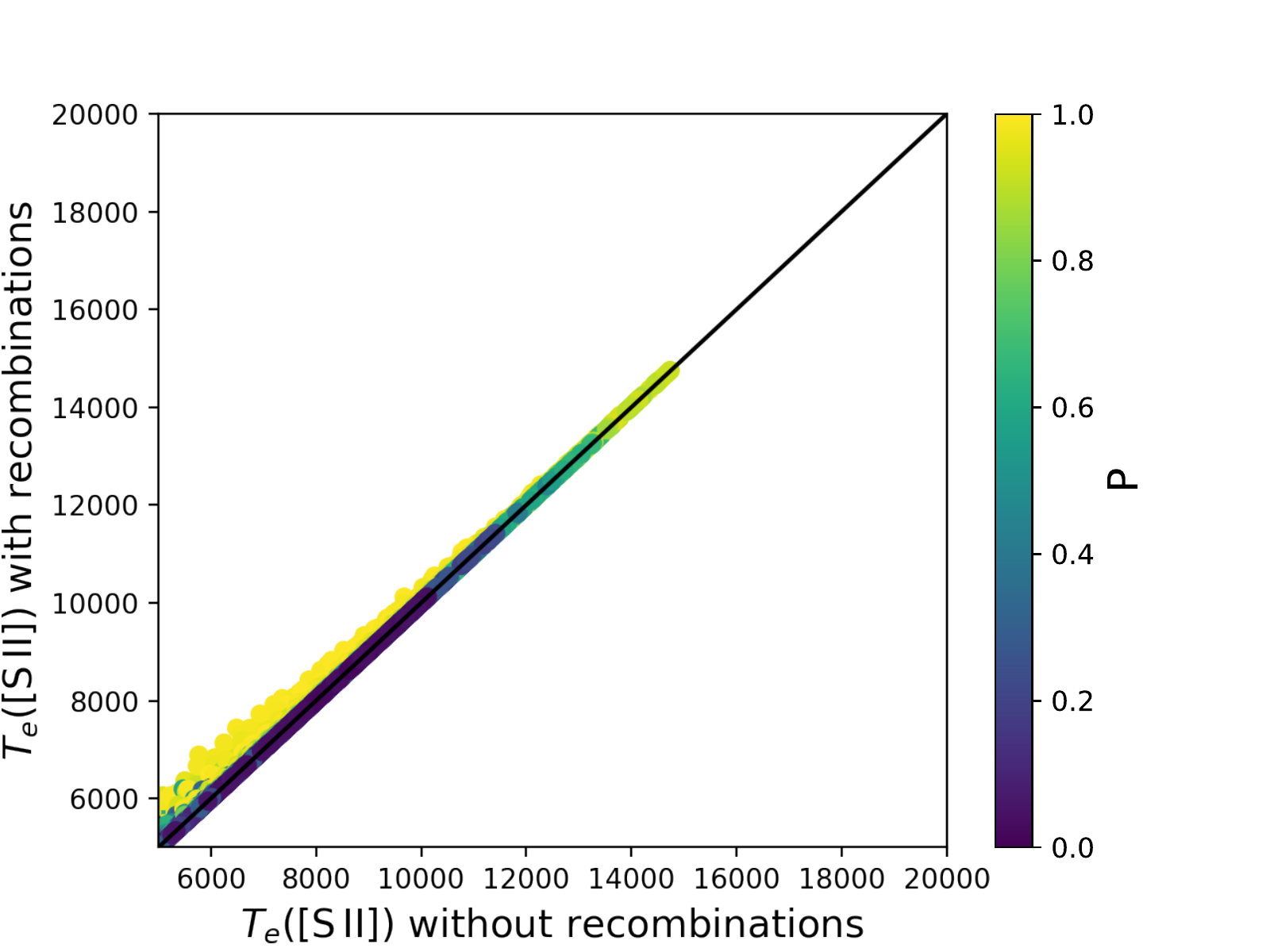}
\caption{Comparison of the derived $T_{\rm e}([\ion{S}{2}])$ without recombination contributions with the hypothetical case of recombination contributions in proportion to those of [\ion{O}{2}]. These predictions are based on photoionization models described in Section~\ref{sec:phot_models}. The color of the symbols represents the value of their $P$ parameter (as in Fig~\ref{fig:TN2S2O2}). The black solid line represents a 1:1 linear relation. The recombination contributions are negligible for $T_{\rm e}([\ion{S}{2}])>7000$ K.}
\label{fig:recs_contrS2}
\end{figure}

To date, there are no evidences of relevant recombination contributions to the [\ion{S}{2}] CELs. Furthermore, there is also a lack of calculations of effective recombination coefficients for this ion. However,  potential recombination contributions to the [\ion{S}{2}] CELs can be tested by assuming that this ion has the same electronic configuration than [\ion{O}{2}]. Therefore, as a first approximation, the recombination contribution to the $^2P$ and $^2D$ levels of [\ion{S}{2}] would be similar to that of [\ion{O}{2}] weighted by the S$^{2+}$/O$^{2+}$ abundance ratio. In Fig.~\ref{fig:recs_contrS2} we show that the potential recombinations under this case are negligible when $T_{\rm e}([\ion{S}{2}])>7000$ K, which is the case for all our data (see Fig.~\ref{fig:TN2S2O2}).

\begin{figure}
\includegraphics[width=.48\textwidth]{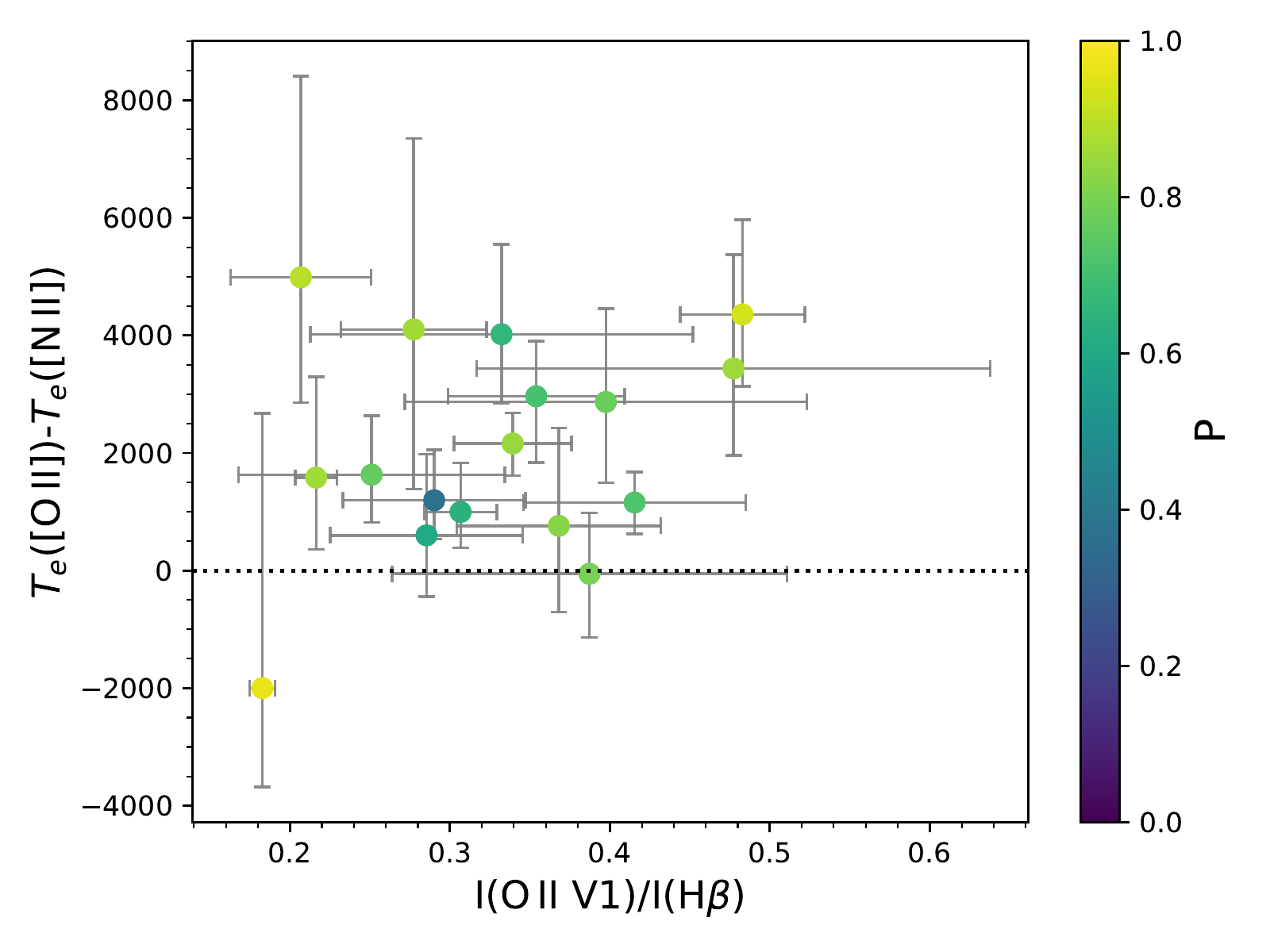}
\caption{$T_{\rm e}([\ion{O}{2}])-T_{\rm e}([\ion{N}{2}])$ difference as a function of the intensity of the \ion{O}{2} recombination multiplet V1. The color of the symbols represents the value of their $P$ parameter (as in Fig~\ref{fig:TN2S2O2}).}
\label{fig:dif_temps}
\end{figure}

Considering the above, if the difference between $T_{\rm e}([\ion{O}{2}])$ and $T_{\rm e}([\ion{N}{2}])$ is actually produced by recombinations, this would increase as a function of the intensity of the \ion{O}{2} RLs. However, Fig.~\ref{fig:dif_temps} demonstrate that $T_{\rm e}([\ion{O}{2}])- T_{\rm e}([\ion{N}{2}])$ and the intensity of the \ion{O}{2} V1 RLs \citep[adopted from][]{mendezdelgado:2023} do not correlate, discarding significant recombination contributions. Most importantly, if $T_{\rm e}([\ion{O}{2}])$ are affected by recombinations, the close 1:1 relation shown by $T_{\rm e}([\ion{O}{2}])$ and $T_{\rm e}([\ion{S}{2}])$ in  Fig.~\ref{fig:TN2S2O2} would be difficult to explain, given the predictions of Fig.~\ref{fig:recs_contrS2}. Therefore, recombination effects on [\ion{S}{2}], [\ion{O}{2}] and [\ion{N}{2}] CELs do not seem to explain the observed differences in their  corresponding temperatures.

\subsubsection{Temperature inhomogeneities?}

\citet{Peimbert:1967} introduced the formalism of internal temperature inhomogeneities in ionized nebulae, quantified by the root mean square temperature fluctuations parameter ($t^2$). In the presence of such fluctuations in the volume where S$^+$, O$^+$ and N$^+$ coexist, we would expect $T_{\rm e}([\ion{O}{2}]) \geq T_{\rm e}([\ion{N}{2}]) \geq T_{\rm e}([\ion{S}{2}])$, as a consequence of the different excitation energies of the atomic levels involved \citep[see equation~15 of][] {Peimbert:1967}. However, this is not the case in the observed trends shown in Fig.~\ref{fig:TN2S2O2}, in agreement with the recent results by \citet{mendezdelgado:2023}. Those authors find that although the effects of $t^2$ are evident in the high-ionization volume of nebulae, they seem to be absent in the low-ionization one.

\subsubsection{Density inhomogeneities}

\begin{figure}
\includegraphics[width=.48\textwidth]{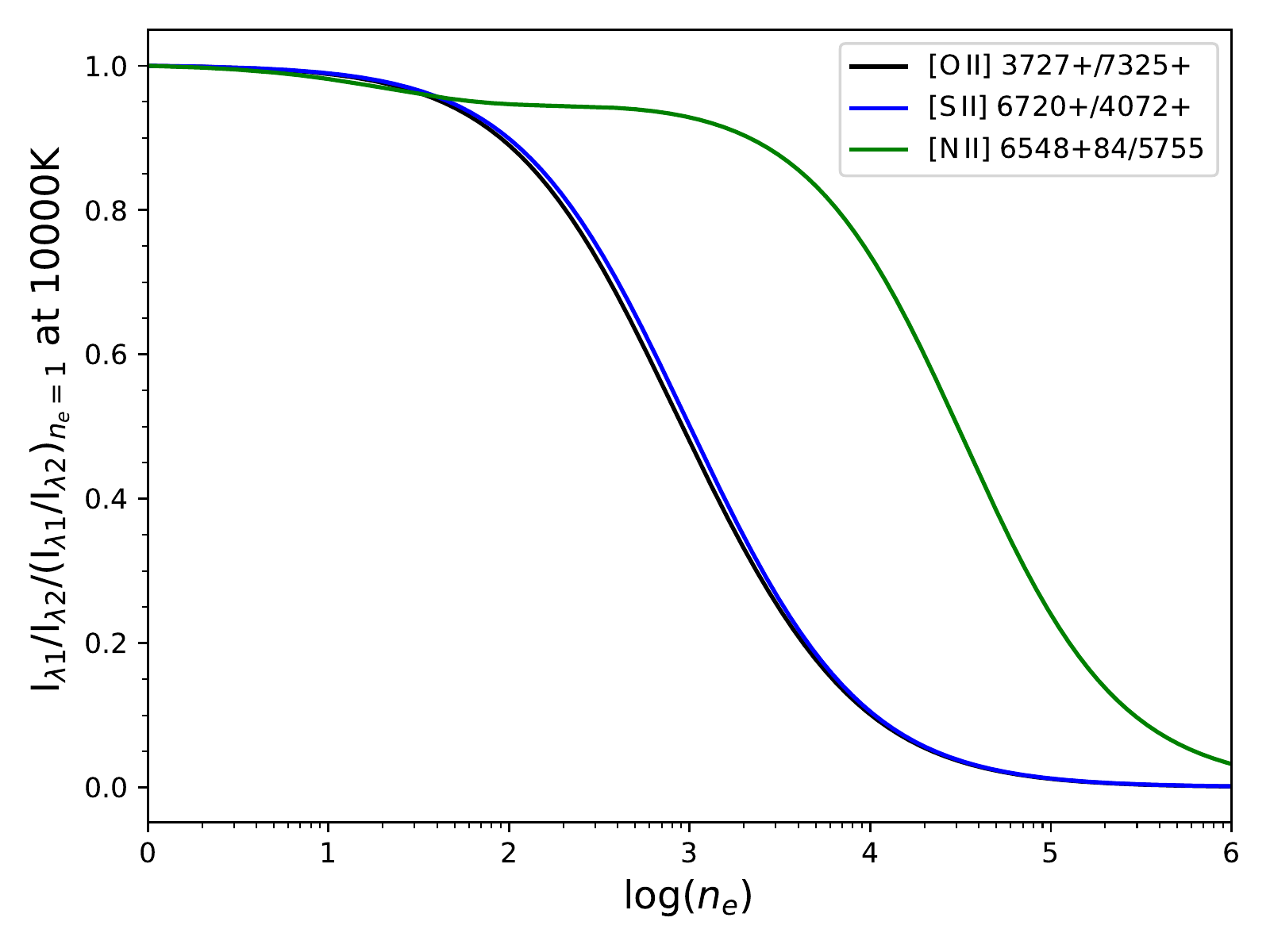}
\caption{Density dependence of the [\ion{O}{2}] $\lambda\lambda$7319+20+30+31/$\lambda\lambda$3726+29, [\ion{S}{2}] $\lambda\lambda$4069+76/$\lambda\lambda$6716+31 and [\ion{N}{2}] $\lambda$5755/$\lambda\lambda$6548+84 line intensity ratios, commonly used to infer $T_{\rm e}([\ion{O}{2}])$, $T_{\rm e}([\ion{S}{2}])$ and $T_{\rm e}([\ion{N}{2}])$, respectively. We have assumed $T_{\rm e} =$ 10,000 K. The line intensity ratios have been normalized with the expected values at $n_{\rm e}=1 \text{ cm}^{-3}$.}
\label{fig:Tem_diags_den_dependence}
\end{figure}

 The [\ion{O}{2}] $\lambda\lambda$7319+20+30+31/$\lambda\lambda$3726+29 and [\ion{S}{2}] $\lambda\lambda$4069+76/$\lambda\lambda$6716+31 line intensity ratios are highly dependent on density as it is shown in Fig.~\ref{fig:density_diagnostics}. When $T_{\rm e}$ is fixed, the $n_{\rm e}$-sensitivity of the aforementioned line intensity ratios is larger than that of [\ion{S}{2}] $\lambda 6731/\lambda 6716$, [\ion{O}{2}] $\lambda 3726/\lambda 3729$, [\ion{Cl}{3}] $\lambda 5538/\lambda 5518$, [\ion{Fe}{3}] $\lambda 4702/\lambda 4658$ and [\ion{Ar}{4}] $\lambda 4740/ \lambda 4711$ in practically the entire range $10^2 \text{ cm}^{-3}<n_{\rm e}<10^6 \text{ cm}^{-3}$. If there are density inhomogeneities in the nebulae, these line ratios would give higher densities than those derived from [\ion{S}{2}] $\lambda 6731/\lambda 6716$ or [\ion{O}{2}] $\lambda 3726/\lambda 3729$  \citep{Peimbert:1971, Rubin:1989}. The presence of high-density clumps biases [\ion{S}{2}] $\lambda 6731/\lambda 6716$ and [\ion{O}{2}] $\lambda 3726/\lambda 3729$ towards lower values of $n_{\rm e}$ and this would impact $T_{\rm e}([\ion{N}{2}])$ determination to a smaller extent than $T_{\rm e}([\ion{O}{2}])$ and $T_{\rm e}([\ion{S}{2}])$. This behavior is illustrated in Fig.~\ref{fig:Tem_diags_den_dependence}, where it can be seen that $T_{\rm e}([\ion{N}{2}])$ is insensitive to density up to $\sim 10^4 \text{ cm}^{-3}$, two orders of magnitude beyond than in the case of $T_{\rm e}([\ion{O}{2}])$ or $T_{\rm e}([\ion{S}{2}])$. If the high-density gas is $10^2 \text{ cm}^{-3}<n_{\rm e}<10^4 \text{ cm}^{-3}$ or if the density is higher but occupies a small fraction of the total ionized volume, $T_{\rm e}([\ion{N}{2}])$ may remain unaffected in contrast to what happens with $T_{\rm e}([\ion{O}{2}])$ and $T_{\rm e}([\ion{S}{2}])$.

\begin{figure}
\includegraphics[width=.48\textwidth]{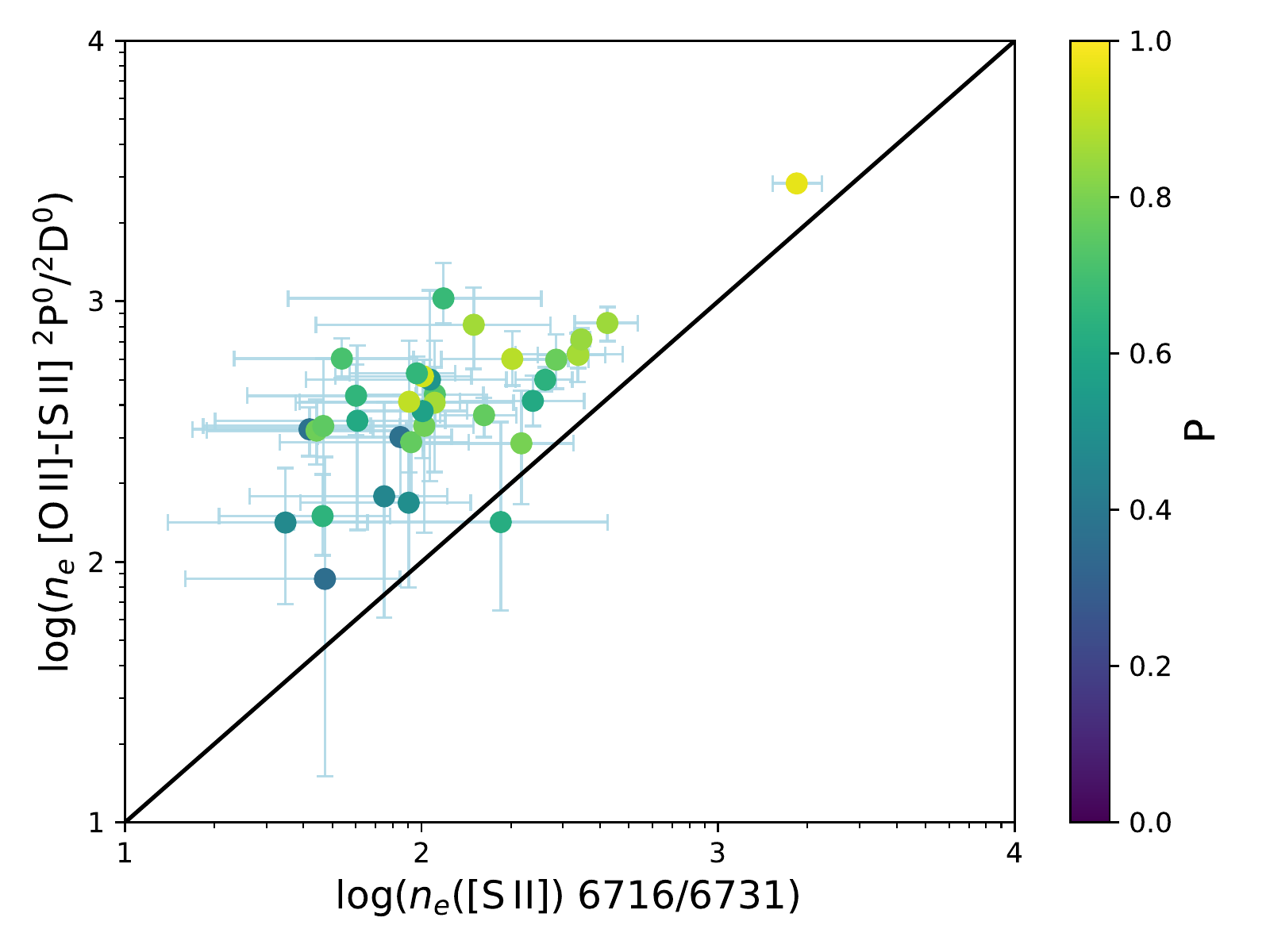}
\caption{Comparison between the average $n_{\rm e}$ obtained from the  [\ion{O}{2}] $\lambda\lambda$7319+20+30+31/$\lambda\lambda$3726+29 and [\ion{S}{2}] $\lambda\lambda$4069+76/$\lambda\lambda$6716+31 line intensity ratios and $n_{\rm e}$([\ion{S}{2}] $\lambda 6731/\lambda 6716$) for extragalactic \ion{H}{2} regions. The color of the symbols represents the value of their $P$ parameter (as in Fig~\ref{fig:TN2S2O2}). The black solid line represents a 1:1 linear relation.}
\label{fig:densis_au_neb}
\end{figure}

As a conclusion, we propose that the presence of high-density inclusions within the volume observed in the spectra of extragalactic \ion{H}{2} regions naturally explains the behavior seen in Fig.~\ref{fig:TN2S2O2}, including the bottom panel, as [\ion{O}{2}] $\lambda\lambda$7319+20+30+31/$\lambda\lambda$3726+29 and [\ion{S}{2}] $\lambda\lambda$4069+76/$\lambda\lambda$6716+31 have a similar dependency on density\footnote{In fact, the $n_{\rm e}$ dependency of [\ion{O}{2}] $\lambda\lambda$7319+20+30+31/$\lambda\lambda$3726+29 is slightly higher, and this may explain the larger number of points below the 1:1 line.}. If we use the [\ion{O}{2}] and [\ion{S}{2}] $^2$P$^{0}$/$^2$D$^{0}$ (auroral to nebular) line intensity ratios as density diagnostics instead of temperature ones by cross-matching them with $T_{\rm e}([\ion{N}{2}])$ ({\it getCrossTemDen} of PyNeb can be used), we obtain densities that are consistent with each other and systematically larger than $n_{\rm e}$([\ion{S}{2}] $\lambda 6731/\lambda 6716$) (or $n_{\rm e}$([\ion{O}{2}] $\lambda 3726/\lambda 3729$)), as shown in Fig.~\ref{fig:densis_au_neb}. $n_{\rm e}$([\ion{S}{2}] $\lambda 6731/\lambda 6716$) underestimate the density by $\sim 300 \text{ cm}^{-3}$ on average, even when $n_{\rm e}$([\ion{S}{2}] $\lambda 6731/\lambda 6716$) $<10^2 \text{ cm}^{-3}$. If $T_{\rm e}([\ion{N}{2}])$ is adopted, this underestimate of $n_{\rm e}$ has a small impact in the calculation of chemical abundances based on optical CELs, except when [\ion{S}{2}] and [\ion{O}{2}]  auroral lines are used. However, the underestimate of density is relevant in the case of ionized gas pressure determinations and the correct interpretation of the properties depending on this quantity. We remark that the presence of high density inclusions and the underestimate of density by $n_{\rm e}$([\ion{S}{2}] $\lambda 6731/\lambda 6716$) and $n_{\rm e}$([\ion{O}{2}] $\lambda 3726/\lambda 3729$) (see Fig.~\ref{fig:densities_plot}) are extremely relevant when using infrared fine structure CELs \citep{Lamarche:2022}. In analogy to what happens with temperature inhomogeneities in the optical CELs, density inhomogeneities may introduce systematic bias in the chemical abundances derived from infrared CELs. If $n_{\rm e}$ is underestimated, ionic abundances are also underestimated.

\begin{figure}
\includegraphics[width=.48\textwidth]{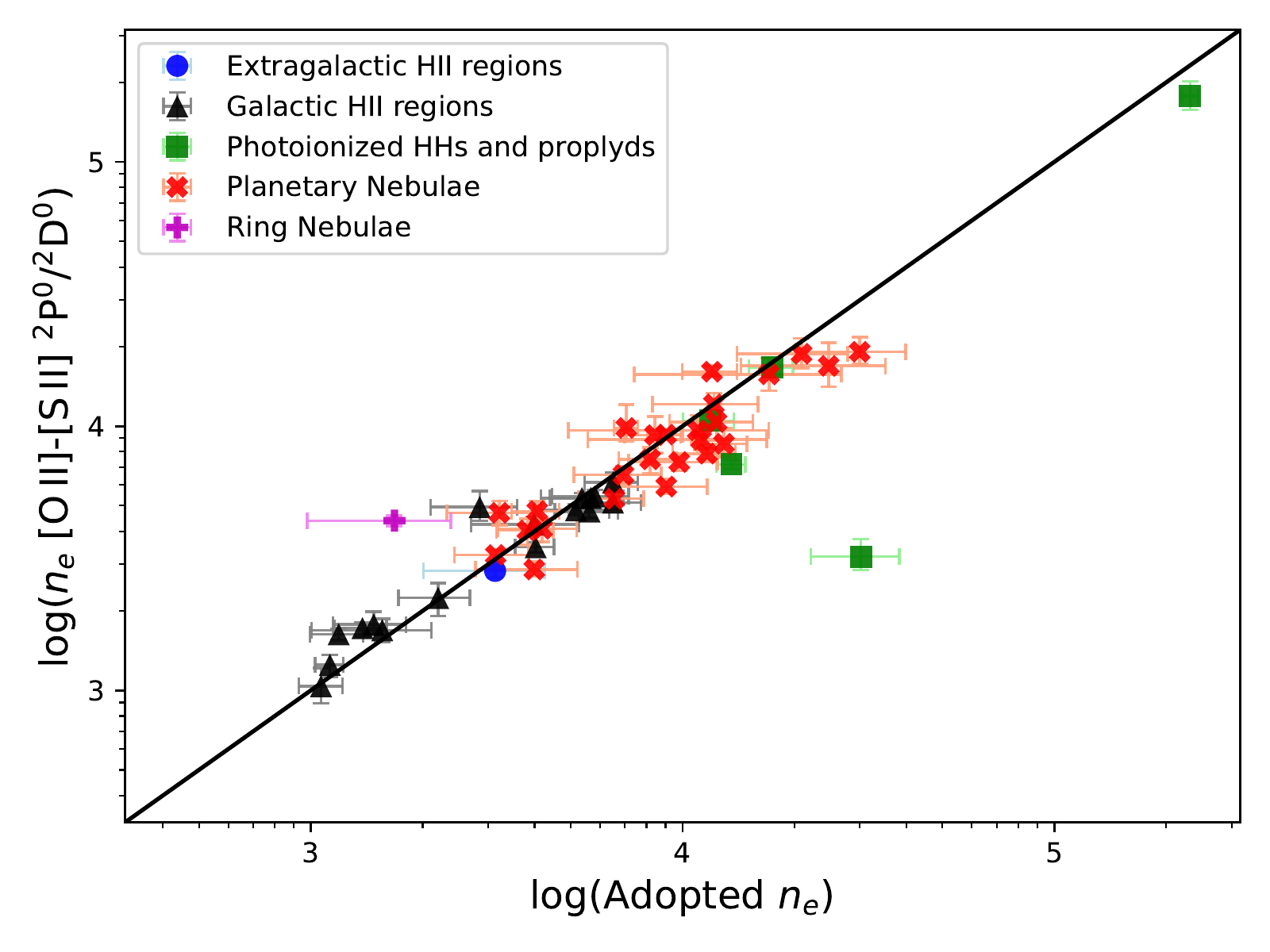}
\caption{Comparison between the average $n_{\rm e}$ obtained from the  [\ion{O}{2}] $\lambda\lambda$7319+20+30+31/$\lambda\lambda$3726+29 and [\ion{S}{2}] $\lambda\lambda$4069+76/$\lambda\lambda$6716+31 line intensity ratios and the adopted density for high density  nebulae ($n_{\rm e}$([\ion{S}{2}] $\lambda 6731/\lambda 6716$)>1000 cm$^{-3}$). The black solid line represents a 1:1 linear relation.}
\label{fig:densis_au_neb_high_dens}
\end{figure}

In the case of the high-density nebulae of the DESIRED sample, where $n_{\rm e}$([\ion{S}{2}] $\lambda 6731/\lambda 6716$) > 1000 cm$^{-3}$, we find good consistency between the adopted density and the values derived from the [\ion{O}{2}] and [\ion{S}{2}] $^2$P$^{0}$/$^2$D$^{0}$ line intensity ratios, as shown in Fig.~\ref{fig:densis_au_neb_high_dens}. It should be noted that in these cases, the adopted densities are mainly weighted by the [\ion{Cl}{3}], [\ion{Fe}{3}] and [\ion{Ar}{4}] density diagnostics. This suggests that although high-density clumps (or density gradients) may be present in all ionized nebulae, the systematic effects on the derived properties can be reduced in those objects showing higher mean densities. In contrast, in low-density nebulae, the presence of high-density clumps can go unnoticed by using $n_{\rm e}$([\ion{S}{2}] $\lambda 6731/\lambda 6716$) or $n_{\rm e}$([\ion{O}{2}] $\lambda 3726/\lambda 3729$) and therefore affecting the reliability of further calculations involving these parameters. A possible solution would be the use of [\ion{O}{2}] $\lambda\lambda$7319+20+30+31/$\lambda\lambda$3726+29 and [\ion{S}{2}] $\lambda\lambda$4069+76/$\lambda\lambda$6716+31 as density indicators together with [\ion{N}{2}] $\lambda 5755/ \lambda 6584$ to determine the temperature. Another conclusion of the discussion carried out so far is that the use of $T_{\rm e}([\ion{O}{2}])$ and $T_{\rm e}([\ion{S}{2}])$ should be avoided when $T_{\rm e}([\ion{N}{2}])$ is available. 

\begin{figure}
\includegraphics[width=.48\textwidth]{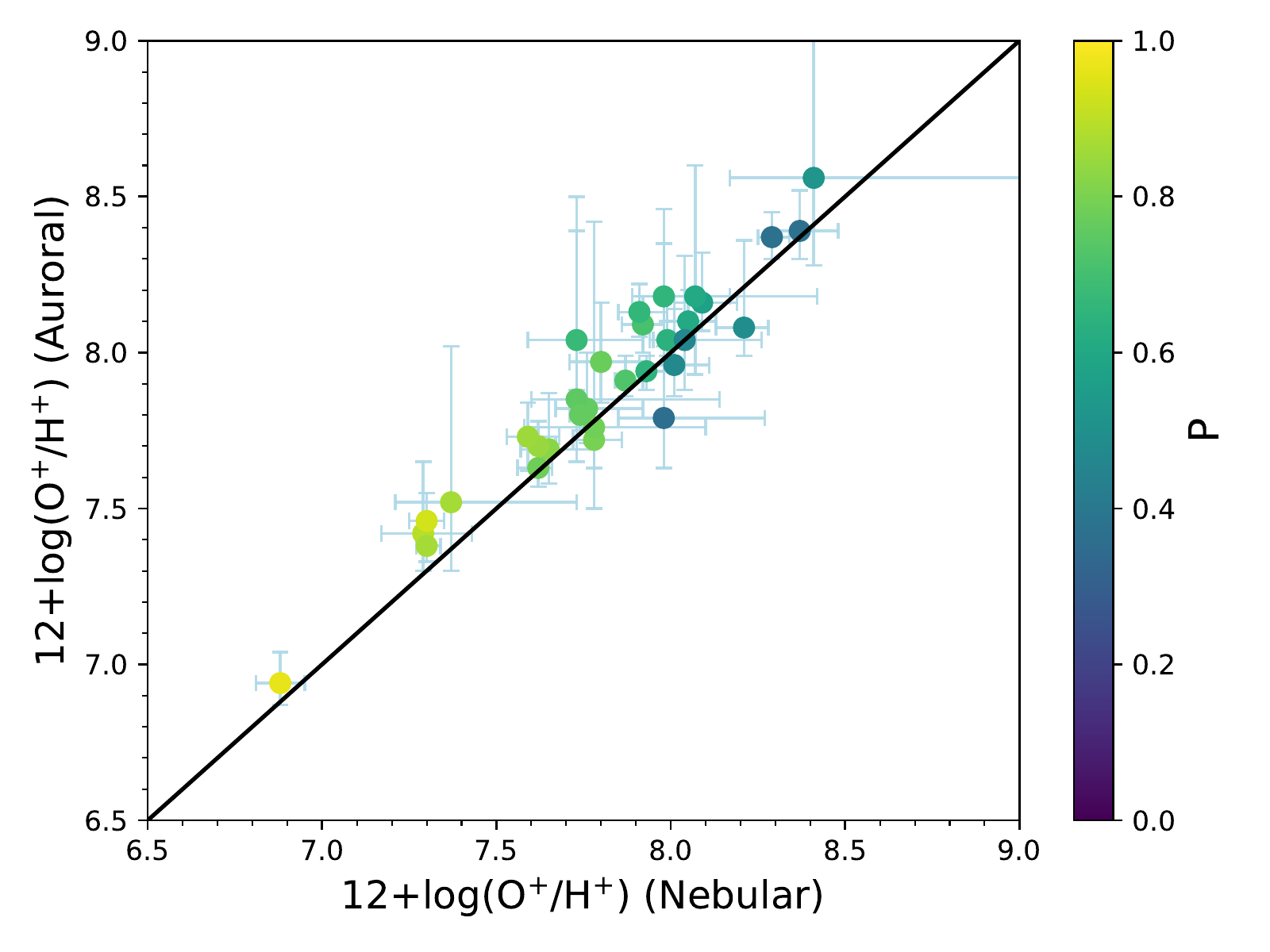}
\includegraphics[width=.48\textwidth]{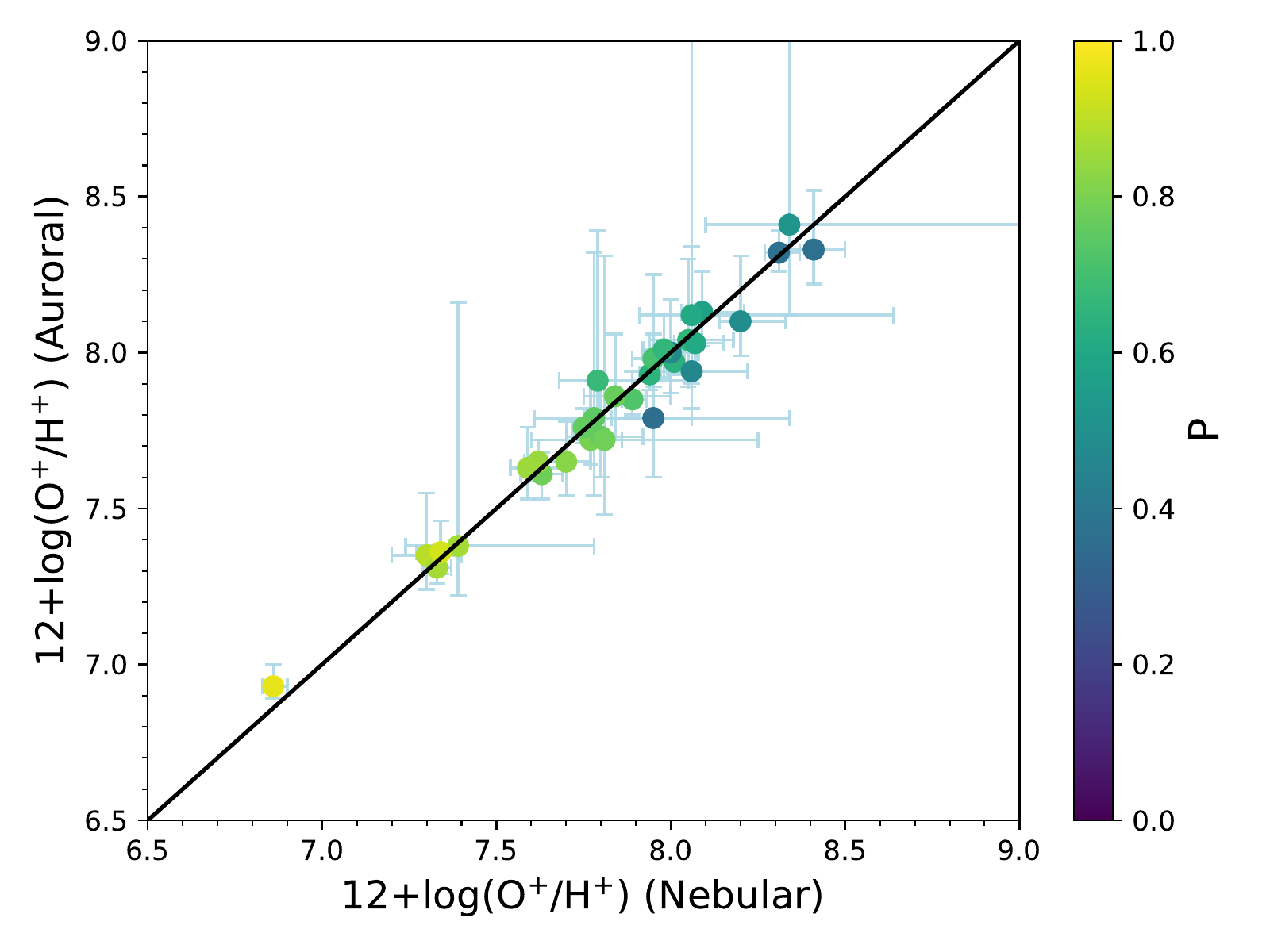}
\caption{Comparison between the O$^{+}$ abundance derived with [\ion{O}{2}] auroral and nebular lines. Upper panel: the physical conditions adopted are $T_{\rm e}([\ion{N}{2}])$ and the average of $n_{\rm e}$([\ion{S}{2}] $\lambda 6731/\lambda 6716$) and $n_{\rm e}$([\ion{O}{2}] $\lambda 3726/\lambda 3729$). Bottom panel: the physical conditions adopted are $T_{\rm e}([\ion{N}{2}])$ and the average of $n_{\rm e}$([\ion{S}{2}] $\lambda\lambda$4069+76/$\lambda\lambda$6716+31) and $n_{\rm e}$([\ion{O}{2}] $\lambda\lambda$7319+20+30+31/$\lambda\lambda$3726+29). The color of the symbols represents the value of their $P$ parameter (as in Fig~\ref{fig:TN2S2O2}). The black solid line represents a 1:1 linear relation. The relation between both quantities is tighter in the bottom panel. }
\label{fig:O_plus_red_blue}
\end{figure}

The different sensitivity of auroral and nebular [\ion{O}{2}] lines to density cause the systematic difference between the O$^+$ abundances determined with the [\ion{O}{2}] auroral and nebular lines; fact that has been described by several authors, especially in the case of PNe \citep{stasinska:1998, Escudero:2004, Rodriguez:2020}. In the case of PNe, there may be other phenomena playing a role. In the upper panel of Fig.~\ref{fig:O_plus_red_blue} we compare the O$^{+}$ abundance derived from [\ion{O}{2}] auroral and nebular lines using $T_{\rm e}([\ion{N}{2}])$ and the average of $n_{\rm e}$([\ion{S}{2}] $\lambda 6731/\lambda 6716$) and $n_{\rm e}$([\ion{O}{2}] $\lambda 3726/\lambda 3729$). In the figure, we can see that the O$^{+}$/H$^+$ ratio derived with the [\ion{O}{2}] auroral lines is up to $\sim$0.1 dex higher, on the average. In the bottom panel of Fig.~\ref{fig:O_plus_red_blue} we show the same comparison but using [\ion{O}{2}] $\lambda\lambda$7319+20+30+31/$\lambda\lambda$3726+29 and [\ion{S}{2}] $\lambda\lambda$4069+76/$\lambda\lambda$6716+31 as density indicators. As we can see, with this approach the systematic difference is removed.

\subsection{ The DESIRED temperature relationships for extragalactic HII regions}
\label{subsec:rest}

The temperature can be stratified within ionized nebulae, which is reflected in differences between the representative values of different ionic species. The most common procedure to consider the temperature stratification when deriving chemical abundances is to adopt  $T_{\rm e}([\ion{O}{3}])$ and $T_{\rm e}([\ion{N}{2}])$ for the high and low ionization volumes, respectively. Other temperature-sensitivity line ratios as $T_{\rm e}$([\ion{S}{3}] $\lambda 6312/\lambda 9069$) arises from zones of intermediate ionization \citep[e.g.][]{Berg:2015}. 

\begin{figure*}
\begin{minipage}{\textwidth}
\centering
\includegraphics[width=.48\textwidth]{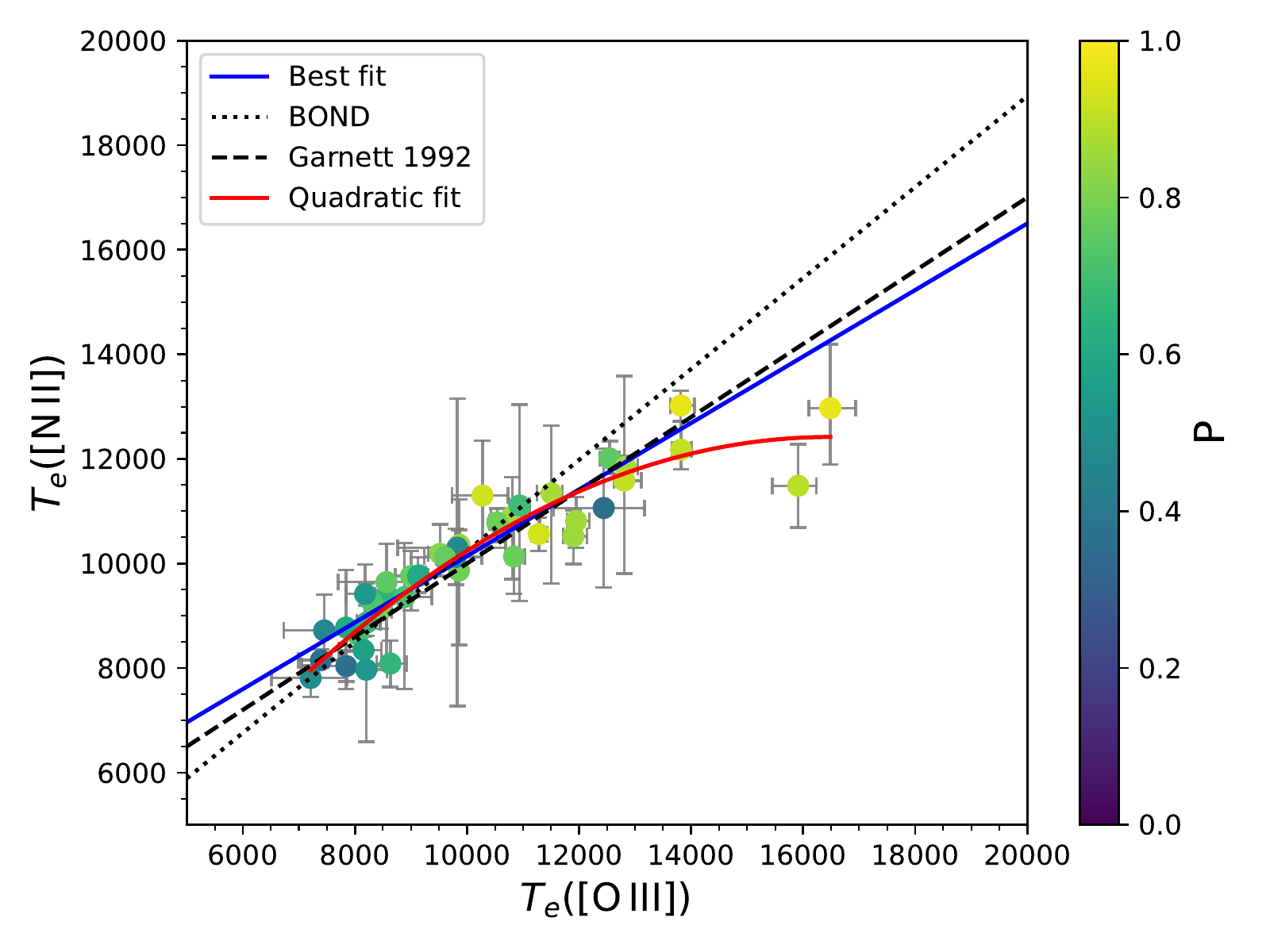}
\includegraphics[width=.48\textwidth]{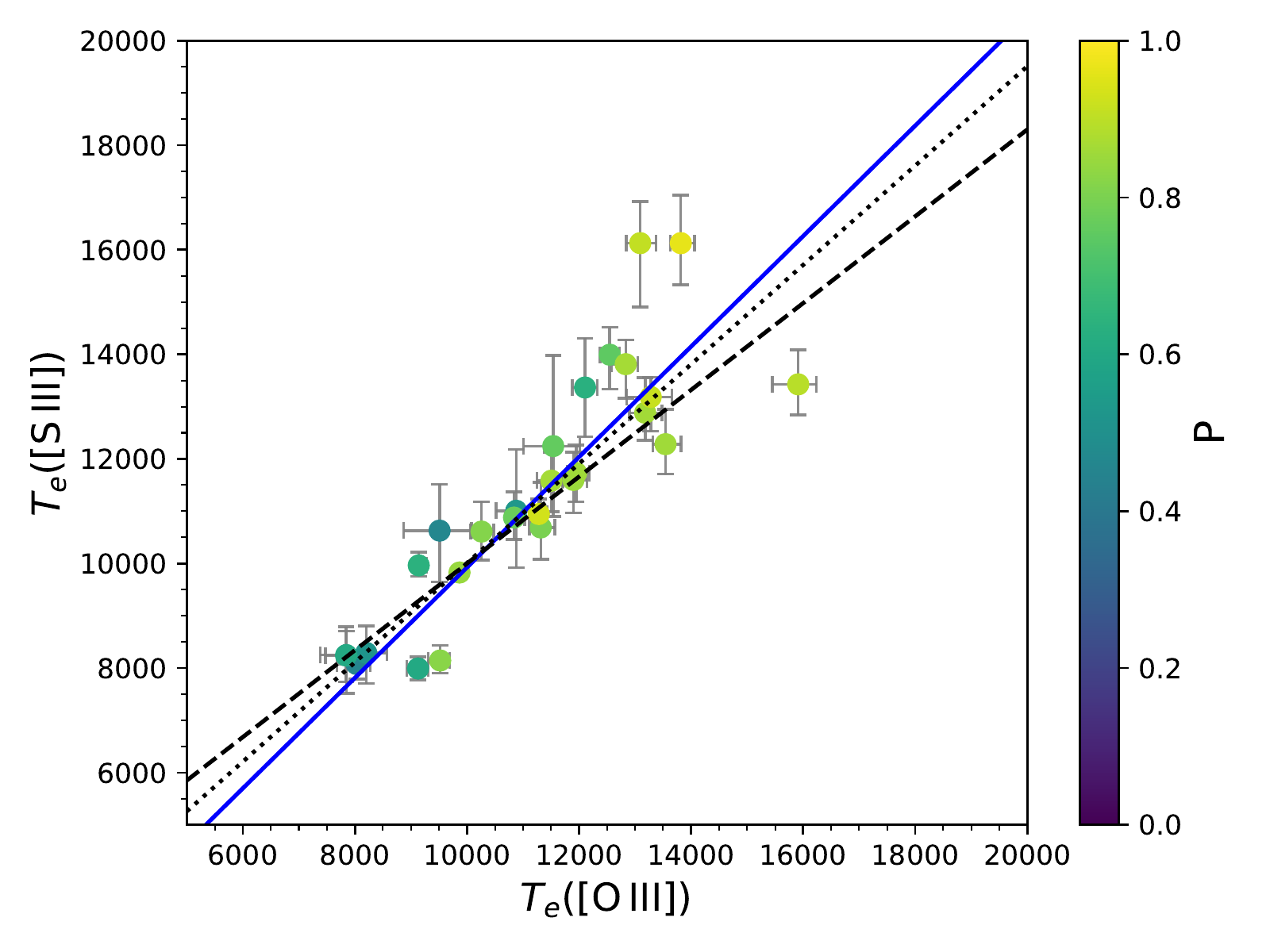}
\includegraphics[width=.48\textwidth]{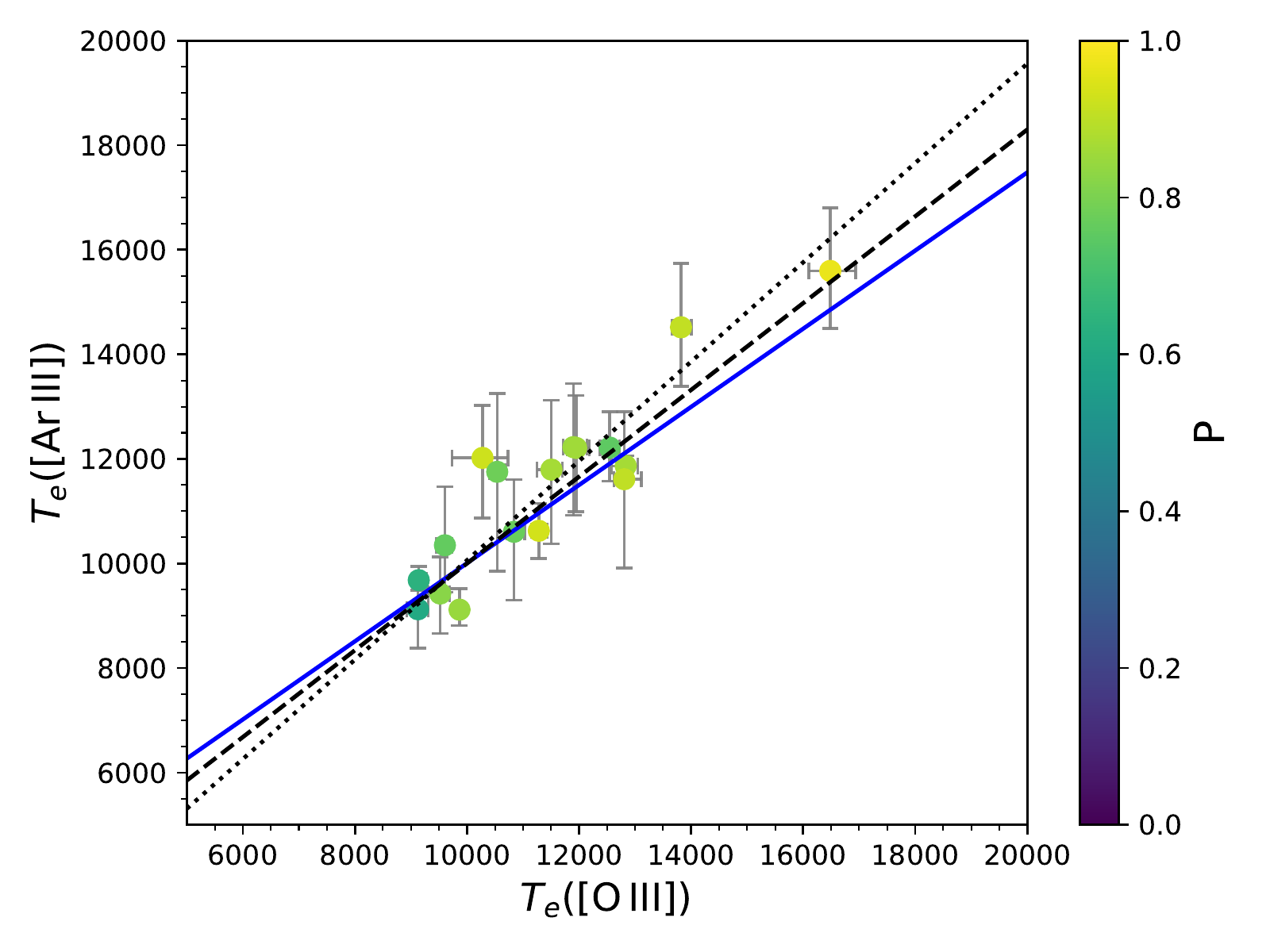}
\includegraphics[width=.48\textwidth]{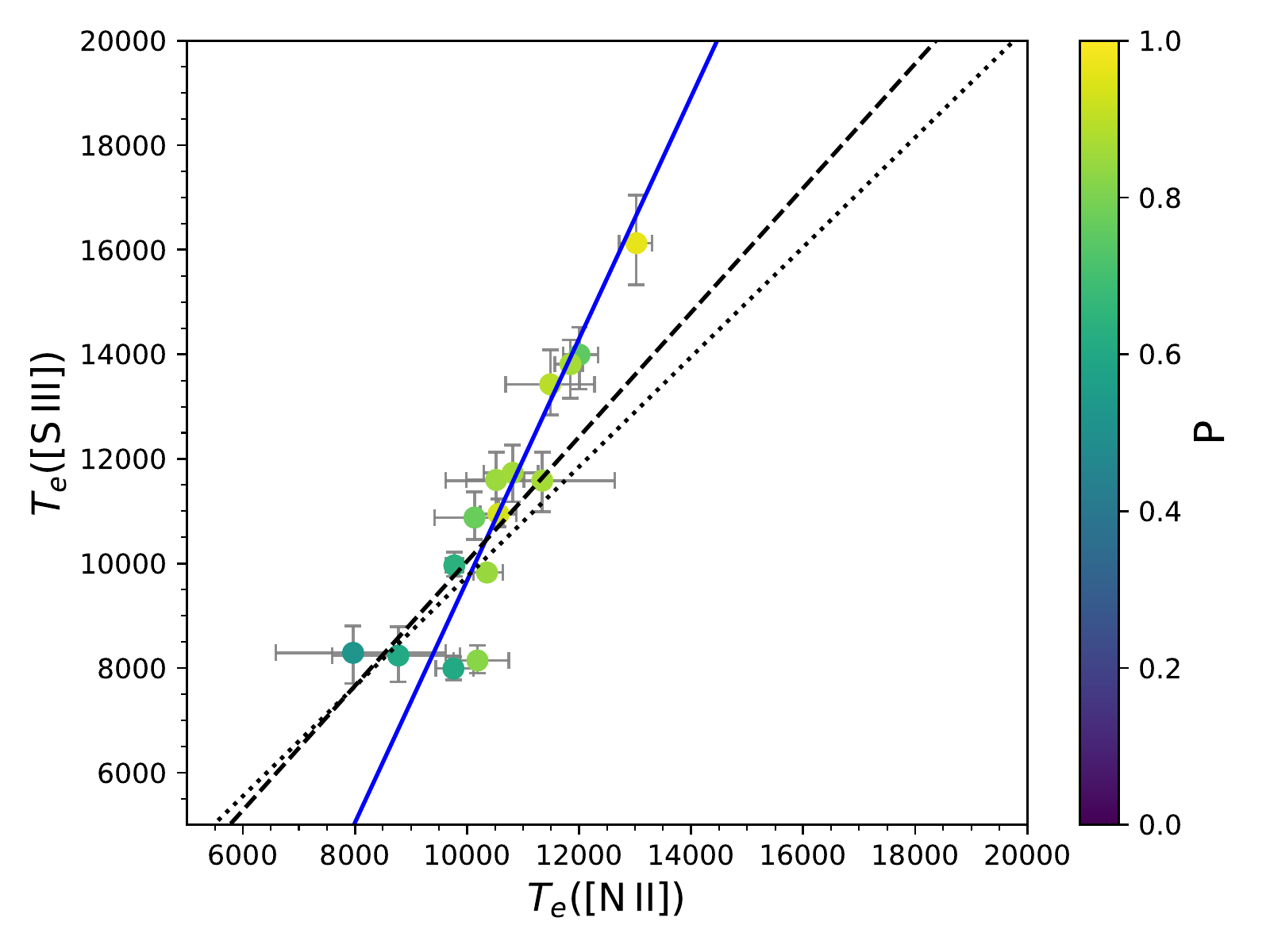}
\includegraphics[width=.48\textwidth]{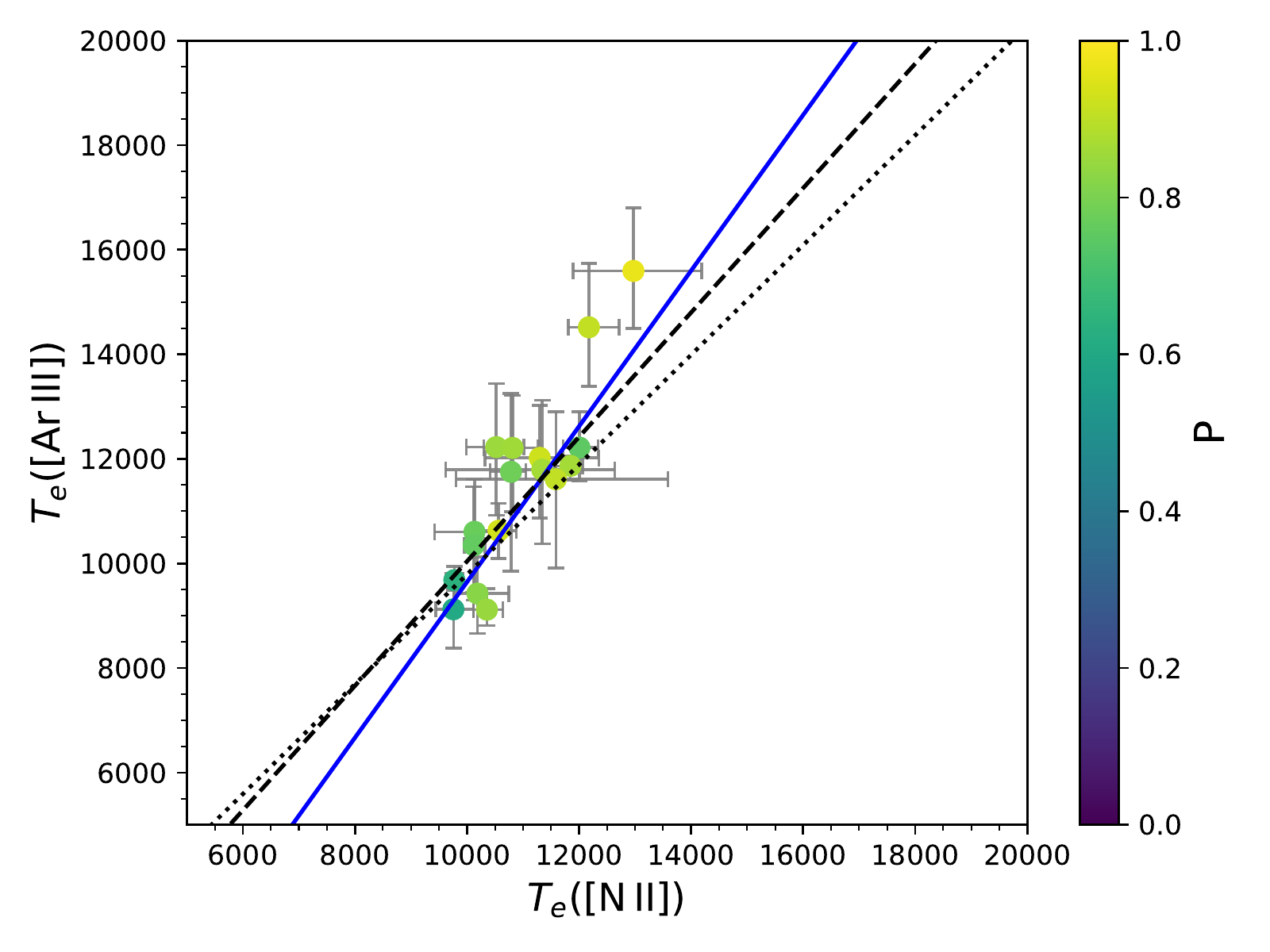}
\includegraphics[width=.48\textwidth]{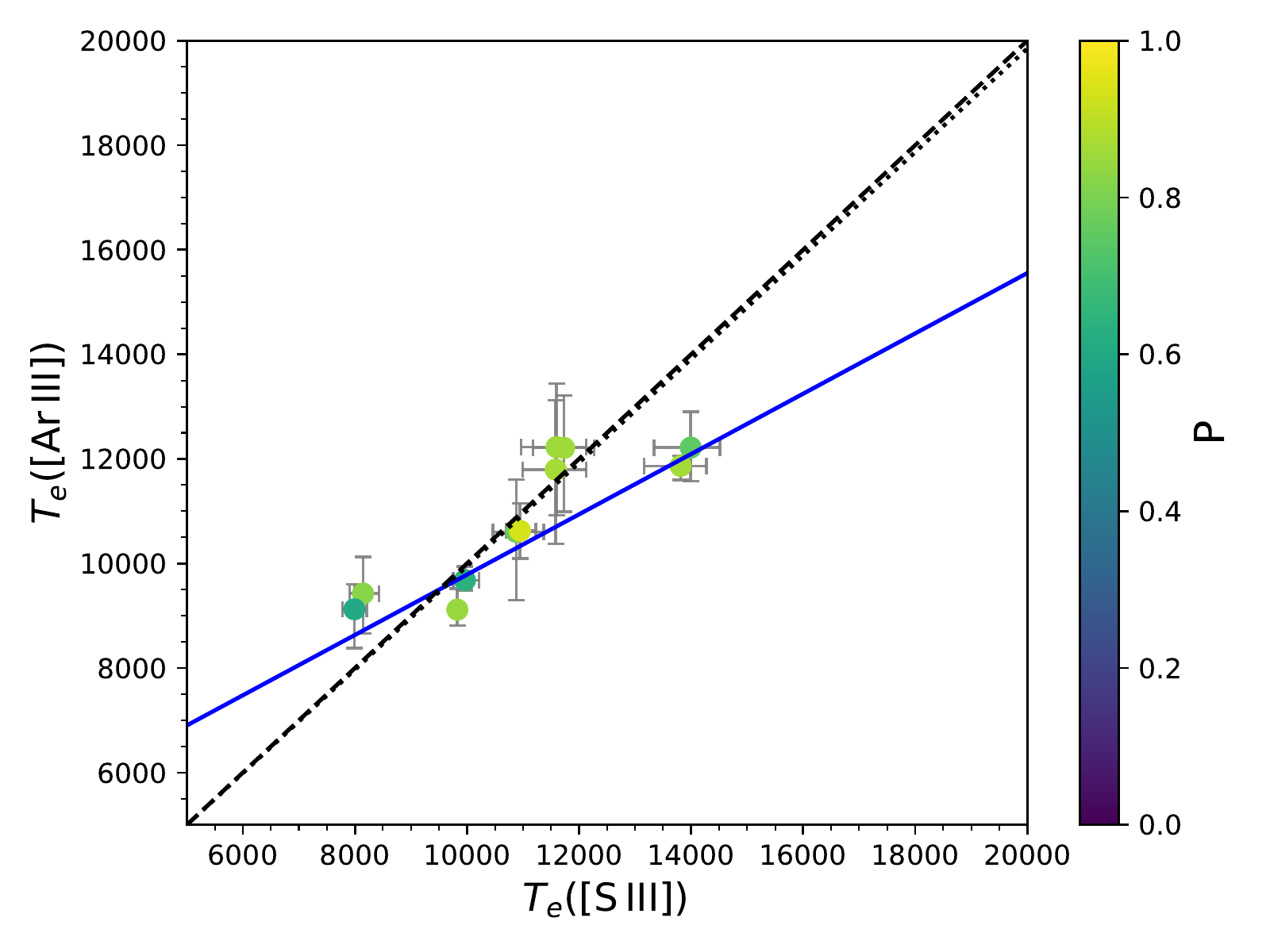}
\end{minipage}
\caption{Temperature relations of the DESIRED extragalactic \ion{H}{2} regions. {\it Top panels:} $T_{\rm e}$[\ion{N}{2}] (left) and $T_{\rm e}$([\ion{S}{3}]) (right) as a function of $T_{\rm e}$([\ion{O}{3}]). {\it Middle panels:} The $T_{\rm e}$([\ion{O}{3}]) - $T_{\rm e}$([\ion{Ar}{3}]) relation (left) and the $T_{\rm e}$([\ion{N}{2}])- $T_{\rm e}$([\ion{S}{3}]) relation (right). {\it Bottom panels:} The $T_{\rm e}$([\ion{N}{2}])-$T_{\rm e}$([\ion{Ar}{3}]) relation (left) and the $T_{\rm e}$([\ion{S}{3}]) and $T_{\rm e}$([\ion{Ar}{3}]) relation (right). 
The solid blue line represent the linear fit of the data. The dashed and dotted lines indicates the model predictions of \citet{Garnett:1992} and the BOND models \citep{ValeAsari:2016}, respectively. The red solid line in the upper left panel represents a second degree polynomial fit. }
\label{fig:temps_rels}
\end{figure*}

Fig.~\ref{fig:temps_rels} shows the DESIRED temperature relationships derived from different diagnostics associated with different ionization volumes of the gas. In each plot, we include the best fit to the data, the predicted linear fit from the BOND models (see Section~\ref{sec:phot_models}) and the model-derived relations of \citet{Garnett:1992}. In Table~\ref{tab:temps_rel} we present the DESIRED temperature relations (column 4) and the scatter and number of objects considered in each case (columns 5-6).

Upper left panel of Fig.~\ref{fig:temps_rels} shows the $T_{\rm e}([\ion{O}{3}])$ \textit{vs.} $T_{\rm e}([\ion{N}{2}])$ relationship defined for the DESIRED extragalactic \ion{H}{2} regions and the linear fit to the data. There is wealth of works devoted to study this relation in the literature \citep[e.g.,][]{Campbell:1986, Garnett:1992, Pagel:1992, Pilyugin:2007, Esteban:2009, arellano:2020b, Berg:2020, Rogers:2021, Rogers:2022}, finding a relatively high scatter. \citet{arellano:2020b} showed that part of the dispersion is due to the effects of metallicity and the degree of ionization and therefore related to nebular properties. With the DESIRED extragalactic \ion{H}{2} regions, we minimize spurious scatter that can occur by using low signal-to-noise spectra and confirm a departure from a linear relationship. This departure becomes larger with the degree of ionization (and lower metallicities) and becomes noticeable when $T_{\rm e}([\ion{O}{3}])>10000 \text{ K}$. Such a deviation from a linear relationship has been reported by several authors previously \citep[e.g.][]{Pilyugin:2007, arellano:2020b}. In Fig.~\ref{fig:temps_rels} we also include a  quadratic fit to the data, which is only valid within $7000\text{ K} < T_{\rm e}([\ion{O}{3}]) < 16,500 \text{ K}$. However, its shape at $T_{\rm e}([\ion{O}{3}]) > 13,000 \text{ K}$ is determined by the position of only two objects in the diagram, NGC~5408 \citep{Esteban:2014} and NGC~2363 \citep{Esteban:2009}. As shown in Table~\ref{tab:temps_rel}, the photoionization models described in Section~\ref{sec:phot_models} are not able to reproduce the curvature observed between $T_{\rm e}([\ion{O}{3}])$ and $T_{\rm e}([\ion{N}{2}])$ in the DESIRED extragalactic  \ion{H}{2} regions. This curvature is not reproduced either if the models are weighted with the methodology proposed by \citet{Amayo:2021} in their equation~(4), considering the observational sample compiled by \citet{Zurita:2021} and \citet{Izotov:2007}.

 \citet[][see their fig.~2 and equation~(4)] {mendezdelgado:2023} derive a tight linear relation between $T_{\rm e}([\ion{N}{2}])$ and the average temperature of the high ionization volume, $T_{\rm 0}(\text{O}^{2+})$, parameter that can be used to estimate the O/H ratio without the bias induced by temperature inhomogeneities which does affect the abundances determined using $T_{\rm e}([\ion{O}{3}])$. Nevertheless, $T_{\rm e}([\ion{N}{2}])$ is usually very difficult to determine in faint low metallicity \ion{H}{2} regions and the only available temperature diagnostic is often $T_{\rm e}([\ion{O}{3}])$. In such cases, it is possible to use the relations presented in Table~\ref{tab:temps_rel} to estimate $T_{\rm e}([\ion{N}{2}])$ and consequently $T_{\rm 0}(\text{O}^{2+})$, using equation~(4) from \citet{mendezdelgado:2023}. 

Fig.~\ref{fig:temps_rels} includes also temperature relations involving the uncommon $T_{\rm e}([\ion{Ar}{3}])$, derived from the [\ion{Ar}{3}] $\lambda 5192/ \lambda 7135$ intensity ratio. DESIRED contains the largest collection of $T_{\rm e}([\ion{Ar}{3}])$ determinations for \ion{H}{2} regions. Considering that the ionization conditions of Ar$^{2+}$ are different than those of S$^{2+}$ or O$^{2+}$, we cannot strictly say that $T_{\rm e}([\ion{Ar}{3}])$ is also representative of the same ionization volume where S$^{2+}$ or O$^{2+}$ lie. 
In the middle left panel of Fig.~\ref{fig:temps_rels}, we can see that $T_{\rm e}([\ion{Ar}{3}])$ follows a rather linear relationship with $T_{\rm e}([\ion{O}{3}])$ for the spectra of extragalactic \ion{H}{2} regions. Lower left panel of Fig.~\ref{fig:temps_rels} shows that the behavior of the $T_{\rm e}([\ion{Ar}{3}])$ {\it vs.} $T_{\rm e}([\ion{N}{2}])$ relationship has certain similarity to the $T_{\rm e}([\ion{O}{3}])$ {\it vs.} $T_{\rm e}([\ion{N}{2}])$ one. The two objects with the highest $T_{\rm e}$ show some deviation towards larger $T_{\rm e}([\ion{Ar}{3}])$ values. This  agrees with the results obtained by \citet{mendezdelgado:2023}, who find that $T_{\rm e}([\ion{Ar}{3}])$ and $T_{\rm e}([\ion{O}{3}])$ seem to be affected by $t^2$ in a similar way.

Upper right panel of Fig.~\ref{fig:temps_rels} shows the $T_{\rm e}([\ion{S}{3}])$ {\it vs.} $T_{\rm e}([\ion{O}{3}])$ relationship defined by the DESIRED spectra of extragalactic \ion{H}{2} regions. The slope of the linear fit to the data is very similar to that obtained from model predictions of \citet{Garnett:1992} and \citet{ValeAsari:2016}. However, the dispersion around the fit is larger for the observational points with higher $T_{\rm e}$ and $P$ parameter values. Those points correspond mainly to spectra of \ion{H}{2} regions of the Magellanic Clouds \citep{dominguez:2022}. As mentioned in Section~\ref{sec:pc}, some of our estimates of $T_{\rm e}([\ion{S}{3}])$ might not be completely free of telluric absorptions in the [\ion{S}{3}] $\lambda 9069, 9531$ lines and this fact may enhance the derived $T_{\rm e}([\ion{S}{3}])$. In the middle right panel of Fig.~\ref{fig:temps_rels} we present the $T_{\rm e}([\ion{S}{3}])$ {\it vs.} $T_{\rm e}([\ion{N}{2}])$ relationship, which follows a linear relation with a remarkably small dispersion except for the spectra with lowest $T_{\rm e}$ values. Our linear fit has a steeper slope compared to the relations found by \citet{Berg:2020} or \citet{Rogers:2021}. This might be due to the larger proportion of DESIRED spectra with $T_{\rm e}([\ion{S}{3}])$ $>$ 12000 compared to the samples of \citet{Berg:2020} or \citet{Rogers:2021}, where the vast majority of objects are below that $T_{\rm e}$ value. The higher slope defined by the DESIRED spectra may be related to the fact that --as it was also noted in the 
$T_{\rm e}([\ion{O}{3}])$ {\it vs.} $T_{\rm e}([\ion{N}{2}])$ and $T_{\rm e}([\ion{Ar}{3}])$ {\it vs.} $T_{\rm e}([\ion{N}{2}])$ relationships-- $T_{\rm e}([\ion{S}{3}])$ tends to be higher than $T_{\rm e}([\ion{N}{2}])$ in spectra with larger $T_{\rm e}$ values. As it has been said before, and following the results by \citet{mendezdelgado:2023}, this indicates that $T_{\rm e}([\ion{S}{3}])$ may also be affected by $t^2$. This possibility, however, requires a verification since the telluric absorptions in the [\ion{S}{3}] $\lambda 9069, 9531$ lines act in the same direction as $t^2$.

\begin{table*}
\caption{The DESIRED temperature relations for extragalactic \ion{H}{2} regions.}
\label{tab:temps_rel}
\begin{tabular}{cccccc}
\hline
& Models (K)  & $\sigma$ (K) & Observations (K) & $\sigma$ (K)& N\\
\hline
\multirow{4}{*}{$x=T_{\rm e}([\ion{O}{3}])$} &$T_{\rm e}([\ion{N}{2}])=0.87x + 1540$ & 600 &$T_{\rm e}([\ion{N}{2}])=0.65 (\pm 0.04) x +3640 (\pm 360)$&400 & 42\\

& $T_{\rm e}([\ion{N}{2}])=-1.11\times 10^{-5} x^2+ 1.06 x + 760 $  & &$T_{\rm e}([\ion{N}{2}])=-5.19\times 10^{-5} x^2+ 1.71 x - 1680 $& & \\

 &$T_{\rm e}([\ion{S}{3}])=0.95x+510$& 200& $T_{\rm e}([\ion{S}{3}])=1.06 (\pm 0.10) x -640(\pm 1070)$& 640 &27\\
 
 & $T_{\rm e}([\ion{Ar}{3}])=0.95x +560$ & 250 &$T_{\rm e}([\ion{Ar}{3}])=0.78 (\pm 0.09) x +2120(\pm 1000)$& 580 & 17\\
 
\hline

\multirow{3}{*}{$x=T_{\rm e}([\ion{S}{3}])$} &$T_{\rm e}([\ion{N}{2}])= 0.92x +1010  $ & 470 &$T_{\rm e}([\ion{N}{2}])=0.43 (\pm 0.06) x +5820(\pm 500)$ & 330 & 15 \\

 &$T_{\rm e}([\ion{Ar}{3}])=0.99 x + 50 $ & 70 &$T_{\rm e}([\ion{Ar}{3}])=0.58 (\pm 0.09) x +4020(\pm 930)$ &380 & 11 \\
 
 &$T_{\rm e}([\ion{O}{3}])= 1.04x -480  $ & 210 &$T_{\rm e}([\ion{O}{3}])=0.95 (\pm0.10 ) x + 600 (\pm 950 )$& 830&27\\

\hline

\multirow{3}{*}{$x=T_{\rm e}([\ion{N}{2}])$} &$T_{\rm e}([\ion{S}{3}])=1.05 x -750  $ & 510 &$T_{\rm e}([\ion{S}{3}])=2.31 (\pm 0.24) x -13470(\pm 2570)$& 920 & 15\\

 &$T_{\rm e}([\ion{Ar}{3}])= 1.05 x -710  $ & 480& $T_{\rm e}([\ion{Ar}{3}])=1.57 (\pm 0.24) x -6070(\pm 2570)$& 790 & 17\\
 
 &$T_{\rm e}([\ion{O}{3}])=1.09 x -1190  $ & 660 &$T_{\rm e}([\ion{O}{3}])=1.55  (\pm 0.09) x- 5620 (\pm 860 )$&710&42 \\

\hline
\end{tabular}
\end{table*}

\section{Discussion and  Conclusions}
\label{sec:dis}

In this paper we present a first study based on DEep Spectra of Ionized REgions Database (DESIRED), a collection of high-quality deep optical spectra of ionized nebulae from the literature. The data were mostly obtained with 8-10m telescopes over more than 20 years by our research group and have been carefully reduced in an homogeneous way. DESIRED contains $\sim29380$ emission lines of 190 spectra of Galactic and extragalactic \ion{H}{2} regions, PNe, RNe as well as photoionized HH objects and one proplyd of the Orion Nebula. The main aim of the study of the DESIRED sample as a whole is to draw attention to and quantify systematic effects that may bias the determination of physical conditions and chemical abundances of ionized gas in the Universe, as well as to better understand the physics of the formation of certain faint emission lines. The philosophy of DESIRED has been to prioritize the quality and depth of the spectra over their quantity in the design of the observations. However, due to the continuity of the project over the years, the number of objects has been increasing substantially, reaching a level comparable to that of a small survey, with the possibility of increasing in the future, especially with observations of low-metallicity (12+log(O/H) $<$ 8.0) objects with very large aperture telescopes. Although formally this is the first paper based on the exploitation of DESIRED, it was also used by \citet{mendezdelgado:2023}, who analyzed the systematic bias introduced by temperature fluctuations in the determination of ionized abundances in \ion{H}{2} regions, a task impossible to perform with any other sample.

In this paper, we explore the density structure of the DESIRED objects as well as the $T_{\rm e}$-$T_{\rm e}$ relations for extragalactic \ion{H}{2} regions. Regarding the density structure, we show that [\ion{Cl}{3}] $\lambda 5538/ \lambda 5518$, [\ion{Fe}{3}] $\lambda 4658/ \lambda 4702$ and [\ion{Ar}{4}] $\lambda 4740/ \lambda 4711$ are good density indicators when $10^3\text{ cm}^{-3}<n_{\rm e}<10^6\text{ cm}^{-3}$, whereas [\ion{S}{2}] $\lambda 6731/\lambda 6716$, [\ion{O}{2}] $\lambda 3726/\lambda 3729$ are density sensitive when $10^2\text{ cm}^{-3}<n_{\rm e}<10^4\text{ cm}^{-3}$. We find good consistency between diagnostics associated to different ionization volumes when the sensitivity ranges are similar. This implies that the sensitivity range of the diagnostics used is a more relevant parameter to obtain good density determinations than their selection attending to the ionization volume in which the abundance is determined. Based on these findings, in Section~\ref{sec:density_structure} we present simple and consistent criteria to derive the representative density for chemical abundance studies in the optical range.

We demonstrate that $n_{\rm e}$([\ion{S}{2}] $\lambda 6731/\lambda 6716$) and $n_{\rm e}$([\ion{O}{2}] $\lambda 3726/\lambda 3729$) are biased towards lower densities in extragalactic \ion{H}{2} regions due to the presence of density inhomogeneities and the non-linear sensitivity of these indicators. This is inferred from the behavior of [\ion{O}{2}] $\lambda \lambda$7319+20+30+31/$\lambda \lambda$ 3726+29 and [\ion{S}{2}] $\lambda \lambda$4069+76/$\lambda \lambda$6716+31 intensity ratios, commonly used to compute $T_{\rm e}$([\ion{O}{2}]) and $T_{\rm e}$([\ion{S}{2}]), respectively. When $T_{\rm e}$([\ion{O}{2}]) and $T_{\rm e}$([\ion{S}{2}]) --derived adopting $n_{\rm e}$([\ion{S}{2}] $\lambda 6731/\lambda 6716$) and $n_{\rm e}$([\ion{O}{2}] $\lambda 3726/\lambda 3729$-- are compared with $T_{\rm e}$([\ion{N}{2}] $\lambda 5755/ \lambda 6584$) they show systematic trends that can not be explained  by observational errors, mismatches between the ionization volumes, recombination contribution or temperature fluctuations, but are explained by the presence of an inhomogeneous density structure. The sensitivity of [\ion{O}{2}] $\lambda \lambda$7319+20+30+31/$\lambda \lambda$ 3726+29 and [\ion{S}{2}] $\lambda \lambda$4069+76/$\lambda \lambda$6716+31 to higher densities -- $10^2\text{ cm}^{-3}<n_{\rm e}<10^6\text{ cm}^{-3}$-- makes them better diagnostics than  $n_{\rm e}$([\ion{S}{2}] $\lambda 6731/\lambda 6716$) or $n_{\rm e}$([\ion{O}{2}] $\lambda 3726/\lambda 3729$) when they are cross-correlated with $T_{\rm e}$([\ion{N}{2}]), since they are sensitive to the presence of high-density clumps.

In the analysis of extragalactic \ion{H}{2} regions, the density underestimate of $n_{\rm e}$([\ion{S}{2}] $\lambda 6731/\lambda 6716$) or $n_{\rm e}$([\ion{O}{2}] $\lambda 3726/\lambda 3729$) is of $\sim 300 \text{ cm}^{-3}$ on the average, even if the aforementioned diagnostics give values consistent with the low density limit ($< 100 \text{ cm}^{-3}$). The implications of this underestimate in the calculation of chemical abundances from optical spectra are rather small, being constrained up to $\sim 0.1 $ dex when O$^{+}$ abundances are estimated with the [\ion{O}{2}] $\lambda \lambda 7319+20+30+31$ CELs. However, the density underestimate is critical for studies based on infrared fine structure CELs. For instance, [\ion{O}{3}] 88$\mu$m decreases its emissivity $\sim$40 per cent when $n_{\rm e}$ changes from 200 cm$^{-3}$ to 500 cm$^{-3}$, implying an increase of the derived chemical abundances of $\sim 70$ per cent. Density diagnostics in the infrared such as [\ion{O}{3}] $88\mu \text{m}/52\mu \text{m}$ are likely to suffer a bias towards lower densities even to a greater extent than $n_{\rm e}$([\ion{S}{2}] $\lambda 6731/\lambda 6716$) or $n_{\rm e}$([\ion{O}{2}] $\lambda 3726/\lambda 3729$) due to their different sensitivity ranges  (see Fig.~\ref{fig:density_diagnostics}).

\text{Finally, we} present the temperature relations for  DESIRED extragalactic \ion{H}{2} regions  considering the $T_{\rm e}$-sensitive [\ion{N}{2}] $\lambda 5755/ \lambda 6584$, [\ion{O}{3}] $\lambda 4363/ \lambda 5007$, [\ion{Ar}{3}] $\lambda 5192/ \lambda 7135$ and [\ion{S}{3}] $\lambda 6312/ \lambda 9069$ intensity ratios. The availability of such a number of different $T_{\rm e}$ diagnostics permits us to calculate chemical abundances considering the stratification of temperature at different ionization volumes. We confirm a departure from a linear fit in the $T_{\rm e}$([\ion{O}{3}]) {\it vs.} $T_{\rm e}$([\ion{N}{2}]) relationship, which is more prominent in regions of lower metallicity. This is consistent with the presence of larger  temperature inhomogeneities in the high ionization volume of these systems, as \citet{mendezdelgado:2023} propose in a recent study. A similar departure from a linear fit seems also to be present in the $T_{\rm e}$([\ion{Ar}{3}]) {\it vs.} $T_{\rm e}$([\ion{N}{2}]) and $T_{\rm e}$([\ion{S}{3}]) {\it vs.} $T_{\rm e}$([\ion{N}{2}]) relationships of the DESIRED spectra of extragalactic \ion{H}{2} regions.

\section*{Acknowledgements}
We thank the referee, Grażyna Stasińska, for her careful revision of the manuscript and useful comments that have contributed to increase the quality of the paper. JEMD thank to A. Amayo for her help regarding the handling of the BOND photoionization models. JEM-D, OE and KK gratefully acknowledge funding from the Deutsche Forschungsgemeinschaft (DFG, German Research Foundation) in the form of an Emmy Noether Research Group (grant number KR4598/2-1, PI Kreckel).  CE and JG-R acknowledge support from the Agencia Estatal de Investigaci\'on del Ministerio de Ciencia e Innovaci\'on (AEI-MCINN) under grant {\it Espectroscop\'ia de campo integral de regiones H\,II locales. Modelos para el estudio de regiones H\,II extragal\'acticas} with reference 10.13039/501100011033. JG-R acknowledges support from an Advanced Fellowship under the Severo Ochoa excellence program CEX2019-000920-S. JG-R and VG-LL acknowledge financial support from the Canarian Agency for Research, Innovation and Information Society (ACIISI), of the Canary Islands Government, and the European Regional Development Fund (ERDF), under grant with reference ProID2021010074. CE, JG-R and VG-LL acknowledge support under grant P/308614 financed by funds transferred from the Spanish Ministry of Science, Innovation and Universities, charged to the General State Budgets and with funds transferred from the General Budgets of the Autonomous Community of the Canary Islands by the MCIU.

\section*{DATA AVAILABILITY}

The original data is public and available in the references cited in Tables \ref{tab:photoionizedHHs}-\ref{tab:Galactic_references}. All our calculations are present in the files of the online material. DESIRED files, although already public, can be shared upon reasonable request.



\bibliographystyle{mnras}
\bibliography{Mendez}



\appendix

\section{Appendix}
\label{appendix}

\begin{table*}
\caption{DESIRED Photoionized Herbig-Haro objects and proplyds.}
\label{tab:photoionizedHHs}
\begin{tabular}{ccccccccccccc}
\hline
Object & Spectrograph & Telescope & Reference \\
\hline
HH202S& \multirow{7}{*}{UVES} & \multirow{7}{*}{VLT} & \citet{MesaDelgado:2009}\\
HH204& & & \citet{mendezdelgado:2021b}\\
HH514I (jet base) & & & \multirow{2}{*}{\citet{mendezdelgado:2022b}}\\
HH514II (knot) & & & \\
HH529II & & & \multirow{2}{*}{\citet{mendezdelgado:2021a}}\\
HH529III & & & \\
Proplyd 170-337& & & \citet{mendezdelgado:2022b}\\
\hline
\end{tabular}
\end{table*}

\begin{table*}
\caption{DESIRED Galactic Ring Nebulae.}
\label{tab:RNs_references}
\begin{tabular}{ccccccccccccc}
\hline
Object & Zone & Ionizing star type & Spectrograph & Telescope & Reference \\
\hline

\multirow{5}{*}{G2.4+1.4} & A1 & \multirow{5}{*}{WO2} & OSIRIS & GTC & \multirow{5}{*}{\citet{Esteban:2016}}\\
 & A2 &  & MagE &Clay Telescope &\\
 & A3 &  & \multirow{3}{*}{OSIRIS} & \multirow{3}{*}{GTC} &\\
 & A4 &  & & &\\
 & A5 &  & & &\\

\hline
\multirow{6}{*}{NGC 6888} & A1 & \multirow{6}{*}{WN6} & \multirow{6}{*}{OSIRIS} & \multirow{6}{*}{GTC} & \multirow{6}{*}{\citet{Esteban:2016}}\\
 & A2 &  & & &\\
 & A3 &  & & &\\
 & A4 &  & & &\\
 & A5 &  & & &\\
 & A6 &  & & &\\
\hline
\multirow{6}{*}{NGC 7635} & A1 & \multirow{6}{*}{O6.5f} & \multirow{6}{*}{OSIRIS} & \multirow{6}{*}{GTC} & \multirow{6}{*}{\citet{Esteban:2016}}\\
 & A2 &  & & &\\
 & A3 &  & & &\\
 & A4 &  & & &\\
 & A5 &  & & &\\
 & A6 &  & & &\\
\hline
RCW 52 & 10:46:02.5 $-$58:38:05 &  O8 & \multirow{2}{*}{MagE} & \multirow{2}{*}{Clay Telescope} & \multirow{2}{*}{\citet{Esteban:2016}}\\
RCW 58 & 11:06:00.1 $-$65:32:06 &  WN8 & &  & \\
Sh~2-298& 07:18:28.10 $-$13:17:19.7 & WN4 & UVES  & VLT & \citet{Esteban:2017}\\
Sh~2-308& 06:53:02.2 $-$23:53:30 & WN4&  MagE  & Clay Telescope & \citet{Esteban:2016}\\
\hline
\end{tabular}
\end{table*}

\begin{table*}
\caption{DESIRED Extragalactic \ion{H}{2} regions.}
\label{tab:extragalactic_references}
\begin{tabular}{ccccccccccccc}
\hline
Nebula & Galaxy &Spectrograph & Telescope & Reference \\
\hline
He 2-10 & He 2-10 & UVES & VLT  &  \citet{Esteban:2014}\\
\hline
30 Doradus & \multirow{5}{*}{LMC}  & \multirow{5}{*}{UVES} & \multirow{5}{*}{VLT} & \citet{Peimbert:2003}\\
IC 2111a  &  &    & &\multirow{4}{*}{\citet{dominguez:2022}}\\
N11Bb &  &   &   & \\
N44Cb&  &   &   & \\
NGC 1714a&  &   &   & \\
\hline
BA289  &\multirow{8}{*}{M31} & \multirow{7}{*}{OSIRIS}  &  \multirow{7}{*}{GTC} & \multirow{7}{*}{\citet{Esteban:2020}}\\
BA310  &  &  &  & \\
BA371  &  &  &  & \\
BA374  &  &  &  & \\
BA379  &  &  &  & \\
K160  &  &  &  & \\
K703  &  &  &  & \\
K932b  &  & HIRES & KECK & \citet{Esteban:2009}\\
\hline
B2011 b5 & \multirow{13}{*}{M33} & \multirow{11}{*}{UVES} & \multirow{11}{*}{VLT} & \multirow{11}{*}{\citet{Toribio:2016}}\\
B2011 b15 &  &   &   &  \\
BCLMP 29 &  &   &   &  \\
BCLMP 88w &  &   &   & \\
BCLMP 290 & & & & \\
BCLMP 626 &  &   &   & \\
IC 131 &  &   &   & \\
IC 132 &  &   &   & \\
NGC 588 &  &   &   & \\
LGC H II3 &  &   &   & \\
LGC H II11 &  &   &   & \\
NGC 595 &  &  \multirow{2}{*}{HIRES} & \multirow{2}{*}{KECK}&\multirow{2}{*}{\citet{Esteban:2009}}\\
NGC 604 &  &   &   & \\
\hline
Nucleus 13 & M83 & HIRES & KECK  &  \citet{Esteban:2009}\\
\hline
H37 &\multirow{14}{*}{M101} & \multirow{11}{*}{OSIRIS}  &  \multirow{11}{*}{GTC} & \multirow{11}{*}{\citet{Esteban:2020}}\\
H219 &  &  &  & \\
H681 &  &  &  & \\
H1146 &  &  &  & \\
H1216 &  &  &  & \\
H1118 &  &  &  & \\
NGC 5447 &  &  &  & \\
NGC 5455 &  &  &  & \\
NGC 5462 &  &  &  & \\
NGC 5471 &  &  &  & \\
SDH323 &  &  &  & \\
NGC 5447 & &ISIS&WHT&\multirow{3}{*}{\citet{Esteban:2009}}\\
H1013g & &\multirow{2}{*}{HIRES} & \multirow{2}{*}{KECK}&\\
NGC 5461  &  &  &  & \\
\hline
Mrk 1271 & Mrk 1271 & UVES & VLT  &  \citet{Esteban:2014}\\
\hline
R2 &\multirow{7}{*}{NGC 300} & \multirow{7}{*}{UVES}  &  \multirow{7}{*}{VLT} & \multirow{7}{*}{\citet{Toribio:2016}}\\
R5  &  &  &  & \\
R14 &  &  &  & \\ 
R20 &  &  &  & \\
R23 &  &  &  & \\
R27 &  &  &  & \\
R76a &  &  &  & \\
\hline
Zone C & NGC 1741 & \multirow{2}{*}{HIRES} & \multirow{2}{*}{KECK}  & \multirow{2}{*}{ \citet{Esteban:2009}}\\
NGC 2363 (Mrk 71) & NGC 2366 &   &   &  \\
\hline

\end{tabular}
\end{table*}

\begin{table*}
\ContinuedFloat
\caption{DESIRED Extragalactic \ion{H}{2} regions (continued).}
\begin{tabular}{ccccccccccccc}
\hline
Nebula & Galaxy &Spectrograph & Telescope & Reference \\
\hline
VS 24d &\multirow{3}{*}{NGC 2403} & \multirow{3}{*}{HIRES}& \multirow{3}{*}{KECK}&\multirow{3}{*}{\citet{Esteban:2009}}\\
VS 38d &  &  &  & \\
VS 44d &  &  &  & \\
\hline
NGC 3125 & NGC 3125 & UVES & VLT  &  \citet{Esteban:2014}\\
Region 70 & NGC 4395 &  \multirow{2}{*}{HIRES} & \multirow{2}{*}{KECK}  &  \multirow{2}{*}{\citet{Esteban:2009}}\\
Brightest \ion{H}{2} region & NGC 4861 & &  & \\
\hline
\ion{H}{2}-1&\multirow{4}{*}{NGC 5253} & \multirow{4}{*}{UVES} &\multirow{4}{*}{VLT}&\multirow{4}{*}{\citet{lopezsanchez:2007}}\\
\ion{H}{2}-2 &  &  &  & \\
UV-1&  &  &  & \\
UV-2&  &  &  & \\
\hline
NGC 5408 & NGC 5408 &  \multirow{4}{*}{UVES} & \multirow{4}{*}{VLT}  &  \multirow{4}{*}{\citet{Esteban:2014}}\\
 NGC 6822 & NGC 6822 & &  &  \\
 POX 4 & POX 4 & &  &  \\
SDSS J1253-0312 & SDSS J1253-0312 & &  & \\
\hline
N66Ab  & \multirow{4}{*}{SMC} & \multirow{4}{*}{UVES} & \multirow{4}{*}{VLT} &\multirow{4}{*}{\citet{dominguez:2022}}\\
N81b  &  &   &   & \\
N88Ab  &  &   &   & \\
N90b &  &   &   & \\
\hline
Tol 1457-262 & Tol 1457-262 & \multirow{2}{*}{UVES} & \multirow{2}{*}{VLT} &  \multirow{2}{*}{\citet{Esteban:2014}}\\
Tol 1924-416 & Tol 1924-416 & &  & \\
\hline
\end{tabular}
\end{table*}

\begin{table*}
\caption{DESIRED Galactic Planetary Nebulae.}
\label{tab:PNes_references}
\begin{tabular}{ccccccccccccc}
\hline
Nebula & Spectrograph & Telescope & Reference \\
\hline
Abell 46 & \multirow{2}{*}{ISIS} & \multirow{2}{*}{WHT} & \multirow{2}{*}{\citet{Corradi:2015}}\\
Abell 63 & & &\\
Cn 1-5 & MIKE & Clay Telescope & \citet{garciarojas:2012}\\
H 1-40 & \multirow{2}{*}{UVES} & \multirow{2}{*}{VLT}&\multirow{2}{*}{\citet{garciarojas:2018}}\\
H 1-50  & & &\\
Hb 4 & MIKE & Clay Telescope & \citet{garciarojas:2012}\\
He 2-73 & UVES& VLT&\citet{garciarojas:2018}\\
He 2-86 & MIKE & Clay Telescope & \citet{garciarojas:2012}\\
He 2-96 & \multirow{2}{*}{UVES} & \multirow{2}{*}{VLT}&\multirow{2}{*}{\citet{garciarojas:2018}}\\
He 2-158 & & &\\
IC 418 &  Blanco echelle &  CTIO Blanco & \citet{Sharpee:2003}\\
IC 2501 &  \multirow{2}{*}{MIKE} & \multirow{2}{*}{Baade Telescope} & \multirow{2}{*}{\citet{Sharpee:2007}}\\
IC 4191 &   &   & \\
IC 4776& UVES & VLT & \citet{Sowicka:2017}\\
M 1-25 &\multirow{2}{*}{MIKE} & \multirow{2}{*}{Clay Telescope} & \multirow{2}{*}{\citet{garciarojas:2012}}\\
M 1-30  &   &   & \\
M 1-31 & UVES& VLT&\citet{garciarojas:2018}\\
M 1-32 & MIKE & Clay Telescope & \citet{garciarojas:2012}\\
M 1-33 &\multirow{2}{*}{UVES} & \multirow{2}{*}{VLT}&\multirow{2}{*}{\citet{garciarojas:2018}}\\
M 1-60 &   &   & \\
M 1-61 & MIKE & Clay Telescope & \citet{garciarojas:2012}\\
M 2-31 & \multirow{2}{*}{UVES} & \multirow{2}{*}{VLT}&\citet{garciarojas:2018}\\
M 2-36 & & & \citet{espiritu:2021}\\
M 3-15 & MIKE & Clay Telescope & \citet{garciarojas:2012}\\
NGC 2440 &  MIKE &  Baade Telescope & \citet{Sharpee:2007}\\
NGC 3918 &  UVES & VLT & \citet{garciarojas:2015}\\
NGC 5189 & MIKE & Clay Telescope & \citet{garciarojas:2012}\\
NGC 5315 & UVES & VLT & \citet{Madonna:2017}\\
NGC 6369 & MIKE & Clay Telescope & \citet{garciarojas:2012}\\
NGC 6778 & FORS2 & VLT & \citet{Jones:2016}\\
NGC 7027 &  Mayall echelle  &  KPNO Mayall Telescope & \citet{Sharpee:2007}\\
Ou5 & ISIS & WHT & \citet{Corradi:2015}\\
PC 14 &\multirow{2}{*}{MIKE} & \multirow{2}{*}{Clay Telescope} & \multirow{2}{*}{\citet{garciarojas:2012}}\\
Pe 1-1  &   &   & \\
\hline
\end{tabular}
\end{table*}

\begin{table*}
\caption{DESIRED Galactic \ion{H}{2} regions.}
\label{tab:Galactic_references}
\begin{tabular}{ccccccccccccc}
\hline
Nebula & Zone & Spectrograph & Telescope & Reference \\
\hline
\multirow{3}{*}{IC 5146} &  93.8'' from  BD+46 3474 (A2) & \multirow{3}{*}{ISIS} & \multirow{3}{*}{WHT} &\multirow{3}{*}{\citet{GarciaRojas:2014}}\\
  &   115.8'' from  BD+46 3474 (A3) &  &  & \\
  &  137.8'' from  BD+46 3474 (A4) &  &  & \\
\hline

M8 & 12'' S of the center of the Hourglass & \multirow{4}{*}{UVES} &\multirow{4}{*}{VLT}  & \citet{GarciaRojas:2007a}\\
M16 & 48'' N and 40'' W of BD-13 4930 & & & \citet{GarciaRojas:2006}\\
M17 & 300'' S and 72'' E of the center of BD-164819 & & & \citet{GarciaRojas:2007a}\\
M20 & 17'' N and 10'' E of HD164492 & & & \citet{GarciaRojas:2006}\\

\hline
\multirow{3}{*}{M43} &  51.55'' from  HD 37061 (A4) & \multirow{3}{*}{ISIS} & \multirow{3}{*}{WHT} &\multirow{3}{*}{\citet{Simon-Diaz:2011}}\\
  &   68.05'' from  HD 37061 (A5) &  &  & \\
  &  84.55'' from  HD 37061 (A6) &  &  & \\
\hline

NIL &  154.3'' from $\theta^1$ Ori C (cut 2) &  \multirow{4}{*}{UVES} &\multirow{4}{*}{VLT} & \citet{mendezdelgado:2021b}\\
NGC 2579 & 5'' N of DENIS J082054.8-361258 & & & \citet{Esteban:2013}\\
NGC 3576 & 24'' N and 65'' W of HD 97499 & & & \citet{garciarojas:2004}\\
NGC 3603 & 12'' N and 116'' E of HD 306201& & & \citet{GarciaRojas:2006}\\
\hline
\multirow{12}{*}{Orion Nebula} &  29.4'' from $\theta^1$ Ori C & \multirow{12}{*}{UVES} & \multirow{12}{*}{VLT} &\citet{Esteban:2004}\\
  &  45.0'' from $\theta^1$ Ori C  &   &    & \multirow{2}{*}{\citet{DelgadoInglada:2016}}\\
  &  109.4'' from $\theta^1$ Ori C (Orion Bar)  &   &    & \\
  &  75.5'' from $\theta^1$ Ori C  &   &    & \citet{MesaDelgado:2009}\\
  &   34.6'' from $\theta^1$ Ori C (cut 1) &  &   &\multirow{4}{*}{\citet{mendezdelgado:2021a}}\\
  &   34.8'' from $\theta^1$ Ori C (cut 2)   &   &    &   \\
   &  35.3'' from $\theta^1$ Ori C (cut 3)  &   &    &   \\
   &  35.8'' from $\theta^1$ Ori C (cut 4) &   &    &   \\
  &   149.4'' from $\theta^1$ Ori C (cut 1) &  &   &\multirow{2}{*}{\citet{mendezdelgado:2021b}}\\
  &   154.3'' from $\theta^1$ Ori C (cut 2)   &   &    &   \\
 &   13.5'' from $\theta^1$ Ori C (cut 2) &  &   &\multirow{2}{*}{\citet{mendezdelgado:2022b}}\\
 &   9.5'' from $\theta^1$ Ori C (cut 3)   &   &    &   \\
\hline

Sh~2-29& 18:09:49.90 $-$24:04:14.5  & \multirow{15}{*}{OSIRIS} &\multirow{15}{*}{GTC}  & \multirow{6}{*}{\citet{arellano:2021}}\\

Sh~2-32& 18:09:52.72 $-$23:38:39.5    & & & \\

Sh~2-47& 18:18:16.02 $-$15:36:17.2  & & &  \\

Sh~2-48& 18:22:12.46 $-$14:36:19.1    & & & \\

Sh~2-53& 18:25:16.21 $-$13:09:12.2    & & & \\

Sh~2-54& 18:17:48.99 $-$11:43:45.4    & & & \\

Sh~2-61& 18:33:21.04 $-$04:57:55.7 & & & \citet{Esteban:2018}\\

Sh~2-82& 19:30:22.80 +18:15:14.4 & & & \citet{arellano:2021}\\

Sh~2-83& 19:24:52.55 +20:47:24.4 & & & \citet{Esteban:2017}\\

Sh~2-88B& 19:46:46.51 $-$25:12:39.4 & & & \citet{arellano:2021}\\

Sh~2-90& 19:49:09.33 +26:51:41.4 & & & \citet{Esteban:2018}\\

Sh~2-93 & 19:54:58.17 +27:12:45.3 & & & \citet{arellano:2021}\\

Sh~2-100&20:02:00.69 +33:29:23.9& & &    \multirow{3}{*}{\citet{Esteban:2017}}\\

Sh~2-127& 21:28:40.95 +54:42:13.9 & & & \\

Sh~2-128& 21:32:49.86 +55:53:16.2 & & & \\

Sh~2-152& 22:58:40.80 +58:46:53.8 & OSIRIS  & GTC & \citet{Esteban:2018}\\

Sh~2-175& 00:27:47.93 64:42:00.3 & \multirow{13}{*}{OSIRIS} &\multirow{13}{*}{GTC} & \citet{Esteban:2018}\\

Sh~2-209& 04:11:25.69 +51:14:33.8 & & & \multirow{2}{*}{\citet{Esteban:2017}}\\

Sh~2-212& 04:40:56.13 +50:26:53.1 & & & \\

Sh~2-219& 04:56:11.69 +47:23:49.4 & & &  \multirow{8}{*}{\citet{Esteban:2018}}\\

Sh~2-235& 05:41:03.12 +35:51:33.1 & & & \\

Sh~2-237& 05:31:26.10 +34:15:06.2 & & & \\

Sh~2-257& 06:12:43.99 +17:58:52.4 & & & \\

Sh~2-266& 06:18:46.14 +15:17:27.6 & & & \\

Sh~2-270& 06:10:12.75 +12:48:44.6 & & & \\

Sh~2-271& 06:14:53.89 +12:21:11.2 & & & \\

Sh~2-285& 06:55:16.49 $-$00:31:01.9 & & & \\

Sh~2-288& 07:08:48.90 +07:08:48.9 & & & \citet{Esteban:2017}\\

Sh~2-297& 07:05:16.03 $-$12:19:45.4  & & & \citet{Esteban:2018}\\

Sh~2-311& 126'' S of HD 64315  & UVES & VLT & \citet{garciarojas:2005}\\

\hline

\end{tabular}
\end{table*}

\begin{table*}
\caption{Average spectral resolution and coverage from the sampled regions. The precise spectral resolutions are found in the reference articles shown in Tables \ref{tab:photoionizedHHs}-\ref{tab:Galactic_references}.}
\label{tab:telescopes_refs}
\begin{tabular}{ccccccccccccc}
\hline
Instrument & Average R  & Spectral coverage & Telescope & No. spectra taken & Ins. Ref.\\
\hline
 \multirow{3}{*}{ISIS} & \multirow{3}{*}{$\sim 3000$} & $\lambda \lambda 4225-5075$ \& $\lambda \lambda 5430-8195$ \AA & \multirow{3}{*}{4.2m WHT} & \multirow{3}{*}{10}& \multirow{3}{*}{\citet{Jorden:1990}}\\
 &  & $\lambda \lambda 3610-9150$ \AA & & & &\\
 &  & $\lambda \lambda 3386-5102$ \& $\lambda \lambda 6064-7585$ \AA & & &\\

 MagE & $\sim 4000$ & $\lambda \lambda 3100-10000$ \AA & 6.5m Magellan Clay Telescope & 4 &\citet{Marshall:2008}\\
 
 \multirow{2}{*}{MIKE}   & $\sim 27000$ & $\lambda \lambda 3350-9400$ \AA & 6.5m Magellan Clay Telescope & 12 & \multirow{2}{*}{\citet{Bernstein:2003}}\\

  & $\sim 25000$ & $\lambda \lambda 3280-7580$ \AA & 6.5m Magellan Baade Telescope & 3 & \\

 Blanco echelle & $\sim 33000$ & $\lambda \lambda 3400-9700$ \AA  & 4m CTIO Blanco & 1 &\\

Mayall echelle & $\sim 15000$ & $\lambda \lambda 4600-9200$ \AA & 4m KPNO Mayall & 1 & \citet{York:1981} \\

HIRES & $\sim 23000$ & $\lambda \lambda 3550-7440$ \AA & 10m KECK I & 13 & \citet{Vogt:1994} \\

 \multirow{2}{*}{OSIRIS} & $\sim 1020$ & $\lambda \lambda 3600-7750$ \AA  & \multirow{2}{*}{10.4m GTC} & \multirow{2}{*}{63} & \multirow{2}{*}{\citet{Cepa:2000,Cepa:2003}} \\

& $\sim 2520$ & $\lambda \lambda 4430-6070$ \AA  & & &\\

UVES & $\sim 20000$ & $\lambda \lambda 3100-10420$ \AA  & 8.2m VLT Kueyen & 82 & \citet{DOdorico:2000}\\

FORS2 & $\sim 4000$ & $\lambda \lambda 3600-7200$ \AA  & 8.2m VLT Antu & 1 & \citet{Appenzeller:1998}\\
\hline
\end{tabular}
\end{table*}

\begin{table*}
\caption{Atomic data set used for collisionally excited lines.}
\label{tab:atomic_data}
\begin{tabular}{lcc}
\hline
\multicolumn{1}{l}{Ion} & \multicolumn{1}{c}{Transition Probabilities} &
\multicolumn{1}{c}{Collision Strengths} \\
\hline

O$^{+}$   &  \citet{FroeseFischer:2004} & \citet{Kisielius:2009}\\
O$^{2+}$  &  \citet{Wiese:1996}, \citet{Storey:2000} & \citet{Storey:2014}\\
N$^{+}$   &  \citet{FroeseFischer:2004} & \citet{Tayal:2011}\\
S$^{+}$   &  \citet{IFF05} & \citet{Tayal:2010}\\
S$^{2+}$  &  \citet{FroeseFischer:2006} & \citet{Grieve:2014}\\
Cl$^{2+}$ &  \citet{Fritzsche:1999} & \citet{Butler:1989}\\
Ar$^{2+}$ &   \citet{Mendoza:1983b}, \citet{Kaufman:1986}  & \citet{Galavis:1995}\\
Ar$^{3+}$ &   \citet{Mendoza:1982b}  & \citet{Ramsbottom:1997}\\
Fe$^{2+}$ & \citet{Quinet:1996}& \citet{Zhang:1996}\\
\hline
\end{tabular}
\end{table*}

\bsp	
\label{lastpage}
\end{document}